\pdfoutput=1
\documentclass[twocolumn,preprintnumbers,superscriptaddress,amsmath,amssymb,nofootinbib]{revtex4}

\usepackage{graphicx}
\usepackage{dcolumn}
\usepackage{bm}
\usepackage{amsmath, amssymb}
\usepackage{braket}

\allowdisplaybreaks[1]

\usepackage[normalem]{ulem}  
\usepackage[dvipdfmx]{color} 

\renewcommand\sout{\bgroup \color{red} \ULdepth=-.5ex \ULset}

\newcommand{\Psfig}[2]{\includegraphics[width=#1]{#2}}
\newcommand{\PsfigII}[2]{\includegraphics[scale=#1]{#2}}

\usepackage{siunitx}

\begin{document}

\preprint{}

\title{\boldmath Short-range baryon-baryon potentials in constituent quark model revisited}

\author{Takayasu Sekihara} 
\email{sekihara@post.j-parc.jp}
\affiliation{Graduate School of Life and Environmental Sciences,
  Kyoto Prefectural University, Sakyo-ku, Kyoto 606-8522, Japan}

\author{Taishi Hashiguchi}
\affiliation{Graduate School of Life and Environmental Sciences,
  Kyoto Prefectural University, Sakyo-ku, Kyoto 606-8522, Japan}

\date{\today}

\begin{abstract}

  We revisit the short-range baryon-baryon potentials in the flavor
  SU(3) sector, using the constituent quark model.  We employ the
  color Coulomb, linear confining, and color magnetic forces between
  two constituent quarks, and solve the three-quark Schr\"{o}dinger
  equation using the Gaussian expansion method to evaluate the wave
  functions of the octet $( N , \Lambda , \Sigma , \Xi )$ and decuplet
  $( \Delta , \Sigma ^{\ast} , \Xi ^{\ast} , \Omega )$ baryons.  We
  then solve the six-quark equation using the resonating group method
  and systematically calculate equivalent local potentials for the
  $S$-wave two-baryon systems which reproduce the relative wave
  functions of two baryons in the resonating group method.  As a
  result, we find that the flavor antidecuplet states with total spin
  $J = 3$, namely, $\Delta \Delta$, $\Delta \Sigma ^{\ast}$, $\Delta
  \Xi ^{\ast}$-$\Sigma ^{\ast} \Sigma ^{\ast}$, and $\Delta
  \Omega$-$\Sigma ^{\ast} \Xi ^{\ast}$ systems, have attractive
  potentials sufficient to generate dibaryon bound states as hadronic
  molecules.  In addition, the $N \Omega$ system with $J = 2$ in
  coupled channels has a strong attraction and forms a bound state.
  We also make a comparison with the baryon-baryon potentials from
  lattice QCD simulations and try to understand the behavior of the
  potentials from lattice QCD simulations.

\end{abstract}

\maketitle

\section{Introduction}

Understanding the baryon-baryon interactions has been an interesting
topic in hadron physics, as they provide important clues to the quark
dynamics inside baryons.  The nuclear force, being the most
extensively studied case, has been investigated through low-energy
nucleon-nucleon ($N N$) scattering data and the properties of the $N
N$ bound state, \textit{i.e.}, the deuteron.  Phenomenological $N N$
potentials, which precisely reproduce the $N N$ data, are known to
have a short-range (relative distance $r < \SI{1}{fm}$) repulsive core
and medium-range ($\SI{1}{fm} < r < \SI{2}{fm}$) and long-range ($r >
\SI{2}{fm}$) attractive parts~\cite{Machleidt:1989tm}.  While meson
exchanges can explain the medium- and long-range parts of the nuclear
force, quark degrees of freedom are expected to be significant in the
short range.  In fact, constituent quark model calculations indicate
that the short-range repulsive core of the nuclear force is governed
by two factors~\cite{Oka:2000wj}: the Pauli exclusion principle among
valence quarks, and the spin-spin interaction of the quarks that
causes the mass splitting between the nucleon and the $\Delta$ baryon.
To confirm this scenario in more general cases, studies of
baryon-baryon interactions with different quark content are desired.

Recently, due to experimental and numerical developments, much
attention has been paid to interactions between two baryons belonging
to the octet ($N$, $\Lambda$, $\Sigma$, and $\Xi$) and decuplet
($\Delta$, $\Sigma ^{\ast}$, $\Xi ^{\ast}$, and $\Omega$).  For
example, high statistics $\Sigma ^{-} p$ and $\Sigma ^{+} p$
scattering experiments were performed in
Refs.~\cite{J-PARCE40:2021qxa} and \cite{J-PARCE40:2022nvq},
respectively, and the nuclear $1 s$ state of the $\Xi$ hypernucleus
${}^{15}_{\Xi} \mathrm{C}$ was discovered in
Ref.~\cite{Yoshimoto:2021ljs}.  Both of these provide us with some
information on the $N \Sigma$ and $N \Xi$ interactions.  In addition
to scattering experiments, we can now use lattice quantum
chromodynamics (QCD) simulations and relativistic ion collisions to
study baryon-baryon interactions.  In lattice QCD simulations, we can
extract baryon-baryon local potentials directly from the quark-gluon
dynamics of QCD using the HAL QCD method~\cite{Aoki:2012tk}, which has
been applied to various systems including decuplet baryons,
\textit{e.g.}, $\Omega \Omega$~\cite{Gongyo:2017fjb}, $N
\Omega$~\cite{HALQCD:2018qyu}, and $\Delta \Delta$ (with heavy pion
mass)~\cite{Gongyo:2020pyy}.  Such baryon-baryon potentials,
especially for unstable baryons, are studied through the analysis of
the correlation functions for any pair of baryons in relativistic ion
collisions~\cite{ALICE:2020mfd}, in which the large number of baryons,
together with theoretical predictions for the correlation
functions~\cite{Morita:2019rph}, enables a detailed determination of
the baryon-baryon interactions.  Furthermore, besides phenomenological
models, baryon-baryon interactions are now theoretically calculated
through chiral effective field theory, in which the degrees of freedom
are tied to QCD symmetries and their realization: 
hyperon-nucleon interactions~\cite{Haidenbauer:2019boi} and
interactions involving decuplet baryons~\cite{Haidenbauer:2017sws}, as
well as the nuclear force~\cite{Machleidt:2011zz}.

Motivated by these studies, in the present paper we aim to
systematically study the baryon-baryon interactions, particularly
focusing on the short-range part, by using a precise wave function for
baryons composed of three nonrelativistic constituent quarks.  In this
sense, our study is an extension of the quark model studies in
Refs.~\cite{Oka:1981ri, Oka:1981rj}, but our calculation covers the
interactions of any pair of the ground-state baryons, \textit{i.e.},
the octet and decuplet baryons.  Similar studies are found in,
\textit{e.g.}, Refs.~\cite{Oka:1986fr, Goldman:1987ma, Oka:1988yq,
  Wang:1995bg, Zhang:1997ny, Li:1999bc, Li:2000cb, Fujiwara:2006yh,
  Park:2016cmg, Park:2019bsz}.  Our study will provide some clues to
understand the mechanism that generates attractive/repulsive force
suggested in experiments and lattice QCD simulations.  Furthermore,
because there are more than one hundred channels of the $S$-wave
two-baryon systems from the ground-state baryons, we may expect
attractive two-baryon potentials that are sufficient to generate
dibaryon bound states in the systematic study.

We evaluate the relative wave function of constituent quarks inside
each baryon as the solution of the three-quark Schr\"{o}dinger equation in the
Gaussian expansion method~\cite{Hiyama:2003cu}. We take into account
the color Coulomb, linear confining, and color magnetic forces between
two constituent quarks.  Then, we employ the resonating group method
(RGM) to calculate the relative wave function of two baryons in $S$
wave.  We translate the relative wave function of two baryons into the
equivalent local potentials that reproduce the relative wave functions
of two baryons in the RGM.  Thanks to this approach, we can compare
our results with the baryon-baryon local potentials deduced in the
lattice QCD simulations, and any research group can utilize the local
potentials for further investigations of two-baryon systems.

The paper is organized as follows.  In Sec.~\ref{sec:2} we formulate
the constituent quark model for one-baryon and two-baryon systems.  We
also explain the method used to evaluate the equivalent local
potentials in this section.  Next, in Sec.~\ref{sec:3} we present our
numerical results of the baryon-baryon potentials and dibaryon bound
states in the present model.  Section~\ref{sec:4} is devoted to the
conclusion of the present study.

\section{\boldmath Formulation}
\label{sec:2}

\subsection{Baryons in the Gaussian expansion method}
\label{sec:2A}

First of all, we construct the wave function of each baryon in the
three-body dynamics of constituent quarks.

In the present study, we employ the color Coulomb, linear confining,
and color magnetic forces between two constituent quarks.  In general,
the $i$-th and $j$-th quarks at a distance $r$ interact via the
potential
\begin{equation}
  V_{i j} ( r )
  = \frac{\vec{\lambda}_{i}}{2} \cdot \frac{\vec{\lambda} _{j}}{2}
  \left [ \frac{\alpha _{i j}}{r}
    - \frac{3}{4} k r + D
    - \frac{2 \pi \alpha _{\rm ss}}{3}
    \frac{\vec{\sigma} _{i} \cdot \vec{\sigma}_{j}}{m_{i} m_{j}}
    \bar{\delta} ( r )
    \right ] ,
  \label{eq:qq_pot}
\end{equation}
where $\vec{\lambda}_{i}$ and $\vec{\sigma}_{i}$ are sets of the
Gell-Mann and Pauli matrices, respectively, acting on the $i$-th
quark, $\alpha _{i j}$ and $\alpha _{\rm ss}$ are the coupling
constants, $k$ is the confining string tension, $D$ is a constant to
reproduce the physical baryon masses, $m_{i}$ is the $i$-th
constituent quark mass, and $\bar{\delta} ( r )$ is a
three-dimensional delta-like function
\begin{equation}
  \bar{\delta} ( r )
  \equiv \left ( \frac{\sigma}{\sqrt{\pi}} \right )^{3}
  \exp \left ( - \sigma ^{2} r^{2} \right ) ,
\end{equation}
with a range parameter $\sigma$.  The coupling constant for the color
Coulomb force $\alpha _{i j}$ is assumed to depend on the reduced mass
of the $i$-$j$ quark pair in the following form:
\begin{equation}
  \alpha _{i j} \equiv \frac{K}{\mu _{i j}} ,
  \quad
  \mu _{i j}
  \equiv \frac{m_{i} m_{j}}{m_{i} + m_{j}} ,
\end{equation}
where $K$ is a constant.  The quark mass dependence of the color
Coulomb coupling constant was initially suggested based on a Lattice
QCD calculation~\cite{Kawanai:2011xb} and has been employed in a quark
model calculation~\cite{Yoshida:2015tia}.  We treat $K$, $\alpha _{\rm
  ss}$, $k$, $\sigma$, and $D$ as model parameters, while keeping the
constituent quark mass $m_{i}$ fixed to reproduce the magnetic moments
of the proton and $\Lambda$: $m_{u} = m_{d} = \SI{336}{MeV}$ and
$m_{s} = \SI{509}{MeV}$, as reported by the Particle Data
Group~\cite{ParticleDataGroup:2022pth}.  Throughout this study we
assume isospin symmetry.

\begin{figure}[!t]
  \centering
  \PsfigII{1.0}{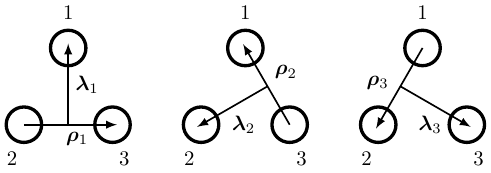}
  \caption{Jacobi coordinates of a three-body system.}
  \label{fig:Jacobi}
\end{figure}

We apply the potential~\eqref{eq:qq_pot} to the one-baryon system
($B$) in the constituent quark model, in which the internal
configurations of three quarks can be described by the Jacobi
coordinates shown in Fig.~\ref{fig:Jacobi}.  Inside a baryon, owing to
the color configuration of constituent quarks, the $i$-th and $j$-th
quarks satisfy the following relation:
\begin{equation}
  \frac{\vec{\lambda}_{i}}{2} \cdot \frac{\vec{\lambda}_{j}}{2}
  = - \frac{2}{3} .
\end{equation}
Hence, two quarks at a distance $\rho$ inside the baryon $B$ interact
via the potential
\begin{equation}
  V_{i j}^{(B)} ( \rho )
  = - \frac{2}{3} \frac{\alpha _{i j}}{\rho}
  + \frac{1}{2} k \rho
  - \frac{2}{3} D 
  + \frac{4 \pi \alpha _{\rm ss}}{9}
  \frac{\vec{\sigma}_{i} \cdot \vec{\sigma}_{j}}{m_{i} m_{j}}
  \bar{\delta} ( \rho ) .
\end{equation}
Then, the Schr\"{o}dinger equation for the quarks inside the baryon $B$ in
the constituent quark model becomes
\begin{align}
  & \left [ m_{1} + m_{2} + m_{3} - \frac{1}{2 \mu _{B}}
    \frac{\partial ^{2}}{\partial \bm{\lambda} ^{2}}
    - \frac{1}{2 \mu _{B}^{\prime}}
    \frac{\partial ^{2}}{\partial \bm{\rho} ^{2}}
    \right .
    \notag \\
    & \left . \phantom{\frac{1}{\mu _{\lambda}^{( B )}}}
    + V_{2 3}^{( B )} ( \rho _{1}) + V_{3 1}^{( B )} ( \rho _{2}) 
    + V_{1 2}^{( B )} ( \rho _{3})
    \right ]
  \Psi ^{( B )} ( \bm{\lambda} , \bm{\rho} )
  \notag \\ &
  = M_{B} \Psi ^{( B )} ( \bm{\lambda} , \bm{\rho} ) ,
  \label{eq:Schr}
\end{align}
where $\Psi ^{( B )} ( \bm{\lambda} , \bm{\rho} )$ is the wave function of
the relative motion of three quarks in the baryon $B$, $\bm{\lambda}
\equiv \bm{\lambda}_{1}$, $\bm{\rho} \equiv \bm{\rho}_{1}$, $M_{B}$ is
the mass of the baryon $B$, and
\begin{equation}
  \mu _{B} \equiv \frac{m_{1} ( m_{2} + m_{3} )}{m_{1} + m_{2} + m_{3}} ,
  \quad
  \mu _{B}^{\prime} \equiv \frac{m_{2} m_{3}}{m_{2} + m_{3}} .
\end{equation}

In the present study, we focus on the ground-state baryons.
Therefore, both the $\lambda$ and $\rho$ modes of the three
constituent quarks have zero orbital angular momenta: $l_{\lambda} =
l_{\rho} = 0$.  To describe this, we employ the Gaussian expansion
method~\cite{Hiyama:2003cu} for the wave function $\Psi ^{( B )} (
\bm{\lambda} , \bm{\rho} )$:
\begin{equation}
  \Psi ^{( B )} ( \bm{\lambda} , \bm{\rho} )
  = \sum _{c = 1}^{3} \sum _{n = 1}^{N} \sum _{n^{\prime} = 1}^{N}
  C_{c, n, n^{\prime}}^{( B )}
  \exp \left ( - \frac{\lambda _{c}^{2}}{r_{n}^{2}}
     - \frac{\rho _{c}^{2}}{r_{n^{\prime}}^{2}} \right ) .
\end{equation}
Here, the index $c$ specifies the Jacobi coordinates in
Fig.~\ref{fig:Jacobi} and range parameters $r_{n}$ ($n = 1$, $\ldots$,
$N$) form a geometric progression:
\begin{equation}
  r_{n} = r_{\rm min} \times
  \left ( \frac{r_{\rm max}}{r_{\rm min}} \right )^{(n - 1) / (N - 1)} ,
\end{equation}
where the minimal and maximal ranges, $r_{\rm min}$ and $r_{\rm max}$,
respectively, are fixed according to the physical condition of the
system.  Then, by using the method summarized in
Ref.~\cite{Hiyama:2003cu}, we numerically solve the Schr\"{o}dinger
equation~\eqref{eq:Schr} and obtain the eigenvector $C_{c, n,
  n^{\prime}}^{( B )}$ as well as the eigenvalue $M_{B}$.

\begin{table}[!t]
  \caption{Model parameters for the baryons.}
  \label{tab:para}
  \centering
  \begin{ruledtabular}
    \begin{tabular}{ll}
      $K$ &
      $\SI{184}{MeV}$
      \\
      $\alpha _{\rm ss}$ &
      $0.785$
      \\
      $k$ &
      $\SI{0.755}{GeV/fm}$
      \\
      $\sigma$ &
      $\SI{3.50}{fm^{-1}}$
      \\
      $D$ &
      $\SI{381}{MeV}$
      \\
      $m_{u} = m_{d}$ &
      $\SI{336}{MeV}$ (fixed)
      \\
      $m_{s}$ & 
      $\SI{509}{MeV}$ (fixed)
      \\
    \end{tabular}
  \end{ruledtabular}
\end{table}

In this study, the model parameters are determined by fitting the
ground-state baryon masses.  The fitted parameters are listed in
Table~\ref{tab:para}, and the resulting baryon masses are listed in
the second column of Table~\ref{tab:mass}, along with their
experimental values~\cite{ParticleDataGroup:2022pth} in parenthesis.
The convergence of the results is found to be good with the number of
the expansion $N = 10$ and the ranges $r_{\rm min} = \SI{0.01}{fm}$
and $r_{\rm max} = \SI{2}{fm}$.

\begin{table}[!t]
  \caption{Properties of the baryons in the present model.  The baryon
    masses reported by the Particle Data
    Group~\cite{ParticleDataGroup:2022pth} are written in
    parenthesis.}
  \label{tab:mass}
  \centering
  \begin{ruledtabular}
    \begin{tabular}{lcc}
      Baryon & $M_{B}$ [MeV] & $\sqrt{\braket{r_{B}^{2}}}$ [fm]
      \\
      \hline
      $N$ & $\phantom{0}$950 $\phantom{0}$(939) & 0.43
      \\
      $\Lambda$ & 1111 (1116) & 0.42
      \\
      $\Sigma$ & 1180 (1193) & 0.44
      \\
      $\Xi$ & 1322 (1318) & 0.41
      \\
      $\Delta$ & 1235 (1232) & 0.51 
      \\
      $\Sigma ^{\ast}$ & 1382 (1385) & 0.49
      \\
      $\Xi ^{\ast}$ & 1530 (1533) & 0.47
      \\
      $\Omega$ & 1679 (1672) & 0.45
      \\
    \end{tabular}
  \end{ruledtabular}
\end{table}

\begin{figure}[!t]
  \centering
  \Psfig{8.6cm}{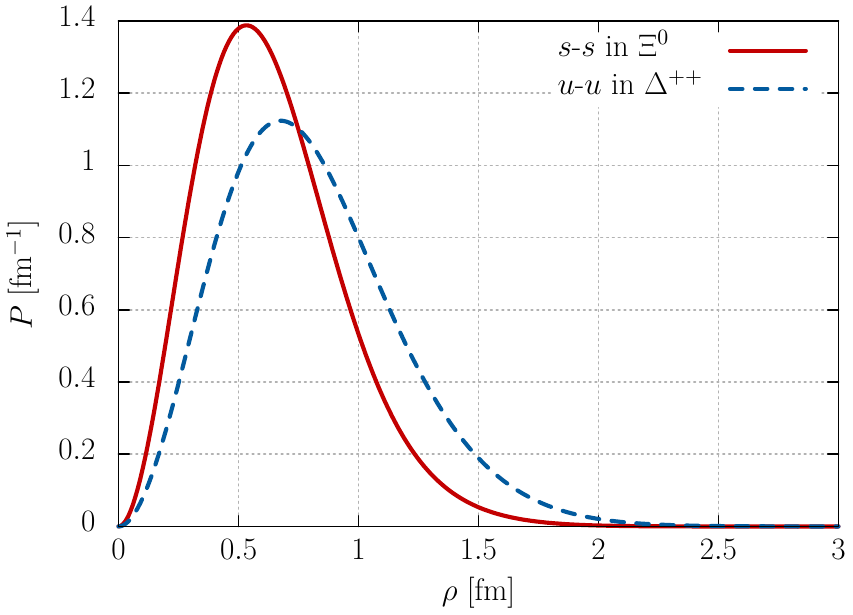}
  \caption{Examples of the density distribution $P ( \rho )$.}
  \label{fig:Prho}
\end{figure}

To evaluate the spatial extension of quarks inside each baryon, we
calculate the mean squared radius of the baryon using the formula
\begin{align}
  \braket{r_{B}^{2}}
  \equiv &
  \frac{1}{3 ( m_{1} + m_{2} + m_{3} )^{2}}
  \left [ ( m_{2} + m_{3} )^{2} \braket{\lambda _{1}^{2}} \right .
  \notag \\ 
  & \quad \left .
  + ( m_{3} + m_{1} )^{2} \braket{\lambda _{2}^{2}} 
  + ( m_{1} + m_{2} )^{2} \braket{\lambda _{3}^{2}} \right ] ,
\end{align}
where $\braket{\lambda _{c}^{2}}$ is the expectation value of $\lambda
_{c}^{2}$:
\begin{equation}
  \braket{\lambda _{c}^{2}}
  \equiv \int d^{3} \rho \int d^{3} \lambda
  \, \lambda _{c}^{2}
  \left | \Psi ^{( B )} ( \bm{\lambda}, \bm{\rho} ) \right | ^{2} .
\end{equation}
The resulting root mean squared radii of baryons, listed in the third
column of Table~\ref{tab:mass}, are smaller than the experimental
values: for instance, the experimental value of the proton charge
radius is about $\SI{0.84}{fm}$~\cite{ParticleDataGroup:2022pth}.
This discrepancy is attributed to the fact that we only consider the
spatial extension of the ``quark core'' and do not take into account
the meson clouds of baryons.  It is instructive to show the
distribution of quark-quark distance inside the baryons, which we
define as
\begin{equation}
  P ( \rho ) = \rho ^{2} \int d \Omega _{\rho} \int d^{3} \lambda
  \left | \Psi ^{( B )}( \bm{\lambda}, \bm{\rho} ) \right | ^{2} .
\end{equation}
Here we choose the distribution of the distance between the $s$-$s$
quarks in the $\Xi ^{0}$ baryon and the $u$-$u$ quarks in the $\Delta
^{++}$ baryon, because the root mean squared radius
$\sqrt{\braket{r_{B}^{2}}}$ of the $\Xi$ baryon has the minimal value
$\SI{0.41}{fm}$, while that of the $\Delta$ baryon has the maximal
value $\SI{0.51}{fm}$.  The resulting distribution is shown in
Fig.~\ref{fig:Prho} as the solid and dashed lines, respectively.  From
the figure, we can see that the quark-quark distance in a baryon is
about less than $\SI{2}{fm}$.  Because three quarks in a baryon form
an almost equilateral triangle, the distribution indicates that the
quarks are distributed within a range $\sim ( 2 / \sqrt{3} ) \,
\si{fm} \approx \SI{1.2}{fm}$ from the center of mass of the baryon.

\subsection{Two-baryon systems in the resonating group method}
\label{sec:2B}

\subsubsection{Creation and annihilation operators of quarks}

Next, we formulate the two-baryon systems in terms of the six-quark
degrees of freedom.  For this purpose, we introduce creation and
annihilation operators of a quark with quantum numbers $\mu \equiv ( f
, s , c )$, where $f$, $s$, and $c$ represent the flavor, spin, and
color, respectively:
\begin{equation}
  \hat{a}_{\mu}^{\dagger} ,
  \quad
  \hat{a}_{\mu} .
  \label{eq:a_oper}
\end{equation}
These operators satisfy the anticommutation relations:
\begin{equation}
  \{ \hat{a}_{\mu ^{\prime}} , \hat{a}_{\mu}^{\dagger} \}
  = \delta _{\mu ^{\prime} , \mu} , 
  \quad 
  \{ \hat{a}_{\mu ^{\prime}} , \hat{a}_{\mu} \}
  = 0 ,
  \quad
  \{ \hat{a}_{\mu ^{\prime}}^{\dagger} , \hat{a}_{\mu}^{\dagger} \}
  = 0 .
\end{equation}
We also introduce creation and annihilation operators of a quark at
the coordinate $\bm{r}$:
\begin{equation}
  \hat{b}_{\mu}^{\dagger} ( \bm{r} ) ,
  \quad
  \hat{b}_{\mu} ( \bm{r} ) ,
\end{equation}
which satisfy the anticommutation relations:
\begin{equation}
  \begin{split}
  & \{ \hat{b}_{\mu ^{\prime}} ( \bm{r}^{\prime} ) ,
  \hat{b}_{\mu}^{\dagger} ( \bm{r} ) \}
  = \delta _{\mu ^{\prime} , \mu} \delta ( \bm{r}^{\prime} - \bm{r} ) , 
  \\
  &
  \{ \hat{b}_{\mu ^{\prime}} ( \bm{r}^{\prime} ) ,
  \hat{b}_{\mu} ( \bm{r} ) \}
  = 0 ,
  \quad
  \{ \hat{b}_{\mu ^{\prime}}^{\dagger} ( \bm{r}^{\prime} ) ,
  \hat{b}_{\mu}^{\dagger} ( \bm{r} ) \}
  = 0 .
  \end{split}
\end{equation}
By using the creation operators of quarks, we can express the ket
vector of the one-baryon $B$ state as:
\begin{align}
  \ket{B} = & \sum _{\vec{\mu}} w_{\vec{\mu}}^{( B )}
  \int d^{3} r_{1} d^{3} r_{2} d^{3} r_{3}
  \psi _{1}^{( B )} ( \bm{r}_{1} ) \psi _{2}^{( B )} ( \bm{r}_{2} )
  \psi _{3}^{( B )} ( \bm{r}_{3} )
  \notag \\
  & \times
  \hat{b}_{\mu _{1}}^{\dagger} ( \bm{r}_{1} )
  \hat{b}_{\mu _{2}}^{\dagger} ( \bm{r}_{2} )
  \hat{b}_{\mu _{3}}^{\dagger} ( \bm{r}_{3} ) \ket{0} ,
\end{align}
where $\vec{\mu} \equiv ( \mu _{1} , \mu _{2} , \mu _{3} )$ is the set
of the quantum numbers of the three quarks, $w_{\vec{\mu}}^{( B )}$ is
the weight for the set $\vec{\mu}$, $\psi _{i}^{( B )} ( \bm{r}_{i} )$
is the spatial wave function of the $i$-th quark, and $\ket{0}$ is the
vacuum.  The weight $w_{\vec{\mu}}^{( B )}$ takes, for example $\Delta
^{++}$ with the third component of the spin $s = 3/2$, value:
\begin{equation}
  w_{\vec{\mu}}^{( \Delta ^{++} ( 3 / 2 ) )}
  =
  \begin{cases}
    1 & \vec{\mu}
    = ( ( u , \uparrow , \mathrm{R} ) ,
    ( u , \uparrow , \mathrm{G} ) , ( u , \uparrow , \mathrm{B} ) ) ,
    \\
    0 & \text{others} .
  \end{cases}
\end{equation}
We assume that the weight is normalized:
\begin{equation}
  \sum _{\vec{\mu}} \left [ w_{\vec{\mu}}^{( B )} \right ] ^{2} = 1 .
  \label{eq:Wnorm}
\end{equation}
Weights for other baryons are summarized in Table~\ref{tab:weight} in
Appendix.  Then, we decompose the product of the spatial wave
functions into the center-of-mass part $\Phi ^{( B )} ( \bm{R} )$ with
the center-of-mass coordinate $\bm{R}$ and the relative part $\Psi ^{(
  B )} ( \bm{\lambda} , \bm{\rho})$ as
\begin{equation}
  \psi _{1}^{( B )} ( \bm{r}_{1} ) \psi _{2}^{( B )} ( \bm{r}_{2} )
  \psi _{3}^{( B )} ( \bm{r}_{3} )
  = \Phi ^{( B )} ( \bm{R} ) \Psi ^{( B )} ( \bm{\lambda} , \bm{\rho} )
\end{equation}
where $\bm{R}$, $\bm{\lambda}$, $\bm{\rho}$ are expressed as
\begin{equation}
  \begin{split}
    & \bm{R} \equiv \frac{m_{1} \bm{r}_{1} + m_{2} \bm{r}_{2} + m_{3} \bm{r}_{3}}
    {m_{1} + m_{2} + m_{3}} ,
    \\
    & \bm{\lambda} \equiv \bm{r}_{1}
    - \frac{m_{2} \bm{r}_{2} + m_{3}\bm{r}_{3}}{m_{2} + m_{3}} ,
    \quad
    \bm{\rho} \equiv \bm{r}_{3} - \bm{r}_{2} .
  \end{split}
\end{equation}
Because the measure of the coordinates satisfies the relation
\begin{equation}
  d^{3} r_{1} d^{3} r_{2} d^{3} r_{3}
  = d^{3} R d^{3} \lambda d^{3} \rho ,
\end{equation}
we rewrite the ket vector of the one-baryon state as
\begin{align}
  \ket{B} = & 
  \int d^{3} R \Phi ^{( B )} ( \bm{R} ) \int d^{3} \lambda d^{3} \rho
  \Psi ^{( B )} ( \bm{\lambda} , \bm{\rho} )
  \notag \\
  & \times
  \hat{W}^{( B ) \dagger} ( \bm{r}_{1} , \bm{r}_{2} , \bm{r}_{3} ) \ket{0} ,
\end{align}
where we introduced an operator
\begin{equation}
  \hat{W}^{( B ) \dagger} ( \bm{r}_{1} , \bm{r}_{2} , \bm{r}_{3} )
  \equiv \sum _{\vec{\mu}} w_{\vec{\mu}}^{( B )}
  \hat{b}_{\mu _{1}}^{\dagger} ( \bm{r}_{1} )
  \hat{b}_{\mu _{2}}^{\dagger} ( \bm{r}_{2} )
  \hat{b}_{\mu _{3}}^{\dagger} ( \bm{r}_{3} ) .
\end{equation}
Provided the normalization of the wave functions 
\begin{equation}
  \int d^{3} R \left | \Phi ^{( B )} ( \bm{R} ) \right | ^{2}
  =
  \int d^{3} \lambda d^{3} \rho
  \left | \Psi ^{( B )} ( \bm{\lambda} , \bm{\rho} ) \right |^{2}
  = 1 ,
\end{equation}
together with the normalization of the weight~\eqref{eq:Wnorm}, the
one-baryon vector is normalized:
\begin{equation}
  \braket{B | B} = 1 .
\end{equation}

\subsubsection{Hamiltonian}

By using the creation and annihilation operators, we can express the
Hamiltonian of the system of quarks.  The Hamiltonian $\hat{H}$ is
composed of the kinetic part $\hat{K}$ and potential part $\hat{V}$:
\begin{equation}
  \hat{H} = \hat{K} + \hat{V} .
\end{equation}
The kinetic part is expressed as
\begin{equation}
  \hat{K} = \sum _{\mu} \int d^{3} r \hat{b}_{\mu}^{\dagger} ( \bm{r} )
  \left ( m_{f} - \frac{1}{2 m_{f}}
  \frac{\partial ^{2}}{\partial \bm{r}^{2}} \right )
  \hat{b}_{\mu} ( \bm{r} ) ,
\end{equation}
where $m_{f}$ is the quark mass of the flavor $f$ and the differential
operator $\partial ^{2} / \partial \bm{r}^{2}$ acts on the wave
functions of quarks.  The potential part, on the other hand, is
composed of the color Coulomb plus linear confining potential and
color magnetic potential:
\begin{equation}
  \hat{V} = \hat{V}_{\rm CL} + \hat{V}_{\rm ss} .
\end{equation}
They are respectively expressed as
\begin{align}
  \hat{V}_{\rm CL}
  = & \frac{1}{2} \sum _{f, f^{\prime}, s, s^{\prime}}
  \int d^{3} r \int d^{3} r^{\prime}
  V_{\rm CL} ( | \bm{r} - \bm{r}^{\prime} | )
  \notag \\
  &
  \times 
  \left [ \hat{b}^{\dagger} ( \bm{r} )
  \frac{\vec{\lambda}}{2} \hat{b} ( \bm{r} ) \right ]_{f, s}
  \cdot
  \left [ \hat{b}^{\dagger} ( \bm{r}^{\prime} )
  \frac{\vec{\lambda}}{2} \hat{b} ( \bm{r}^{\prime} ) \right ]_{f^{\prime}, s^{\prime}}
  ,
\end{align}
\begin{align}
  \hat{V}_{\rm ss}
  = & \frac{1}{2} \sum _{f, f^{\prime}} \int d^{3} r \int d^{3} r^{\prime}
  V_{\rm ss} ( | \bm{r} - \bm{r}^{\prime} | )
  \notag \\
  &
  \times 
  \left [ \hat{b}^{\dagger} ( \bm{r} )
  \frac{\vec{\lambda}}{2} \vec{\sigma} \hat{b} ( \bm{r} ) \right ]_{f}
  \cdot
  \left [ \hat{b}^{\dagger} ( \bm{r}^{\prime} )
  \frac{\vec{\lambda}}{2} \vec{\sigma} \hat{b} ( \bm{r}^{\prime} ) \right ]_{f^{\prime}}
  ,
\end{align}
where
\begin{equation}
  V_{\rm CL} ( r )
  \equiv \frac{K}{\mu _{f f^{\prime}} r}
  - \frac{3}{4} k r + D ,
  \quad
  V_{\rm ss} ( r )
  \equiv - \frac{2 \pi \alpha _{\rm ss}}{3 m_{f} m_{f^{\prime}}}
  \bar{\delta} ( r ) ,
  \label{eq:CplusL}
\end{equation}
\begin{equation}
  \left [ \hat{b}^{\dagger} ( \bm{r} )
    \frac{\vec{\lambda}}{2} \hat{b} ( \bm{r} ) \right ]_{f, s}
  \equiv \sum _{c^{\prime}, c}
  \hat{b}_{f, s, c^{\prime}}^{\dagger} ( \bm{r} )
  \frac{\vec{\lambda}_{c^{\prime} c}}{2}
  \hat{b}_{f, s, c} ( \bm{r} ) ,
\end{equation}
\begin{equation}
  \left [ \hat{b}^{\dagger} ( \bm{r} )
  \frac{\vec{\lambda}}{2} \vec{\sigma} \hat{b} ( \bm{r} ) \right ]_{f}
  \equiv \sum _{s^{\prime}, s, c^{\prime}, c}
  \hat{b}_{f, s^{\prime}, c^{\prime}}^{\dagger} ( \bm{r} )
  \frac{\vec{\lambda}_{c^{\prime} c}}{2}
  \vec{\sigma}_{s^{\prime} s}
  \hat{b}_{f, s, c} ( \bm{r} ) .
\end{equation}
Then, we can show that the Hamiltonian $\hat{H}$ acting on the
ket vector $\ket{B}$ becomes
\begin{align}
  \hat{H} \ket{B} = &
  \int d^{3} R \left ( M_{B} - \frac{1}{2 M_{B}}
  \frac{\partial ^{2}}{\partial \bm{R}^{2}} \right )
  \Phi ^{( B )} ( \bm{R} )
  \notag \\
  & \times \int d^{3} \lambda d^{3} \rho
  \Psi ^{( B )} ( \bm{\lambda} , \bm{\rho} )
  \hat{W}^{( B ) \dagger} ( \bm{r}_{1} , \bm{r}_{2} , \bm{r}_{3} )
  \ket{0} ,
  \label{eq:HatB}
\end{align}
where we used the Schr\"{o}dinger equation~\eqref{eq:Schr} and the relation
of the differential operators
\begin{align}
  & \frac{1}{2 m_{1}} \frac{\partial ^{2}}{\partial \bm{r}_{1}^{2}}
  + \frac{1}{2 m_{2}} \frac{\partial ^{2}}{\partial \bm{r}_{2}^{2}}
  + \frac{1}{2 m_{3}} \frac{\partial ^{2}}{\partial \bm{r}_{3}^{2}}
  \notag \\
  & = \frac{1}{2 ( m_{1} + m_{2} + m_{3} )}
  \frac{\partial ^{2}}{\partial \bm{R} ^{2}}
  + \frac{1}{2 \mu _{B}}
  \frac{\partial ^{2}}{\partial \bm{\lambda} ^{2}}
  + \frac{1}{2 \mu _{B}^{\prime}}
  \frac{\partial ^{2}}{\partial \bm{\rho} ^{2}}
  \notag \\
  & \simeq \frac{1}{2 M_{B}}
  \frac{\partial ^{2}}{\partial \bm{R} ^{2}}
  + \frac{1}{2 \mu _{B}}
  \frac{\partial ^{2}}{\partial \bm{\lambda} ^{2}}
  + \frac{1}{2 \mu _{B}^{\prime}}
  \frac{\partial ^{2}}{\partial \bm{\rho} ^{2}} .
\end{align}
In the last line we used an approximation $m_{1} + m_{2} + m_{3}
\simeq M_{B}$.

\subsubsection{Two-baryon states and the resonating group method}

We can straightforwardly extend the ket vector to express the
two-baryon $B_{a} B_{b}$ state as
\begin{align}
  \ket{ B_{a} B_{b} }
  = & 
  \int d^{3} R_{a} \Phi ^{( B_{a} )} ( \bm{R}_{a} )
  \int d^{3} \lambda _{a} d^{3} \rho _{a}
  \Psi ^{( B_{a} )} ( \bm{\lambda}_{a} , \bm{\rho}_{a} )
  \notag \\
  & \times \int d^{3} R_{b} \Phi ^{( B_{b} )} ( \bm{R}_{b} )
  \int d^{3} \lambda _{b} d^{3} \rho _{b}
  \Psi ^{( B_{b} )} ( \bm{\lambda}_{b} , \bm{\rho}_{b} )
  \notag \\
  &
  \times
  \hat{W}^{( B_{a} ) \dagger} ( \bm{r}_{a1} , \bm{r}_{a2} , \bm{r}_{a3} )
  \hat{W}^{( B_{b} ) \dagger} ( \bm{r}_{b1} , \bm{r}_{b2} , \bm{r}_{b3} )
  \ket{0} .
\end{align}
In the usual manner, we can decompose the product of the spatial wave
functions of the two baryons $\Phi ^{( B_{a} )} ( \bm{R}_{a} ) \Phi
^{( B_{b} )} ( \bm{R}_{b} )$ into the center-of-mass part $\phi (
\bm{R}_{\rm tot} )$ and the relative part $\psi ( \bm{r} )$ as
\begin{equation}
  \Phi ^{( B_{a} )} ( \bm{R}_{a} ) \Phi ^{( B_{b} )} ( \bm{R}_{b} )
  = \phi ( \bm{R}_{\rm tot} ) \psi ( \bm{r} )
\end{equation}
with
\begin{equation}
  \bm{R}_{\rm tot} \equiv
  \frac{M_{B_{a}} \bm{R}_{B_{a}} + M_{B_{b}} \bm{R}_{B_{b}}}{M_{B_{a}} + M_{B_{b}}} ,
  \quad
  \bm{r} \equiv
  \bm{R}_{B_{b}} - \bm{R}_{B_{a}} .
\end{equation}
Then we rewrite the two-baryon state as
\begin{align}
  & \ket{ B_{a} B_{b} }
  \notag \\
  & = 
  \int d^{3} R_{\rm tot} \phi ( \bm{R}_{\rm tot} )
  \int d^{3} r \psi ( \bm{r} )
  \notag \\
  & \phantom{=} \times
  \int d^{3} \lambda _{a} d^{3} \rho _{a}
  \Psi ^{( B_{a} )} ( \bm{\lambda}_{a} , \bm{\rho}_{a} )
  \int d^{3} \lambda _{b} d^{3} \rho _{b}
  \Psi ^{( B_{b} )} ( \bm{\lambda}_{b} , \bm{\rho}_{b} )
  \notag \\
  & \phantom{=}
  \times
  \hat{W}^{( B_{a} ) \dagger} ( \bm{r}_{a1} , \bm{r}_{a2} , \bm{r}_{a3} )
  \hat{W}^{( B_{b} ) \dagger} ( \bm{r}_{b1} , \bm{r}_{b2} , \bm{r}_{b3} )
  \ket{0} ,
\end{align}
where we used the relation of the measure
\begin{equation}
  d^{3} R_{a} d^{3} R_{b}
  = d^{3} R_{\rm tot} d^{3} r .
\end{equation}
In addition, we introduce the two-baryon vector in which the
separation is fixed to be $\bm{r} = \bm{r}_{0}$:
\begin{align}
  & \ket{ B_{a} B_{b} ( \bm{r}_{0} )}
  \notag \\
  & =
  \int d^{3} R_{\rm tot} \phi ( \bm{R}_{\rm tot} )
  \int d^{3} r \delta ( \bm{r} - \bm{r}_{0} )
  \notag \\
  & \phantom{=} \times
  \int d^{3} \lambda _{a} d^{3} \rho _{a}
  \Psi ^{( B_{a} )} ( \bm{\lambda}_{a} , \bm{\rho}_{a} )
  \int d^{3} \lambda _{b} d^{3} \rho _{b}
  \Psi ^{( B_{b} )} ( \bm{\lambda}_{b} , \bm{\rho}_{b} )
  \notag \\
  &
  \phantom{=} \times
  \hat{W}^{( B_{a} ) \dagger} ( \bm{r}_{a1} , \bm{r}_{a2} , \bm{r}_{a3} )
  \hat{W}^{( B_{b} ) \dagger} ( \bm{r}_{b1} , \bm{r}_{b2} , \bm{r}_{b3} )
  \ket{0} .
\end{align}

Now, we derive an equation which the two-baryon $B_{a} B_{b}$ system
obeys.  Suppose that the two-baryon state $\ket{B_{a} B_{b}}$ is an
eigenstate of the Hamiltonian $\hat{H}$ with the eigenenergy $E$:
\begin{equation}
  \hat{H} \ket{B_{a} B_{b}}
  = E \ket{B_{a} B_{b}} .
\end{equation}
When we multiply the bra vector $\bra{B_{c} B_{d} ( \bm{r} )}$ to the
both sides of this equation, we obtain
\begin{equation}
  \bra{ B_{c} B_{d} ( \bm{r} )} \hat{H} \ket{ B_{a} B_{b}}
  = E \braket{ B_{c} B_{d} ( \bm{r} ) | B_{a} B_{b}} .
\end{equation}

The braket $\braket{ B_{c} B_{d} ( \bm{r} ) | B_{a} B_{b}}$ can be
calculated in the usual manner for the creation and annihilation
operators:
\begin{align}
  & \braket{ B_{c} B_{d} ( \bm{r} ) | B_{a} B_{b}}
  \notag \\
  & =
  \int d^{3} R_{\rm tot} \left | \phi ( \bm{R}_{\rm tot} ) \right |^{2}
  \notag \\
  & \phantom{=} \times
  \sum _{\vec{\mu}_{a} , \vec{\mu}_{b} , \vec{\mu}_{c} , \vec{\mu}_{d}}
  ( - 1 )^{P}
  w_{\vec{\mu}_{a}}^{( B_{a} )} w_{\vec{\mu}_{b}}^{( B_{b} )}
  w_{\vec{\mu}_{c}}^{( B_{c} )} w_{\vec{\mu}_{d}}^{( B_{d} )}
  \notag \\
  & \phantom{=} \times
  \int d^{3} \lambda _{c} d^{3} \rho _{c} d^{3} \lambda _{d} d^{3} \rho _{d}
  \left [ \Psi ^{( B_{a} )} ( \lambda _{a} , \rho _{a} )
  \Psi ^{( B_{b} )} ( \lambda _{b} , \rho _{b} ) \right .
  \notag \\
  & \phantom{= \int} \left .
  \times \Psi ^{( B_{c} )} ( \lambda _{c} , \rho _{c} ) ^{\ast}
  \Psi ^{( B_{d} )} ( \lambda _{d} , \rho _{d} ) ^{\ast}
  \psi ( \bm{r}^{\prime} ) \right ]_{B_{a} B_{b} \to B_{c} B_{d}} ,
\end{align}
where $P$ is the total number of permutations of the creation and
annihilation operators, and the subscript ``$B_{a} B_{b} \to
B_{c} B_{d}$'' restricts the summation to the case where the six
creation operators from $B_{a} B_{b}$ are exactly removed by the six
annihilation operators from $B_{c} B_{d}$.  In such a case, the
coordinates $\bm{\lambda}_{a}$, $\bm{\rho}_{a}$, $\bm{\lambda}_{b}$,
$\bm{\rho}_{b}$, and $\bm{r}^{\prime}$ in the $B_{a} B_{b}$ system are
fixed by the coordinates $\bm{\lambda}_{c}$, $\bm{\rho}_{c}$,
$\bm{\lambda}_{d}$, $\bm{\rho}_{d}$, which are the integral variables,
and $\bm{r}$ in the $B_{c} B_{d}$ system.  The wave function for
the center-of-mass motion is normalized as
\begin{equation}
  \int d^{3} R_{\rm tot} \left | \phi ( \bm{R}_{\rm tot} ) \right |^{2}
  = 1 .
\end{equation}
Then, we can express the braket $\braket{ B_{c} B_{d} ( \bm{r} ) |
  B_{a} B_{b}}$ by using the normalization kernel $N ( \bm{r} ,
\bm{r}^{\prime} )$ as
\begin{equation}
  \braket{ B_{c} B_{d} ( \bm{r} ) | B_{a} B_{b}}
  \equiv \int d^{3} r^{\prime} N ( \bm{r} , \bm{r}^{\prime} )
  \psi ( \bm{r}^{\prime} ) .
\end{equation}

\begin{figure}[!t]
  \centering
  \PsfigII{1.0}{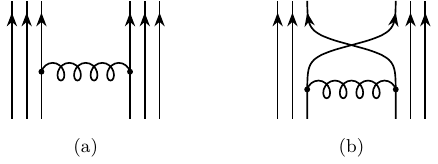}
  \caption{Examples of diagrams depicting the baryon-baryon
    interactions in our model.  Solid lines represent quarks, and
    curled lines denote quark-quark interactions.  (a) Contributions
    without quark shuffling amount to zero.  (b) Quark shuffling
    contributes to the baryon-baryon interactions.}
  \label{fig:shuffle}
\end{figure}

On the other hand, to calculate $\bra{ B_{c} B_{d} ( \bm{r} )} \hat{H}
\ket{ B_{a} B_{b}}$ we use the relation~\eqref{eq:HatB}.  Namely, when
all the annihilation operators in $\hat{H}$ act on $B_{a}$ (or on
$B_{b}$), we can use the relation~\eqref{eq:HatB}.  Additionally,
because the potential part $\hat{V}$ contains the product of two
annihilation operators, $\hat{V}$ can simultaneously act on both
$B_{a}$ and $B_{b}$ as well.  Therefore, we have
\begin{align}
  & \hat{H} \ket{ B_{a} B_{b} }
  \notag \\
  & = 
  \int d^{3} R_{\rm tot} \int d^{3} r
  \left [ M_{B_{a}} + M_{B_{b}} - \frac{1}{2 \mu _{a b}}
    \frac{\partial ^{2}}{\partial \bm{r}^{2}} \right .
  \notag \\
  & \phantom{= \int}
  \left . - \frac{1}{2 ( M_{B_{a}} + M_{B_{b}} )}
  \frac{\partial ^{2}}{\partial \bm{R}_{\rm tot}^{2}} \right ]
  \phi ( \bm{R}_{\rm tot} )
  \psi ( \bm{r} )
  \notag \\
  & \phantom{=} \times
  \int d^{3} \lambda _{a} d^{3} \rho _{a}
  \Psi ^{( B_{a} )} ( \bm{\lambda}_{a} , \bm{\rho}_{a} )
  \int d^{3} \lambda _{b} d^{3} \rho _{b}
  \Psi ^{( B_{b} )} ( \bm{\lambda}_{b} , \bm{\rho}_{b} )
  \notag \\
  & \phantom{=}
  \times
  \hat{W}^{( B_{a} ) \dagger} ( \bm{r}_{a1} , \bm{r}_{a2} , \bm{r}_{a3} )
  \hat{W}^{( B_{b} ) \dagger} ( \bm{r}_{b1} , \bm{r}_{b2} , \bm{r}_{b3} )
  \ket{0}
  \notag \\
  & \phantom{=}
  + \hat{V} \ket{B_{a} B_{b}}_{\rm int} .
  \label{eq:HBaBb}
\end{align}
Here we used the relation
\begin{align}
  & \frac{1}{2 M_{B_{a}}}
  \frac{\partial ^{2}}{\partial \bm{R}_{a}^{2}}
  + \frac{1}{2 M_{B_{b}}}
  \frac{\partial ^{2}}{\partial \bm{R}_{b}^{2}}
  \notag \\
  & = \frac{1}{2 \mu _{a b}}
  \frac{\partial ^{2}}{\partial \bm{r}^{2}}
  + \frac{1}{2 ( M_{B_{a}} + M_{B_{b}} )}
  \frac{\partial ^{2}}{\partial \bm{R}_{\rm tot}^{2}} ,
\end{align}
where $\mu _{a b}$ is the reduced mass for the $B_{a} B_{b}$ system
\begin{equation}
  \mu _{a b} = \frac{M_{B_{a}} M_{B_{b}}}{M_{B_{a}} + M_{B_{b}}} ,
\end{equation}
and the subscript ``int'' of $\hat{V} \ket{B_{a} B_{b}}_{\rm int}$
denotes the inter-baryon contributions to the
potential term, \textit{i.e.}, the potential between one quark from $B_{a}$ and
the other from $B_{b}$.  We are not interested in the center-of-mass
motion, so we neglect the center-of-mass kinetic energy in
Eq.~\eqref{eq:HBaBb}.  Because the first term in Eq.~\eqref{eq:HBaBb}
has the same structure of operators as the $\ket{B_{a} B_{b}}$ state,
we have
\begin{align}
  & \bra{B_{c} B_{d} ( \bm{r} ) } \hat{H} \ket{B_{a} B_{b}}
  \notag \\
  & = \int d^{3} r^{\prime} N ( \bm{r} , \bm{r}^{\prime} )
  \left ( M_{B_{a}} + M_{B_{b}} - \frac{1}{2 \mu _{a b}}
    \frac{\partial ^{2}}{\partial \bm{r}^{\prime \, 2}} \right )
  \psi ( \bm{r}^{\prime} )
  \notag \\
  & \phantom{=} +
  \bra{ B_{c} B_{d} ( \bm{r} )} \hat{V} \ket{B_{a} B_{b}}_{\rm int} .
\end{align}
The potential term $\bra{ B_{c} B_{d} ( \bm{r} )} \hat{V} \ket{B_{a}
  B_{b}}_{\rm int}$ can be calculated in the usual manner for the
creation and annihilation operators as well, and can be expressed by
the non-local potential $V_{\rm int} ( \bm{r} , \bm{r}^{\prime} )$ as
\begin{equation}
  \bra{ B_{c} B_{d} ( \bm{r} )} \hat{V} \ket{B_{a} B_{b}}_{\rm int}
  \equiv \int d^{3} r^{\prime} V_{\rm int} ( \bm{r} , \bm{r}^{\prime} )
  \psi ( \bm{r}^{\prime} ) .
\end{equation}
We note that contributions without quark shuffling between baryons
[Fig.~\ref{fig:shuffle}(a)] amount to zero for the non-local potential
$V_{\rm int} ( \bm{r} , \bm{r}^{\prime} )$ in the quark model due to
the properties of quark color inside baryons and Gell-Mann matrices
$\vec{\lambda}$.  Physically, this means that the gluon cannot mediate
between color singlet states.  Therefore, $V_{\rm int} ( \bm{r} ,
\bm{r}^{\prime} )$ necessarily contains the shuffling of quarks
between baryons, such as shown in Fig.~\ref{fig:shuffle}(b).

As a consequence, we obtain the equation which the $B_{a} B_{b} \to
B_{c} B_{d}$ process should satisfy:
\begin{align}
  & \int d^{3} r^{\prime} \left [ N ( \bm{r} , \bm{r}^{\prime} )
    \left ( - \frac{1}{2 \mu _{a b}}
    \frac{\partial ^{2}}{\partial \bm{r}^{\prime \, 2}} \right )
    + V_{\rm int} ( \bm{r} , \bm{r}^{\prime} )
    \right ]
  \psi ( \bm{r}^{\prime} )
  \notag \\
  & = \mathcal{E} \int d^{3} r^{\prime} 
  N ( \bm{r} , \bm{r}^{\prime} ) \psi ( \bm{r}^{\prime} )
  \label{eq:RGM}
\end{align}
with the eigenenergy
\begin{equation}
  \mathcal{E} \equiv E - M_{B_{a}} - M_{B_{b}} .
  \label{eq:Eigen}
\end{equation}
This integro-differential equation is the resonating group method
(RGM) equation.  Because all the parameters in the present model are
fixed to reproduce the baryon masses, the RGM equation has no free
parameters.  The RGM equation~\eqref{eq:RGM} automatically covers
coupled-channels cases, but we will not take into account the
coupled-channels effects unless explicitly stated.

\subsubsection{Equivalent local potentials}

While the RGM equation contains the non-local potential $V_{\rm int} (
\bm{r} , \bm{r}^{\prime} )$ between two baryons together with the
normalization kernel $N ( \bm{r} , \bm{r}^{\prime} )$, local
potentials are desired for practical studies.  In fact, from the
RGM equation, we can extract equivalent local potentials between two
baryons.  Our strategy is to calculate a local potential that
generates the same wave function as that of the two baryon state in
the RGM equation.

However, the wave function $\psi ( \bm{r} )$ in the RGM
equation~\eqref{eq:RGM} may contain unphysical forbidden states by the
Pauli exclusion principle, which are zero eigenstates of the
normalization kernel $N ( \bm{r} , \bm{r}^{\prime} )$.  To eliminate
these zero modes, we ``reduce'' the wave function in the following
manner:
\begin{equation}
  \psi _{\rm R} ( \bm{r} )
  = \int d^{3} r^{\prime} N^{1/2} ( \bm{r} , \bm{r}^{\prime} ) \psi ( \bm{r} )
  \label{eq:psiR}
\end{equation}
where $N^{1/2} ( \bm{r} , \bm{r}^{\prime} )$ satisfies
\begin{equation}
  N ( \bm{r} , \bm{r}^{\prime} )
  = \int d^{3} r^{\prime \prime} N^{1/2} ( \bm{r} , \bm{r}^{\prime \prime} )
  N^{1/2} ( \bm{r}^{\prime \prime} , \bm{r}^{\prime} ) .
\end{equation}
Now, we can calculate the local potentials for two baryons as follows:
\begin{enumerate}
\item Calculate $N ( \bm{r} , \bm{r}^{\prime} )$, $V_{\rm int} (
  \bm{r} , \bm{r}^{\prime} )$ and solve the RGM equation.  Because we
  are interested in the low-energy behavior of the two-baryon system,
  we focus on the ground state in $S$ wave.  The eigenenergy of the
  two-baryon system $\mathcal{E}$ is given by
  \begin{equation}
    \mathcal{E}_{\rm G} =
    \begin{cases}
      - B & \text{if a bound state exists} ,
      \\
      0 & \text{else} ,
    \end{cases}
  \end{equation}
  where $B$ is the binding energy of the bound state.  Note that the
  angular dependence of $N ( \bm{r} , \bm{r}^{\prime} )$ and $V_{\rm
    int} ( \bm{r} , \bm{r}^{\prime} )$ is irrelevant in the present
  study, because we focus on the $S$-wave state.  For the reduced mass
  $\mu _{a b}$ in the RGM equation and baryon masses in the
  eigenenergy $\mathcal{E}$~\eqref{eq:Eigen}, we use the values in our
  constituent quark model, \textit{i.e.}, the eigenvalues in
  Eq.~\eqref{eq:Schr}.

\item From the wave function $\psi ( r )$ in the RGM equation,
  calculate the reduced wave function $\psi _{\rm R} ( r )$ according
  to Eq.~\eqref{eq:psiR}.

\item Calculate
  \begin{equation}
    \chi _{\rm R} ( r ) \equiv r \psi _{\rm R} ( r ) 
  \end{equation}
  and derive the equivalent local potential $V_{\rm eq} ( r )$ that
  generates the same wave function $\chi _{\rm R}$ with the
  eigenenergy $\mathcal{E}_{\rm G}$~\cite{Oka:1981ri}:
\begin{equation}
  V_{\rm eq} ( r )
  \equiv \mathcal{E}_{\rm G} + \frac{1}{2 \mu _{a b} \chi _{\rm R} ( r )}
  \frac{d^{2} \chi _{\rm R}}{d r^{2}} .
  \label{eq:Veq}
\end{equation}

\end{enumerate}

We note that the equivalent local potential~\eqref{eq:Veq} depends on
the energy $\mathcal{E}$.  In the present study, we fix the energy to
be the ground-state energy, because we are interested in the
low-energy behavior of the two-baryon system, and evaluate the
potential at this energy.

We also note that this strategy works when the wave function $\chi
_{\rm R}$ has no nodes.  However, if the wave function has a node
$\chi _{\rm R} = 0$ and at this point $d^{2} \chi _{\rm R} / d r^{2}
\neq 0$, the equivalent local potential becomes singular.  Indeed, the
wave function in the RGM equation may have nodes so as to make the
wave function orthogonal to the unphysical forbidden states of the RGM
equation due to the Pauli exclusion principle for quarks (see
Ref.~\cite{Oka:1981rj}).  One could remove such contributions to
obtain nonsingular potentials as in, \textit{e.g.},
Ref~\cite{Michel:1998} and references therein for the local
$\alpha$-$\alpha$ potential.  Still, in the present study, we simply
discard singular equivalent local potentials and show only nonsingular
ones.

\section{Numerical results and discussions}
\label{sec:3}

\begin{table}[!t]
  \caption{Parameters for the color Coulomb plus linear confining potential.}
  \label{tab:CplusL}
  \centering
  \begin{ruledtabular}
    \begin{tabular}{rlrrr}
      \multicolumn{1}{c}{$n$}
      & \multicolumn{1}{c}{$x_{n}$ [fm]}
      & \multicolumn{1}{c}{$A_{n}$ [fm$^{-1}$]}
      & \multicolumn{1}{c}{$B_{n}$ [fm]}
      & \multicolumn{1}{c}{$C_{n}$}
      \\
      \hline
      1 & 0.05 &
      $24.0065$ & 
      $-0.3426$ & 
      $-0.06515$
      \\
      2 & 0.06851 &
      $-10.9718$ & 
      $0.8172$ & 
      $0.16491$
      \\
      3 & 0.09387 &
      $16.5001$ & 
      $-1.4085$ & 
      $-0.27838$
      \\
      4 & 0.12862 &
      $-6.3166$ & 
      $1.9560$ & 
      $0.39680$
      \\
      5 & 0.17624 &
      $8.1792$ & 
      $-2.6370$ & 
      $-0.52322$
      \\
      6 & 0.24148 &
      $-2.6114$ & 
      $3.2980$ & 
      $0.67149$
      \\
      7 & 0.33087 &
      $3.7594$ & 
      $-4.3881$ & 
      $-0.87098$
      \\
      8 & 0.45335 &
      $-0.9834$ & 
      $5.8023$ & 
      $1.18519$
      \\
      9 & 0.62118 &
      $1.7175$ & 
      $-8.8336$ & 
      $-1.76209$
      \\
      10 & 0.85113 &
      $-0.2202$ & 
      $14.3552$ & 
      $2.94624$
      \\
      11 & 1.16622 &
      $0.3157$ & 
      $-26.3552$ & 
      $-5.37162$
      \\
      12 & 1.59794 &
      $1.1646$ & 
      $43.6361$ & 
      $9.24036$
      \\
      13 & 2.18948 &
      $-1.6494$ & 
      $-52.4052$ & 
      $-11.51550$
      \\
      14 & 3 &
      $1.4912$ & 
      $26.4737$ & 
      $6.77163$
      \\
    \end{tabular}
  \end{ruledtabular}
\end{table}

In this section, we present our numerical results for the
baryon-baryon potentials in our model and discuss their properties.
We first focus on the single-channel cases without coupled-channels
effects in Sec.~\ref{sec:3A}, and then we consider the
coupled-channels effects in several systems in Sec.~\ref{sec:3B}.

Before presenting the results, we would like to mention two technical
details that allow us to speed up the numerical calculations.  First,
we reduce the number of terms in the Gaussian expansion, denoted by
$N$.  In Sec.~\ref{sec:2A} we used $N = 10$ to achieve certain
convergence.  However, we have verified that, for all the ground-state
baryons, the wave function $\Psi ^{( B )} ( \bm{\lambda} , \bm{\rho}
)$ with $N = 2$ (with $r_{\rm min}$ and $r_{\rm max}$ tuned) deviates
from that with $N = 10$ by only about
$\SI{1}{\percent}$.\footnote{Tuned values of $( r_{\rm min} , r_{\rm
  max} )$ in the $N = 2$ case are: $( 0.454 , 0.931 )$ for $N$, $(
0.443 , 0.896 )$ for $\Lambda$, $( 0.460 , 0.921 )$ for $\Sigma$, $(
0.452 , 0.880 )$ for $\Xi$, $( 0.619 , 1.064 )$ for $\Delta$, $( 0.594
, 1.025 )$ for $\Sigma ^{\ast}$, $( 0.573 , 0.988 )$ for $\Xi
^{\ast}$, and $( 0.554 , 0.947 )$ for $\Omega$, in units of fm.}
Therefore, in this section, we employ $N = 2$ instead of $N = 10$,
which leads to potential errors of about $\sim \SI{1}{\percent}$.
Second, we approximate terms in the color Coulomb plus linear
confining potential $V_{\rm CL} ( r )$~\eqref{eq:CplusL} by sums of 14
Gaussians:
\begin{equation}
  \frac{1}{r} 
  = \sum _{n = 1}^{14} A_{n}
  \exp \left ( - \frac{r^{2}}{x_{n}^{2}} \right ) ,
\end{equation}
\begin{equation}
  r
  = \sum _{n = 1}^{14} B_{n}
  \exp \left ( - \frac{r^{2}}{x_{n}^{2}} \right ) ,
\end{equation}
\begin{equation}
  1
  = \sum _{n = 1}^{14} C_{n}
  \exp \left ( - \frac{r^{2}}{x_{n}^{2}} \right ) .
\end{equation}
The range parameters $x_{n}$ are chosen in a geometric progression
\begin{equation}
  x_{n} = 60^{(n - 1) / 13} \times \SI{0.05}{fm} ,
\end{equation}
while the coefficients $A_{n}$, $B_{n}$, and $C_{n}$ are fixed by
fitting to $1/r$, $r$, and $1$, respectively.  By using the parameter
values listed in Table~\ref{tab:CplusL}, we can reproduce each term
reasonably well over the range $[ \SI{0.05}{fm} , \SI{3}{fm} ]$, which
covers the range of the baryon-baryon interactions of interest in this
study.

\begin{table*}[!t]
  \caption{Values of $N_{33}$.  Two-baryon threshold values, expressed
    in units of MeV, are provided in parentheses in the format
    (model/experimental values).  $S$ is the strangeness of the
    system.  The channels marked with a dagger ($\dagger$) indicate
    that the equivalent local potentials become singular.}
  \label{tab:N33}
  \centering
  \begin{ruledtabular}
    \begin{tabular}{ccccc}
    \begin{minipage}[t]{0.18\hsize}
      \begin{tabular}{lc}
        \multicolumn{2}{c}{$S = 0$}
        \\
        \multicolumn{2}{c}{$N N$ (1899/1878)}
        \\
        \hline
        $( 1 , 0 )$  &  $5/9$
        \\
        $( 0 , 1 )$  &  $5/9$
        \\
        \hline
        \\
        \multicolumn{2}{c}{$N \Delta$ (2184/2171)}
        \\
        \hline
        $( 2 , 2 ) \dagger$  &  $0$
        \\
        $( 2 , 1 )$  &  $4/9$
        \\
        $( 1 , 2 )$  &  $4/9$
        \\
        $( 1 , 1 ) \dagger$  &  $0$
        \\
        \hline
        \\
        \multicolumn{2}{c}{$\Delta \Delta$ (2469/2464)}
        \\
        \hline
        $( 3 , 2 ) \dagger$  &  $0$
        \\
        $( 3 , 0 )$  &  $1$
        \\
        $( 2 , 3 ) \dagger$  &  $0$
        \\
        $( 2 , 1 )$  &  $5/9$
        \\
        $( 1 , 2 )$  &  $5/9$
        \\
        $( 1 , 0 )$  &  $4/9$
        \\
        $( 0 , 3 )$  &  $1$
        \\
        $( 0 , 1 )$  &  $4/9$
        \\
        \hline
        \\
        \\
        \multicolumn{2}{c}{$S = -1$}
        \\
        \multicolumn{2}{c}{$N \Lambda$ (2060/2055)}
        \\
        \hline
        $( 1 , 1/2 )$  &  $1/2$
        \\
        $( 0 , 1/2 )$  &  $1/2$
        \\
        \hline
        \\
        \multicolumn{2}{c}{$N \Sigma$ (2130/2132)}
        \\
        \hline
        $( 1 , 3/2 ) \dagger$  &  $1/9$
        \\
        $( 1 , 1/2 )$  &  $1/2$
        \\
        $( 0 , 3/2 )$  &  $5/9$
        \\
        $( 0 , 1/2 ) \dagger$  &  $1/18$
        \\
        \hline
        \\
        \multicolumn{2}{c}{$N \Sigma ^{\ast}$ (2332/2323)}
        \\
        \hline
        $( 2 , 3/2 ) \dagger$  &  $1/18$
        \\
        $( 2 , 1/2 )$  &  $5/9$
        \\
        $( 1 , 3/2 )$  &  $1/2$
        \\
        $( 1 , 1/2 ) \dagger$  &  $1/9$
        \\
        \hline
        \\
        \multicolumn{2}{c}{$\Delta \Lambda$ (2345/2348)}
        \\
        \hline
        $( 2 , 3/2 ) \dagger$  &  $1/4$
        \\
        $( 1 , 3/2 ) \dagger$  &  $1/4$
        \\
        \hline
        \\
        \\
        \\
        \\
        \\
        \\
      \end{tabular}
    \end{minipage}
    &
    \begin{minipage}[t]{0.18\hsize}
      \begin{tabular}{lc}
        \\
        \multicolumn{2}{c}{$\Delta \Sigma$ (2415/2425)}
        \\
        \hline
        $( 2 , 5/2 ) \dagger$  &  $0$
        \\
        $( 2 , 3/2 ) \dagger$  &  $5/36$
        \\
        $( 2 , 1/2 )$  &  $8/9$
        \\
        $( 1 , 5/2 )$  &  $4/9$
        \\
        $( 1 , 3/2 )$  &  $7/12$
        \\
        $( 1 , 1/2 )$  &  $4/9$
        \\
        \hline
        \\
        \multicolumn{2}{c}{$\Delta \Sigma ^{\ast}$ (2617/2617)}
        \\
        \hline
        $( 3 , 5/2 ) \dagger$  &  $0$
        \\
        $( 3 , 3/2 ) \dagger$  &  $0$
        \\
        $( 3 , 1/2 )$  &  $1$
        \\
        $( 2 , 5/2 ) \dagger$  &  $0$
        \\
        $( 2 , 3/2 )$  &  $5/9$
        \\
        $( 2 , 1/2 )$  &  $5/9$
        \\
        $( 1 , 5/2 )$  &  $5/9$
        \\
        $( 1 , 3/2 )$  &  $5/9$
        \\
        $( 1 , 1/2 )$  &  $4/9$
        \\
        $( 0 , 5/2 )$  &  $1$
        \\
        $( 0 , 3/2 )$  &  $4/9$
        \\
        $( 0 , 1/2 )$  &  $4/9$
        \\
        \hline
        \\
        \\
        \multicolumn{2}{c}{$S = -2$}
        \\
        \multicolumn{2}{c}{$\Lambda \Lambda$ (2221/2231)}
        \\
        \hline
        $( 0 , 0 )$  &  $1/2$
        \\
        \hline
        \\
        \multicolumn{2}{c}{$N \Xi$ (2272/2257)}
        \\
        \hline
        $( 1 , 1 )$  &  $10/27$
        \\
        $( 1 , 0 )$  &  $4/9$
        \\
        $( 0 , 1 ) \dagger$  &  $2/9$
        \\
        $( 0 , 0 )$  &  $2/3$
        \\
        \hline
        \\
        \multicolumn{2}{c}{$\Lambda \Sigma$ (2291/2309)}
        \\
        \hline
        $( 1 , 1 )$  &  $1/3$
        \\
        $( 0 , 1 ) \dagger$  &  $1/3$
        \\
        \hline
        \\
        \multicolumn{2}{c}{$\Sigma \Sigma$ (2361/2386)}
        \\
        \hline
        $( 1 , 1 )$  &  $11/27$
        \\
        $( 0 , 2 )$  &  $5/9$
        \\
        $( 0 , 0 )$  &  $7/18$
        \\
        \hline
        \\
        \\
        \\
        \\
        \\
        \\
      \end{tabular}
    \end{minipage}
    &
    \begin{minipage}[t]{0.18\hsize}
      \begin{tabular}{lc}
        \\
        \multicolumn{2}{c}{$N \Xi ^{\ast}$ (2479/2472)}
        \\
        \hline
        $( 2 , 1 )$  &  $2/9$
        \\
        $( 2 , 0 )$  &  $2/3$
        \\
        $( 1 , 1 )$  &  $14/27$
        \\
        $( 1 , 0 )$  &  $2/9$
        \\
        \hline
        \\
        \multicolumn{2}{c}{$\Lambda \Sigma ^{\ast}$ (2493/2500)}
        \\
        \hline
        $( 2 , 1 )$  &  $1/3$
        \\
        $( 1 , 1 )$  &  $1/3$
        \\
        \hline
        \\
        \multicolumn{2}{c}{$\Delta \Xi$ (2557/2550)}
        \\
        \hline
        $( 2 , 2 )$  &  $1/3$
        \\
        $( 2 , 1 )$  &  $5/9$
        \\
        $( 1 , 2 ) \dagger$  &  $1/9$
        \\
        $( 1 , 1 )$  &  $17/27$
        \\
        \hline
        \\
        \multicolumn{2}{c}{$\Sigma \Sigma ^{\ast}$ (2562/2578)}
        \\
        \hline
        $( 2 , 2 ) \dagger$  &  $1/9$
        \\
        $( 2 , 1 )$  &  $1/3$
        \\
        $( 2 , 0 )$  &  $7/9$
        \\
        $( 1 , 2 ) \dagger$  &  $1/3$
        \\
        $( 1 , 1 )$  &  $11/27$
        \\
        $( 1 , 0 )$  &  $1/3$
        \\
        \hline
        \\
        \multicolumn{2}{c}{$\Delta \Xi ^{\ast}$ (2764/2765)}
        \\
        \hline
        $( 3 , 2 ) \dagger$  &  $0$
        \\
        $( 3 , 1 )$  &  $2/3$
        \\
        $( 2 , 2 )$  &  $1/3$
        \\
        $( 2 , 1 )$  &  $5/9$
        \\
        $( 1 , 2 )$  &  $5/9$
        \\
        $( 1 , 1 )$  &  $13/27$
        \\
        $( 0 , 2 )$  &  $2/3$
        \\
        $( 0 , 1 )$  &  $4/9$
        \\
        \hline
        \\
        \multicolumn{2}{c}{$\Sigma ^{\ast} \Sigma ^{\ast}$ (2764/2769)}
        \\
        \hline
        $( 3 , 1 )$  &  $1/3$
        \\
        $( 2 , 2 )$  &  $2/9$
        \\
        $( 2 , 0 )$  &  $5/9$
        \\
        $( 1 , 1 )$  &  $14/27$
        \\
        $( 0 , 2 )$  &  $7/9$
        \\
        $( 0 , 0 )$  &  $4/9$
        \\
        \hline
        \\
        \\
        \\
        \\
        \\
        \\
      \end{tabular}
    \end{minipage}
    &
    \begin{minipage}[t]{0.18\hsize}
      \begin{tabular}{lc}
        \multicolumn{2}{c}{$S = -3$}
        \\
        \multicolumn{2}{c}{$\Lambda \Xi$ (2433/2434)}
        \\
        \hline
        $( 1 , 1/2 )$  &  $5/18$
        \\
        $( 0 , 1/2 )$  &  $1/2$
        \\
        \hline
        \\
        \multicolumn{2}{c}{$\Sigma \Xi$ (2503/2511)}
        \\
        \hline
        $( 1 , 3/2 )$  &  $5/9$
        \\
        $( 1 , 1/2 )$  &  $5/18$
        \\
        $( 0 , 3/2 )$  &  $5/9$
        \\
        $( 0 , 1/2 ) \dagger$  &  $1/18$
        \\
        \hline
        \\
        \multicolumn{2}{c}{$N \Omega$ (2629/2611)}
        \\
        \hline
        $( 2 , 1/2 )$  &  $1/2$
        \\
        $( 1 , 1/2 )$  &  $1/2$
        \\
        \hline
        \\
        \multicolumn{2}{c}{$\Lambda \Xi ^{\ast}$ (2640/2649)}
        \\
        \hline
        $( 2 , 1/2 )$  &  $1/4$
        \\
        $( 1 , 1/2 )$  &  $17/36$
        \\
        \hline
        \\
        \multicolumn{2}{c}{$\Sigma ^{\ast} \Xi$ (2704/2703)}
        \\
        \hline
        $( 2 , 3/2 ) \dagger$  &  $2/9$
        \\
        $( 2 , 1/2 )$  &  $2/9$
        \\
        $( 1 , 3/2 ) \dagger$  &  $2/9$
        \\
        $( 1 , 1/2 )$  &  $2/3$
        \\
        \hline
        \\
        \multicolumn{2}{c}{$\Sigma \Xi ^{\ast}$ (2710/2727)}
        \\
        \hline
        $( 2 , 3/2 )$  &  $2/9$
        \\
        $( 2 , 1/2 )$  &  $17/36$
        \\
        $( 1 , 3/2 ) \dagger$  &  $2/9$
        \\
        $( 1 , 1/2 )$  &  $1/4$
        \\
        \hline
        \\
        \multicolumn{2}{c}{$\Delta \Omega$ (2914/2904)}
        \\
        \hline
        $( 3 , 3/2 )$  &  $1/2$
        \\
        $( 2 , 3/2 )$  &  $1/2$
        \\
        $( 1 , 3/2 )$  &  $1/2$
        \\
        $( 0 , 3/2 )$  &  $1/2$
        \\
        \hline
        \\
        \multicolumn{2}{c}{$\Sigma ^{\ast} \Xi ^{\ast}$ (2912/2918)}
        \\
        \hline
        $( 3 , 3/2 )$  &  $1/2$
        \\
        $( 3 , 1/2 ) \dagger$  &  $0$
        \\
        $( 2 , 3/2 ) \dagger$  &  $1/18$
        \\
        $( 2 , 1/2 )$  &  $5/9$
        \\
        $( 1 , 3/2 )$  &  $1/2$
        \\
        $( 1 , 1/2 )$  &  $5/9$
        \\
        $( 0 , 3/2 )$  &  $17/18$
        \\
        $( 0 , 1/2 )$  &  $4/9$
        \\
        \hline
        \\
        \\
      \end{tabular}
    \end{minipage}
    &
    \begin{minipage}[t]{0.18\hsize}
      \begin{tabular}{lc}
        \multicolumn{2}{c}{$S = -4$}
        \\
        \multicolumn{2}{c}{$\Xi \Xi$ (2645/2637)}
        \\
        \hline
        $( 1 , 0 ) \dagger$  &  $1/9$
        \\
        $( 0 , 1 )$  &  $5/9$
        \\
        \hline
        \\
        \multicolumn{2}{c}{$\Lambda \Omega$ (2790/2788)}
        \\
        \hline
        $( 2 , 0 ) \dagger$  &  $0$
        \\
        $( 1 , 0 )$  &  $2/3$
        \\
        \hline
        \\
        \multicolumn{2}{c}{$\Xi \Xi ^{\ast}$ (2852/2852)}
        \\
        \hline
        $( 2 , 1 ) \dagger$  &  $1/9$
        \\
        $( 2 , 0 ) \dagger$  &  $0$
        \\
        $( 1 , 1 )$  &  $1/3$
        \\
        $( 1 , 0 )$  &  $2/3$
        \\
        \hline
        \\
        \multicolumn{2}{c}{$\Sigma \Omega$ (2859/2866)}
        \\
        \hline
        $( 2 , 1 )$  &  $1/3$
        \\
        $( 1 , 1 ) \dagger$  &  $1/9$
        \\
        \hline
        \\
        \multicolumn{2}{c}{$\Sigma ^{\ast} \Omega$ (3061/3057)}
        \\
        \hline
        $( 3 , 1 ) \dagger$  &  $0$
        \\
        $( 2 , 1 )$  &  $1/3$
        \\
        $( 1 , 1 )$  &  $5/9$
        \\
        $( 0 , 1 )$  &  $2/3$
        \\
        \hline
        \\
        \multicolumn{2}{c}{$\Xi ^{\ast} \Xi ^{\ast}$ (3059/3067)}
        \\
        \hline
        $( 3 , 0 ) \dagger$  &  $0$
        \\
        $( 2 , 1 )$  &  $2/9$
        \\
        $( 1 , 0 )$  &  $5/9$
        \\
        $( 0 , 1 )$  &  $7/9$
        \\
        \hline
        \\
        \\
        \multicolumn{2}{c}{$S = -5$}
        \\
        \multicolumn{2}{c}{$\Xi \Omega$ (3001/2991)}
        \\
        \hline
        $( 2 , 1/2 ) \dagger$  &  $0$
        \\
        $( 1 , 1/2 )$  &  $4/9$
        \\
        \hline
        \\
        \multicolumn{2}{c}{$\Xi ^{\ast} \Omega$ (3209/3206)}
        \\
        \hline
        $( 3 , 1/2 ) \dagger$  &  $0$
        \\
        $( 2 , 1/2 ) \dagger$  &  $0$
        \\
        $( 1 , 1/2 )$  &  $5/9$
        \\
        $( 0 , 1/2 )$  &  $1$
        \\
        \hline
        \\
        \\
        \multicolumn{2}{c}{$S = -6$}
        \\
        \multicolumn{2}{c}{$\Omega \Omega$ (3358/3345)}
        \\
        \hline
        $( 2 , 0 ) \dagger$  &  $0$
        \\
        $( 0 , 0 )$  &  $1$
        \\
        \hline
      \end{tabular}
    \end{minipage}

    \end{tabular}
  \end{ruledtabular}
\end{table*}

\begin{figure*}[!t]
  \centering
  \Psfig{8.6cm}{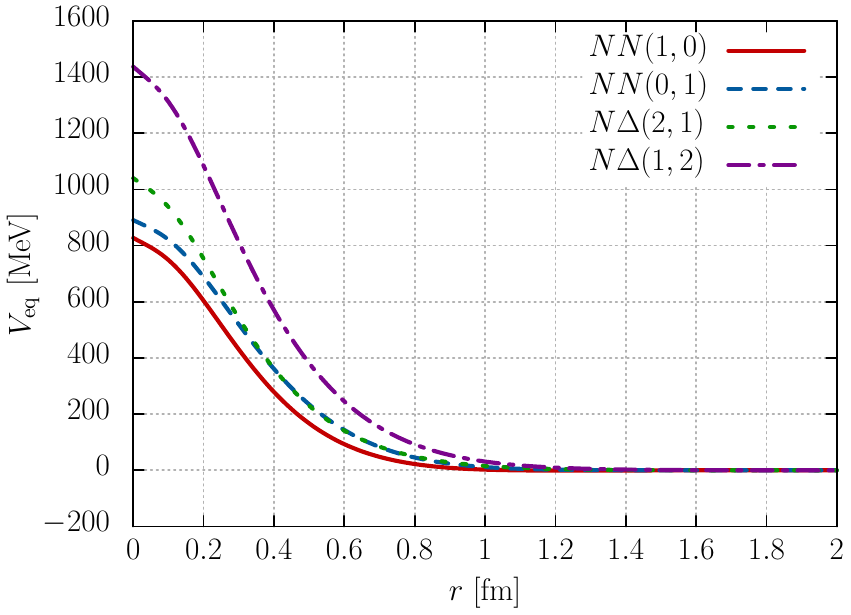}
  ~ ~
  \Psfig{8.6cm}{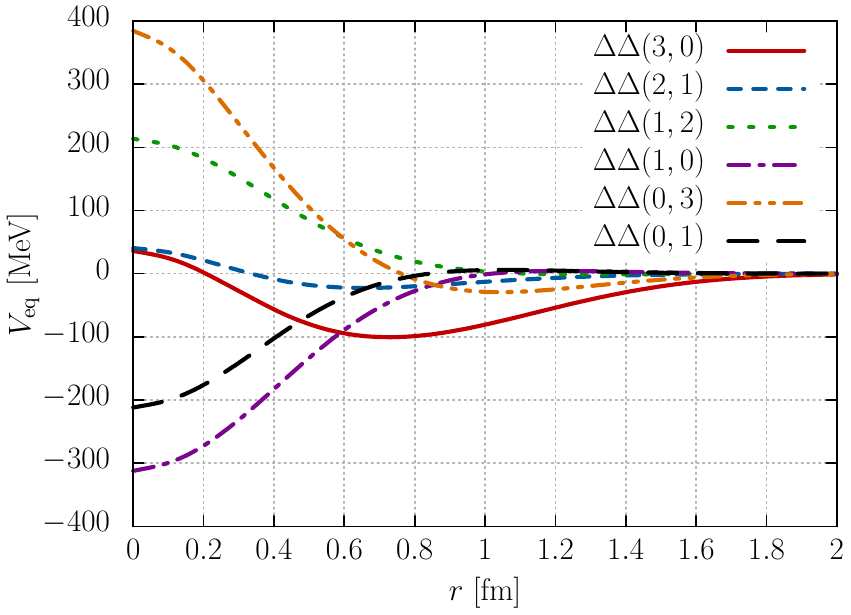}
  \caption{Equivalent local potentials for the baryon-baryon systems:
    strangeness $S = 0$ sector.}
  \label{fig:V_S0}
\end{figure*}

\begin{figure}[!t]
  \centering
  \Psfig{8.6cm}{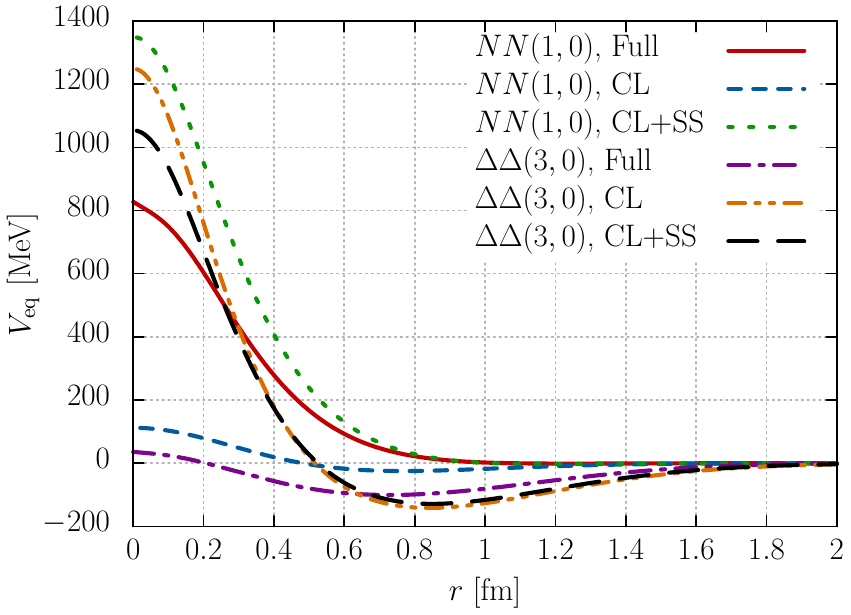}
  \caption{Decomposition of the potentials for the $N N ( 1 , 0 )$ and
    $\Delta \Delta ( 3 , 0 )$ systems.}
  \label{fig:decomp}
\end{figure}

\begin{figure}[!t]
  \centering
  \Psfig{8.6cm}{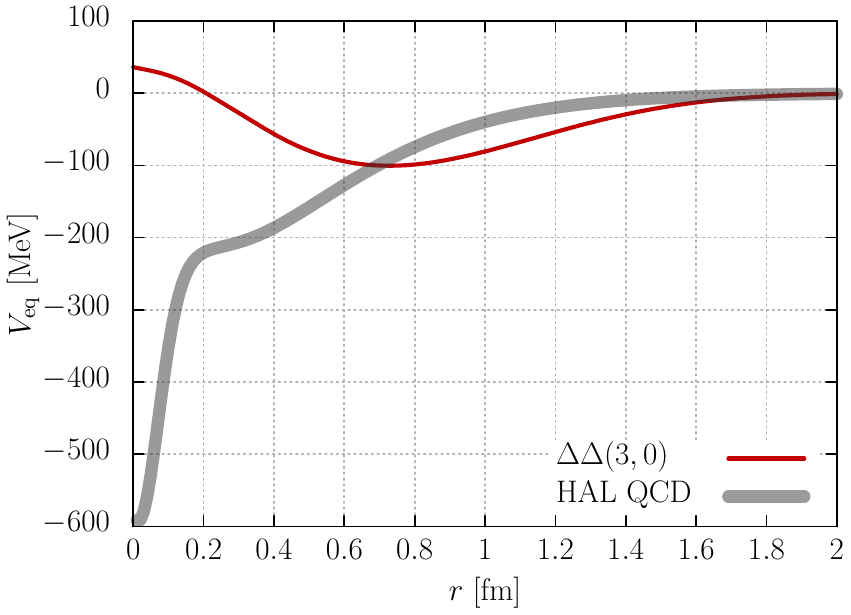}
  \caption{Comparison with the HAL QCD potential for the $\Delta
    \Delta ( 3 , 0 )$ system~\cite{Gongyo:2020pyy}.}
  \label{fig:V_DeltaDelta_HAL}
\end{figure}

\begin{figure*}[!t]
  \centering
  \Psfig{8.6cm}{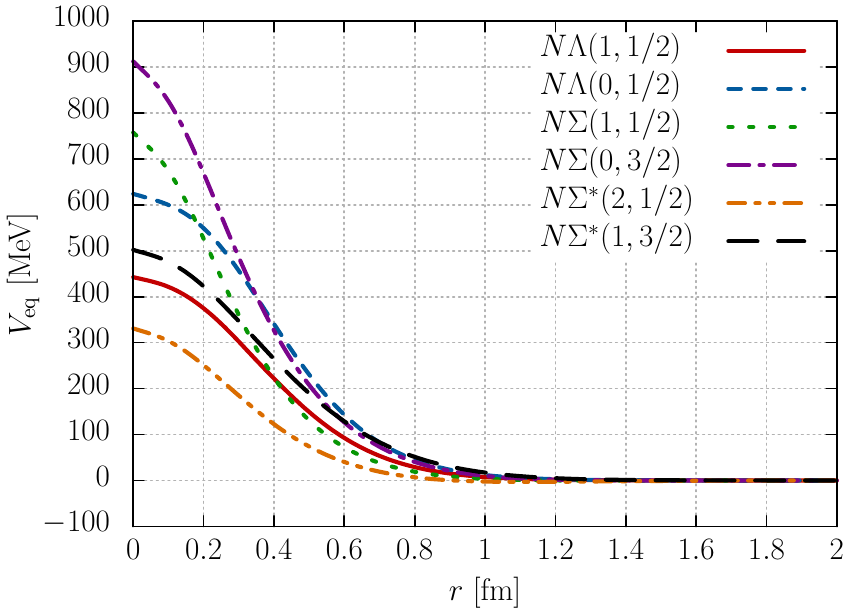}
  ~ ~
  \Psfig{8.6cm}{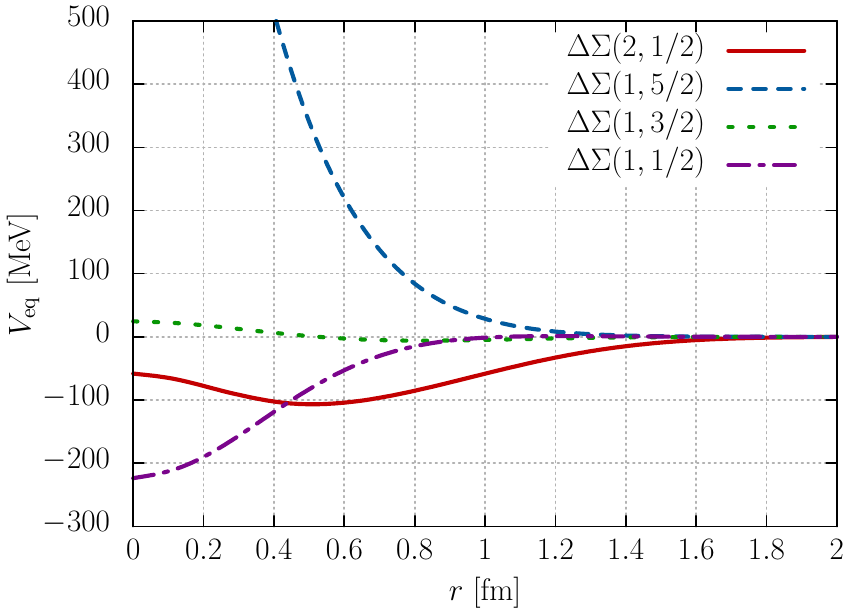}
  \\
  \Psfig{8.6cm}{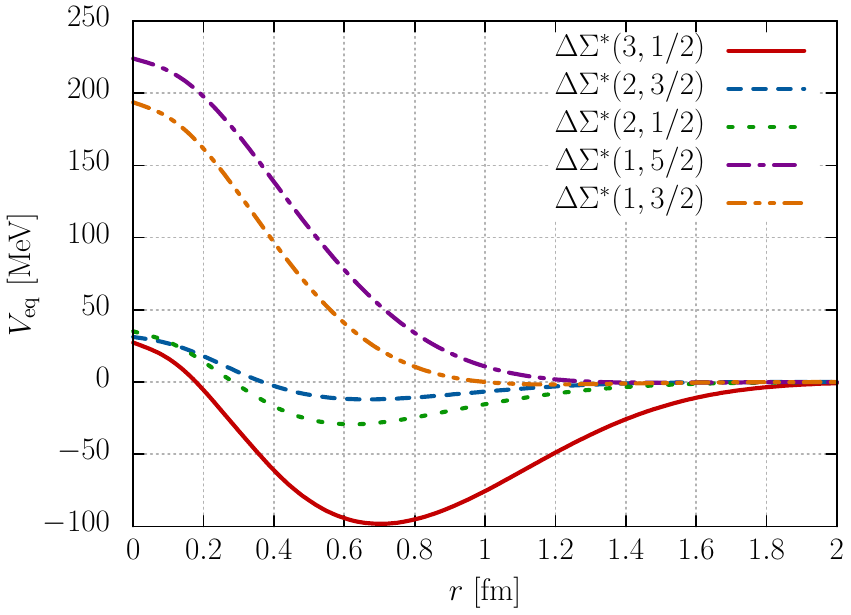}
  ~ ~ 
  \Psfig{8.6cm}{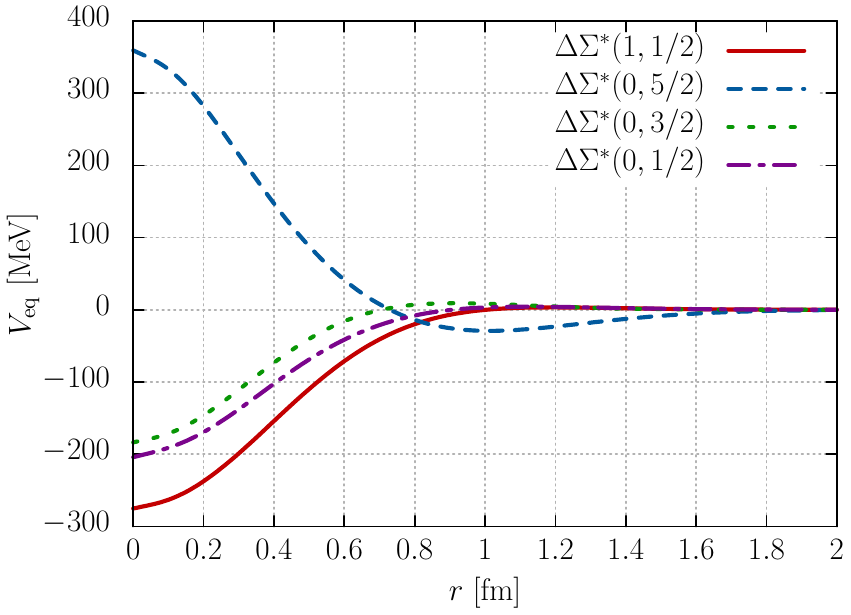}
  \caption{Equivalent local potentials for the baryon-baryon systems:
    strangeness $S = -1$ sector.}
  \label{fig:V_S-1}
\end{figure*}

\begin{figure*}[!t]
  \centering
  \Psfig{8.6cm}{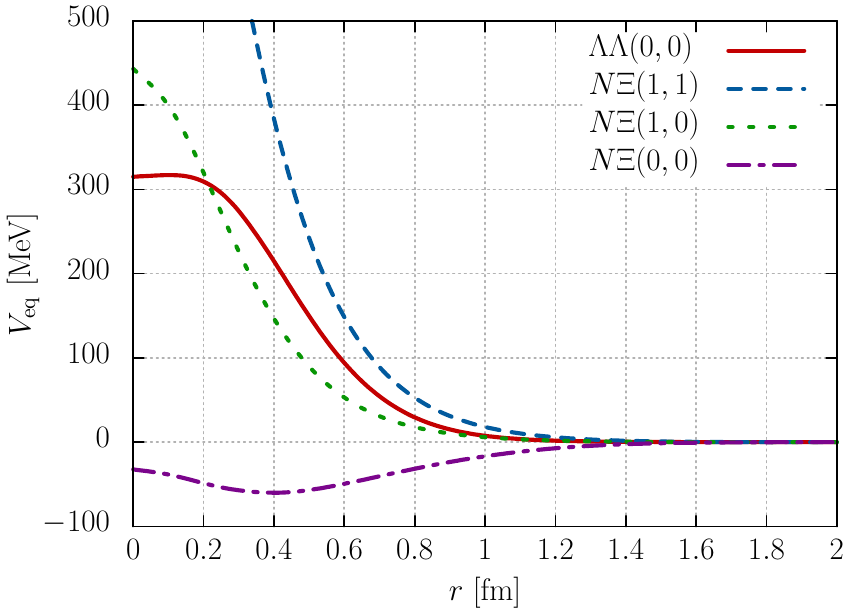}
  ~ ~
  \Psfig{8.6cm}{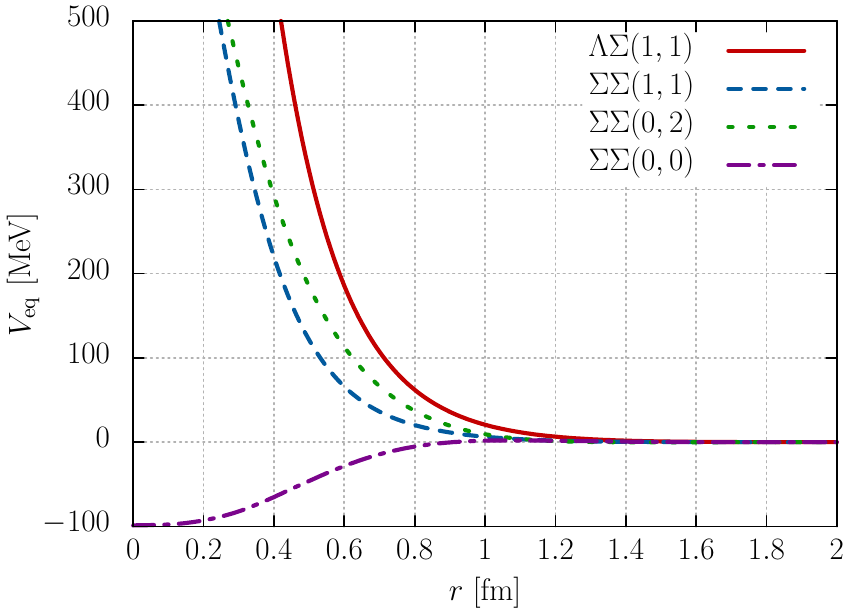}
  \\
  \Psfig{8.6cm}{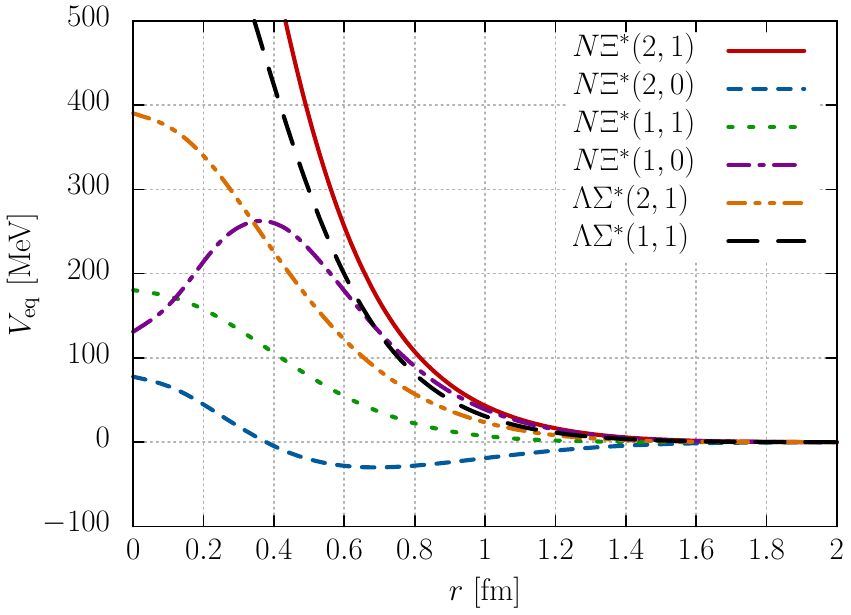}
  ~ ~  
  \Psfig{8.6cm}{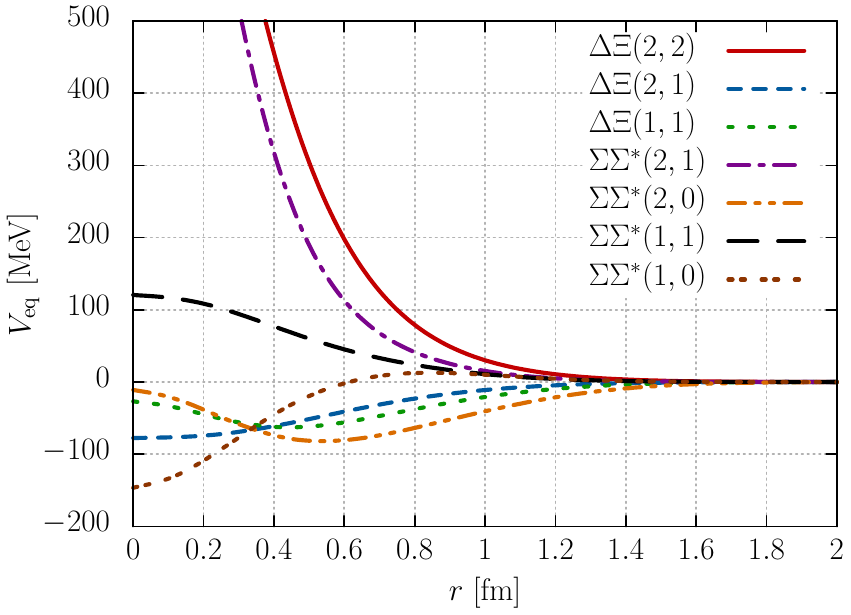}
  \\
  \Psfig{8.6cm}{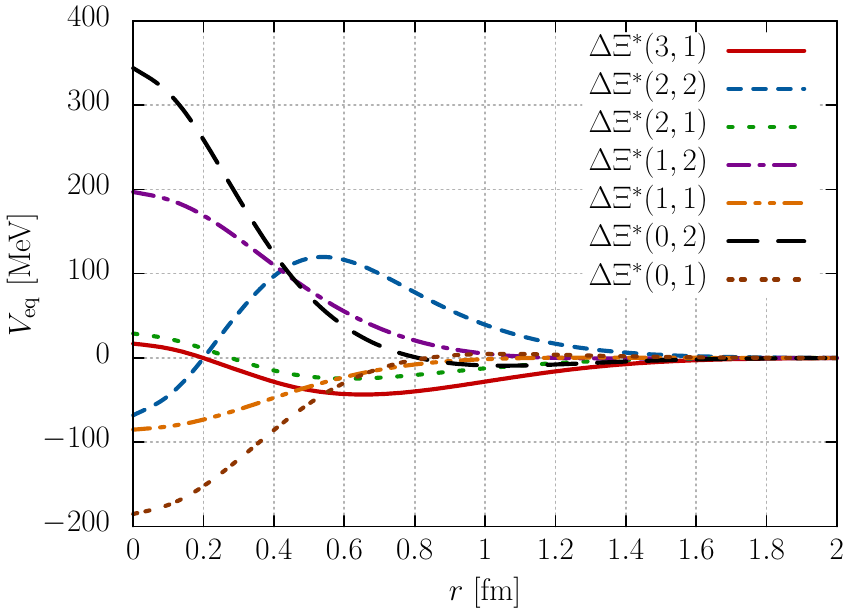}
  ~ ~
  \Psfig{8.6cm}{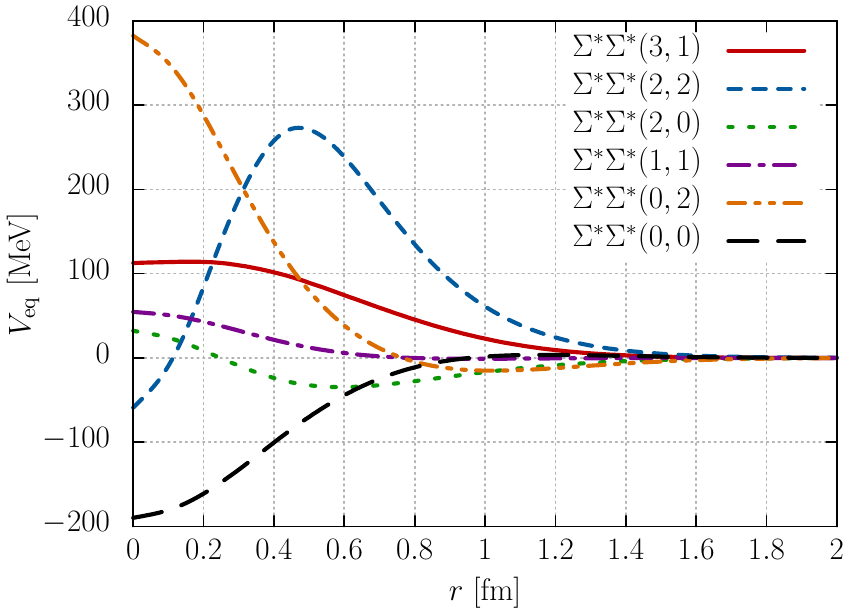}
  \caption{Equivalent local potentials for the baryon-baryon systems:
    strangeness $S = -2$ sector.}
  \label{fig:V_S-2}
\end{figure*}

\begin{figure*}[!t]
  \centering
  \Psfig{8.6cm}{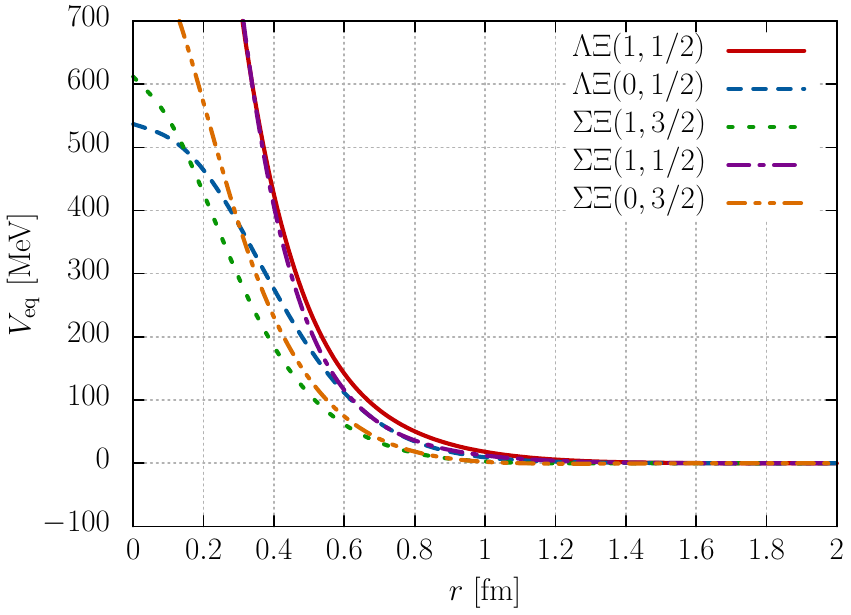}
  ~ ~
  \Psfig{8.6cm}{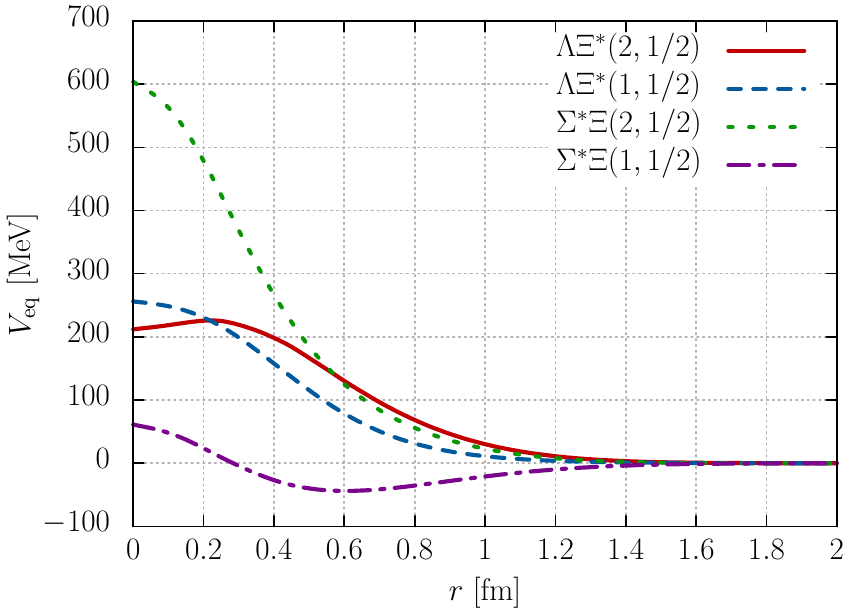}
  \\
  \Psfig{8.6cm}{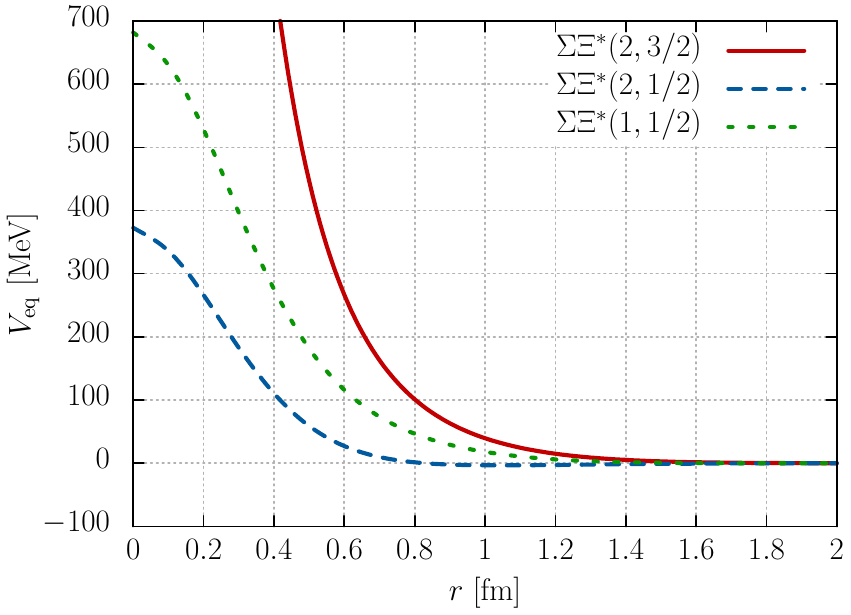}
  ~ ~ 
  \Psfig{8.6cm}{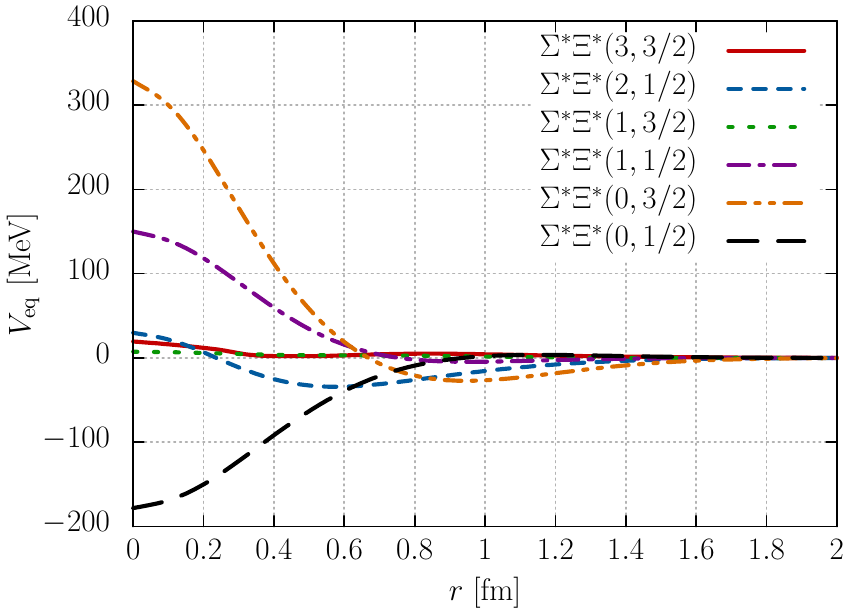}
  \caption{Equivalent local potentials for the baryon-baryon systems:
    strangeness $S = -3$ sector.}
  \label{fig:V_S-3}
\end{figure*}

\begin{figure*}[!t]
  \centering
  \Psfig{8.6cm}{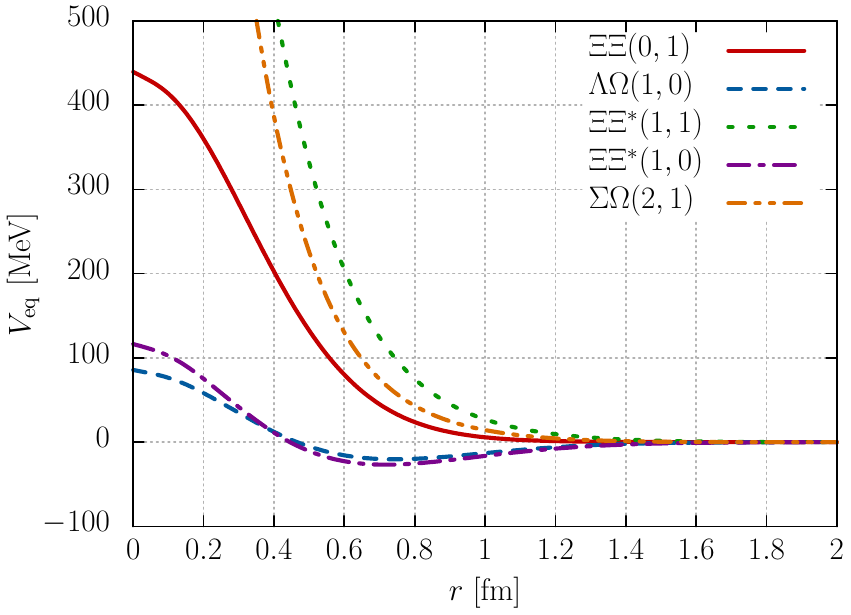}
  ~ ~
  \Psfig{8.6cm}{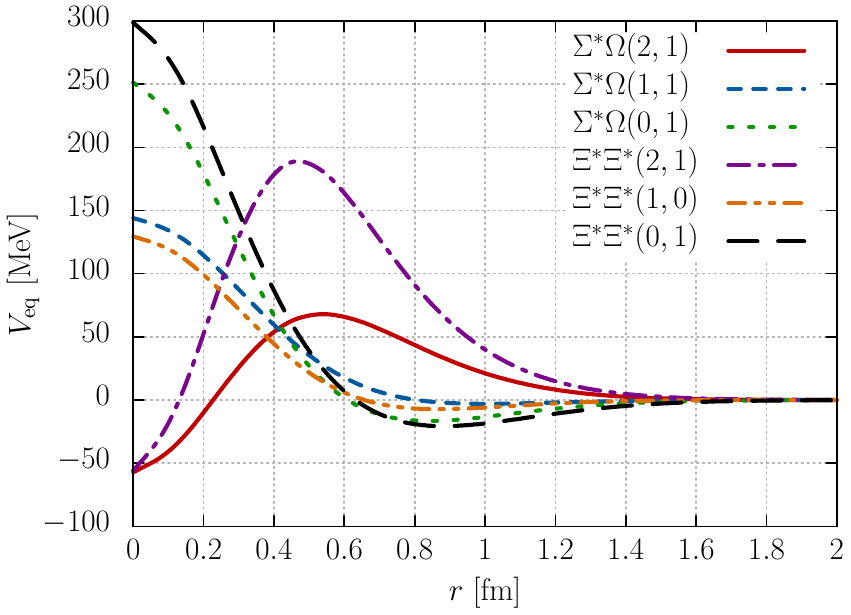}
  \caption{Equivalent local potentials for the baryon-baryon systems:
    strangeness $S = -4$ sector.}
  \label{fig:V_S-4}
\end{figure*}

\begin{figure}[!t]
  \centering
  \Psfig{8.6cm}{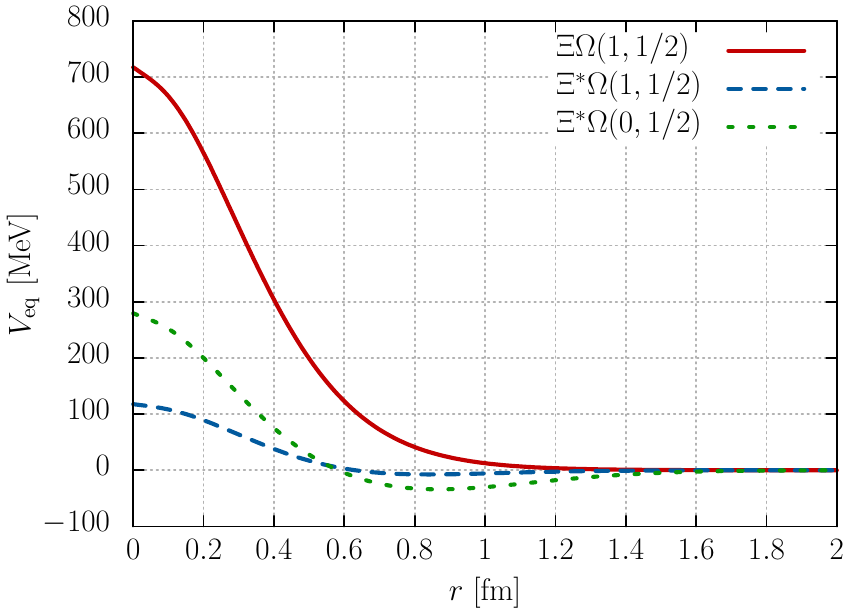}
  \caption{Equivalent local potentials for the baryon-baryon systems:
    strangeness $S = -5$ sector.}
  \label{fig:V_S-5}
\end{figure}

\begin{figure}[!t]
  \centering
  \Psfig{8.6cm}{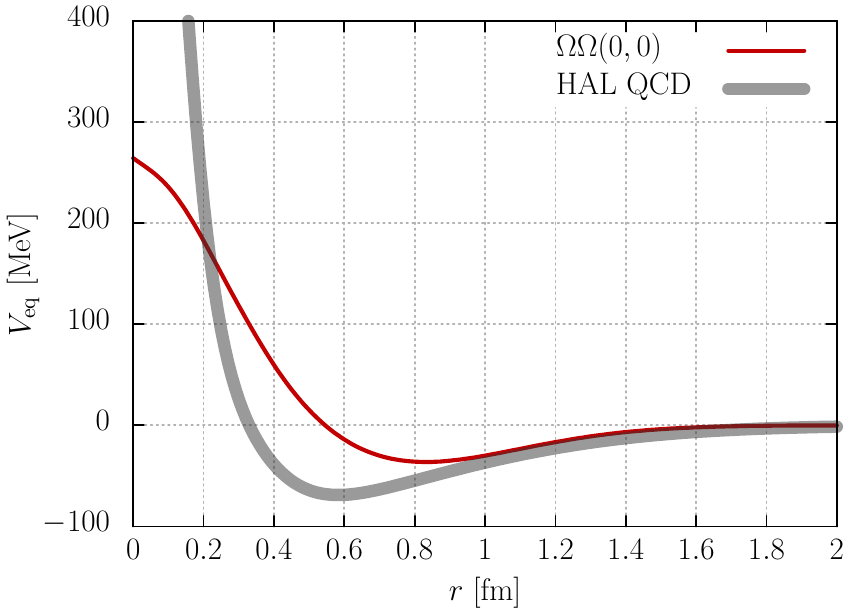}
  \caption{Equivalent local potentials for the baryon-baryon systems:
    strangeness $S = -6$ sector.  We also plot the HAL QCD potential for
    the $\Omega \Omega ( 0, 0 )$ system~\cite{Gongyo:2017fjb}.}
  \label{fig:V_S-6}
\end{figure}

\subsection{\boldmath Single-channel cases}
\label{sec:3A}

\subsubsection{Two-baryon channels}

Firstly, we summarize in Table~\ref{tab:N33} the two-baryon channels
composed of the ground-state baryons in $S$ wave.  In this study, we
specify the quantum numbers of the two-baryon channels using the total
spin $J$ and isospin $I$ as $( J , I )$.  We perform projection onto
the $( J , I )$ state using the Clebsch--Gordan coefficients in the
usual manner.  The same Table also lists the values of the spin-flavor
[33] components for the two-baryon channels, denoted as $N_{33}$.
This measures the contribution of totally antisymmetric states of six
quarks for two ground-state baryons in $S$ wave.  Therefore, a smaller
$N_{33}$ indicates stronger repulsion due to the Pauli exclusion
principle for quarks.  The spin-flavor [33] component $N_{33}$
corresponds to an eigenvalue of the normalization kernel $N ( \bm{r} ,
\bm{r}^{\prime} )$ associated with the eigenvector $\varphi ( \bm{r}
)$~\cite{Oka:2000wj}:
\begin{equation}
  \int d^{3} r^{\prime} N ( \bm{r} , \bm{r}^{\prime} )
  \varphi ( \bm{r}^{\prime} )
  = 2 N_{33} \varphi ( \bm{r} ) .
\end{equation}
We can calculate $N_{33}$ using the creation and annihilation
operators of baryons.  Namely, we have the formula
\begin{equation}
  N_{33} = \frac{1}{2}
  \braket{B_{a} B_{b} ( J , I )_{0} |
    B_{a} B_{b} ( J , I )_{0}} ,
\end{equation}
where $\ket{B_{a} B_{b} ( J , I )_{0}}$ is the two-baryon state with
the quantum numbers $( J , I )$ but without the coordinates of quarks:
\begin{align}
  & \ket{B_{a} B_{b} ( J , I )_{0} }
  \notag \\
  & \equiv
  \sum _{s_{a}, i_{a}} \braket{ J , J | S_{a} , s_{a} , S_{b} , J - s_{a}}
  \braket{ I , I | I_{a} , i_{a} , I_{b} , I - i_{a}}
  \notag \\
  & \phantom{=}
  \times \hat{w}^{( B_{a} ( s_{a} , i_{a} ) ) \dagger} 
  \hat{w}^{( B_{b} ( J - s_{a} , I - i_{a} ) ) \dagger} 
  \ket{0} ,
\end{align}
\begin{equation}
  \hat{w}^{( B ( s, i ) ) \dagger} 
  \equiv \sum _{\vec{\mu}} w_{\vec{\mu}}^{( B ( s , i ) )}
  \hat{a}_{\mu _{1}}^{\dagger} 
  \hat{a}_{\mu _{2}}^{\dagger} 
  \hat{a}_{\mu _{3}}^{\dagger} .
\end{equation}
Here, $S_{a}$ and $I_{a}$ are the spin and isospin values of $B_{a}$,
respectively, $B ( s , i )$ refers to the baryon $B$ with the third
components of spin $s$ and isospin $i$, $\braket{ J , j | S_{a} ,
  s_{a} , S_{b} , s_{b}}$ is the Clebsch--Gordan coefficient, and the
operator $\hat{a}_{\mu}^{\dagger}$ was introduced in
Eq.~\eqref{eq:a_oper}.  As shown in Table~\ref{tab:N33}, the values of
$N_{33}$ are scattered between zero and unity.  In particular, when
only single-channel cases are considered, the channels with $N_{33}$
close to unity usually contain decuplet baryons.  We expect that these
channels may avoid repulsive potentials due to the Pauli exclusion
principle for quarks.

\subsubsection{Strangeness $S = 0$}

Now, let us present the equivalent local potentials for the
baryon-baryon systems with strangeness $S = 0$ in
Fig.~\ref{fig:V_S0}.\footnote{All of the explicit potential values in
our model are provided in the ancillary files.}  As seen in the figure,
significant repulsive cores are observed in the potentials for the $N
N ( 1, 0 )$, $N N ( 0, 1 )$, $N \Delta ( 2, 1 )$, and $N \Delta ( 1, 2
)$ systems.  These cores begin to rise at approximately $\SI{1}{fm}$.
Because the spatial extension of each baryon in our model is typically
less than $\SI{0.5}{fm}$ (see Table~\ref{tab:mass}), our results
indicate that the interaction becomes significant when the distance
between two baryons reaches the sum of their respective radii.  The
values of the four potentials at the origin amount to more than
$\SI{800}{MeV}$, and the repulsion is strongest in the $N \Delta (1,
2)$ system, followed by $N \Delta (2, 1)$, $N (0, 1)$, and $N (1, 0)$.
On the other hand, the $\Delta \Delta$ systems have moderate repulsion
or even attractive cores at a short range.  In particular, thanks to
the attraction, the $\Delta \Delta ( 3 , 0 )$ system generates a bound
state, whose properties will be presented later.  The behavior of
these potentials is quantitatively similar to the results in
Ref.~\cite{Oka:1981rj}.  Therefore, our model with more precise wave
functions strengthens the discussions in Ref.~\cite{Oka:1981rj}.
Here, we note that the $N \Delta ( 2 , 2 )$, $N \Delta ( 1 , 1 )$,
$\Delta \Delta ( 3 , 2 )$, and $\Delta \Delta ( 2 , 3 )$ systems have
singular equivalent local potentials because the wave functions have
nodes caused by unphysical forbidden states due to the Pauli exclusion
principle for quarks.  We mark these channels with a dagger
($\dagger$) in Table~\ref{tab:N33} and do not show these singular
equivalent local potentials.

To confirm the mechanism of attraction/repulsion, we decompose the
potentials by considering the following cases:
\begin{itemize}
\item Case of the color Coulomb plus linear confining force (CL): only
  the color Coulomb plus linear confining potential $V_{\rm CL} ( r )$
  is considered in the RGM equation~\eqref{eq:RGM}, while
  $V_{\rm ss} ( r ) = 0$ and $N ( \bm{r} , \bm{r}^{\prime} ) = \delta
  ( \bm{r} - \bm{r}^{\prime} )$.

\item Case of the color Coulomb, linear confining, and color magnetic
  forces (CL+SS): both the color Coulomb plus linear confining
  potential $V_{\rm CL} ( r )$ and the color magnetic potential
  $V_{\rm ss} ( r )$ are considered in the RGM
  equation~\eqref{eq:RGM}, while $N ( \bm{r} , \bm{r}^{\prime} ) =
  \delta ( \bm{r} - \bm{r}^{\prime} )$.

\item Case of the full calculation (Full).
\end{itemize}
The resulting equivalent local potentials for the $N N ( 1 , 0 )$ and
$\Delta \Delta ( 3 , 0 )$ systems are plotted in
Fig.~\ref{fig:decomp}.  The $N N ( 1 , 0 )$ system has a moderate
interaction due to the color Coulomb plus linear confining force (CL,
the dashed line in Fig.~\ref{fig:decomp}), but when the color magnetic
force is included (CL+SS, the dotted line in Fig.~\ref{fig:decomp}),
it generates a highly repulsive potential.  However, the repulsion
becomes weaker when the normalization kernel $N ( \bm{r} ,
\bm{r}^{\prime} )$ is considered (Full, the solid line in
Fig.~\ref{fig:decomp}).  On the other hand, the $\Delta \Delta ( 3 , 0
)$ system in the CL case has an attraction in the medium range
$\SI{1}{fm} < r < \SI{2}{fm}$ that is strong enough to produce a bound
state (the double dot-dashed line in Fig.~\ref{fig:decomp}).  This
attraction grows when the color magnetic force is included
(the long dashed line in Fig.~\ref{fig:decomp}), and the short-range
repulsion becomes moderate when the normalization kernel is applied
(the dot-dashed line in Fig.~\ref{fig:decomp}).  We emphasize here
that the strength of the attraction/repulsion generated by the color
Coulomb plus linear confining force is correlated with the values of
$N_{33}$: a larger $N_{33}$ generates stronger attraction in the CL
case.  Indeed, the $\Delta \Delta ( 0 , 3 )$ system has the same
equivalent local potential in the CL case as the $\Delta \Delta ( 3 ,
0 )$ system, but the repulsive color magnetic interaction in the
$\Delta \Delta ( 0 , 3 )$ system distorts the potential, resulting in
a repulsive potential in the full calculation (the double dot-dashed
line in the right panel of
Fig.~\ref{fig:V_S0}).  This implies that, even if both the
$\Delta \Delta ( 3 , 0 )$ and $\Delta \Delta ( 0 , 3 )$ dibaryon
states exist as predicted in Ref.~\cite{Dyson:1964xwa}, their nature
will be qualitatively different from each other.  In summary, as
discussed in Ref.~\cite{Oka:1981rj}, both the Pauli exclusion
principle for quarks and color magnetic interactions are essential for
the behavior of baryon-baryon interactions at short distances.

In Fig.~\ref{fig:V_DeltaDelta_HAL} we compare our result for the
equivalent local potential in the $\Delta \Delta ( 3 , 0 )$ system
with the analytic form of the HAL QCD potential with a heavy pion mass
$m_{\pi} = \SI{679}{MeV}$~\cite{Gongyo:2020pyy}.  Both potentials
provide sufficient attraction to generate a $\Delta \Delta ( 3 , 0 )$
bound state, but the details differ.  In particular, at the origin,
while no repulsive core is observed in the HAL QCD
potential, the potential in our quark model exhibits weak repulsion,
which originates from the color Coulomb plus linear force (see
Fig.~\ref{fig:decomp}).  Such a discrepancy can be discussed, for
example, by adding meson exchange contributions to our potential, as a
previous study has shown that the exchanges of scalar and pseudoscalar
mesons can be superposed on the quark-model potential without
introducing a double-counting problem~\cite{Yazaki:1989rh}.  In
addition, the quark mass dependence of the potential in both quark
models and lattice QCD simulations would be important.  However, it
should be noted that the potential itself is not observable and the
value of the potential at the origin is not crucial for the generation
of the bound state.  Therefore, to evaluate the strength of the
potentials, we introduce a quantity
\begin{equation}
  \beta
  \equiv - \frac{16 \mu _{a b}}{\pi ^{2}}
  \int _{0}^{\infty} d r \, r V_{\rm eq} ( r ) .
  \label{eq:beta}
\end{equation}
This quantity is motivated by the condition that a three-dimensional
potential well generates a bound state.  Namely, a three-dimensional
potential well $V ( r ) = - V_{0} \theta ( a - r )$, with a potential
depth $V_{0}$, range $a$, and the Heaviside step function $\theta ( x
)$, generates a bound state if $\beta \ge 1$.  The $\Delta \Delta ( 3
, 0 )$ potential in our quark model provides $\beta = 1.93$, while the
HAL QCD potential $\beta = 2.51$ with heavy $\Delta$ mass $M_{\Delta}
= \SI{1677}{MeV}$~\cite{Gongyo:2020pyy}.  These values imply that the
HAL QCD potential for the $\Delta \Delta ( 3 , 0 )$ system with the
heavy $\Delta$ mass is stronger than that in our model with the almost
physical $\Delta$ mass.

\subsubsection{Strangeness $S < 0$}

The equivalent local potentials with strangeness $S = -1$, $-2$, $-3$,
$-4$, $-5$, and $-6$ are plotted in Figs.~\ref{fig:V_S-1},
\ref{fig:V_S-2}, \ref{fig:V_S-3}, \ref{fig:V_S-4}, \ref{fig:V_S-5},
and \ref{fig:V_S-6}, respectively.  We note that the $N \Omega$ and
$\Delta \Omega$ interactions are absent in the single-channel cases
because the shuffling of quarks associated with the quark-quark
interaction inevitably leads to the transition to inelastic channels
[see Fig.~\ref{fig:shuffle}(b)].  As shown in the figures, the
potentials exhibit both attractive and repulsive behavior, depending
on the channels.\footnote{In the strangeness $S = - 2$ sector, the
baryon-baryon interaction energy in the constituent quark model was
calculated in a static way by subtracting out isolated baryon masses
and relative kinetic energy of two baryons from the total energy of a
compact six-quark state~\cite{Park:2019bsz}.  On the other hand, in
the present study, we calculate the baryon-baryon potentials in a
dynamical way by solving the RGM equation.}  The potentials start to
deviate from zero at about $\SI{1}{fm}$ in almost all channels, which
again indicate that the interaction becomes significant when the
distance between two baryons reaches the sum of their respective
radii.  The exception is the cases of $N_{33} \approx 1$ [for example
  $\Delta \Sigma ^{\ast} ( 3 , 1/2 )$] and $N_{33} \approx 0$ [for
  example $\Delta \Xi ^{\ast} ( 2 , 2 )$], in which the cancellation
among the contributions in the color Coulomb plus linear confining
forces does not occur.  Consequently, the potentials become detectable
even at distances with $r > \SI{1}{fm}$, which corresponds to the
situation that just the tails of the single-baryon wave functions
start to overlap (see Fig.~\ref{fig:Prho}): the potentials at longer
ranges exhibits negative (positive) growth in the $N_{33} \approx 1$
($N_{33} \approx 0$) case.


In Fig.~\ref{fig:V_S-6}, we also plot the analytic form of the $\Omega
\Omega ( 0 , 0 )$ potential in the HAL QCD method.  The behavior of
the potentials in the two approaches is qualitatively consistent.  The
mechanism of the $\Omega \Omega ( 0 , 0 )$ potential in our quark
model is similar to that of the $\Delta \Delta ( 0 , 3 )$: the
repulsion near the origin is the sum of the contributions from the
color Coulomb plus linear confining force and repulsive color magnetic
force, while the medium-range attraction originates from the color
Coulomb plus linear confining force.  Because the strange quark is
heavier than the up and down quarks, the repulsive color magnetic
force, which is proportional to $1 / m_{s}^{2}$ [see
  Eq.~\eqref{eq:CplusL}], becomes weaker in the $\Omega \Omega ( 0 , 0
)$ system, and hence the medium-range attraction persists.  However,
the attraction is not sufficient in our model to generate a bound
state in contrast to the HAL QCD potential.  Indeed, the strength of
the potential $\beta$ in Eq.~\eqref{eq:beta} amounts to $\beta =
0.248$ ($0.937$) in our model (HAL QCD method).  This discrepancy
could be compensated for by including exchange forces of mesons such
as the $\eta$ and $\sigma$ mesons.

\subsubsection{Bound states}

\begin{table}[!t]
  \caption{Properties of the bound states.}
  \label{tab:BS}
  \centering
  \begin{ruledtabular}
    \begin{tabular}{lcc}
      System & $B$ [MeV] & $\sqrt{\braket{r_{\rm D}^{2}}}$ [fm]
      \\
      \hline
      $\Delta \Delta ( 3 , 0 )$ & 13.1 & 1.70
      \\
      $\Delta \Sigma ( 2 , 1/2 )$ & \phantom{0}6.9 & 2.03
      \\
      $\Delta \Sigma ^{\ast} ( 3 , 1/2 )$ & 12.6 & 1.68
      \\
      $\Sigma \Sigma ^{\ast} ( 2 , 0 )$ & \phantom{0}0.7 & 5.24
    \end{tabular}
  \end{ruledtabular}
\end{table}

\begin{figure}[!t]
  \centering
  \Psfig{8.6cm}{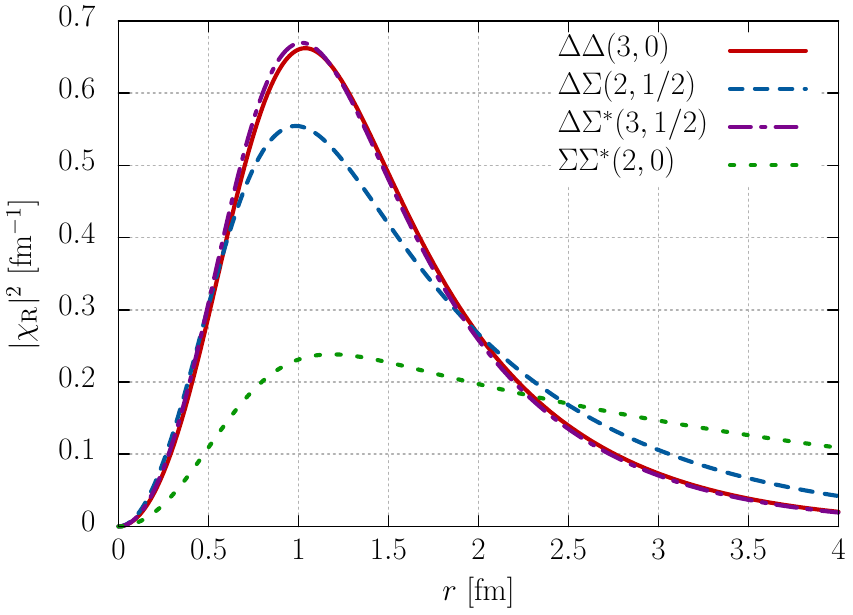}
  \caption{Square of the relative wave functions for the dibaryon
    bound states.}
  \label{fig:chi2}
\end{figure}

In our quark model calculations in single-channel cases, we find bound
states of the $\Delta \Delta ( 3 , 0 )$, $\Delta \Sigma ( 2 , 1/2 )$,
$\Delta \Sigma ^{\ast} ( 3 , 1/2 )$, and $\Sigma \Sigma ^{\ast} ( 2 ,
0 )$ systems.  Table~\ref{tab:BS} shows the binding energies $B$ and
root mean squared distances between two baryons $\sqrt{\braket{r_{\rm
      D}^{2}}}$ of these bound states, which are defined using the
normalized wave functions of the relative motion as
\begin{equation}
  \int _{0}^{\infty} d r | \chi _{\rm R} ( r ) |^{2} = 1 ,
  \quad
  \braket{r_{\rm D}^{2}}
  \equiv \int _{0}^{\infty} d r \, r^{2} | \chi _{\rm R} ( r ) |^{2} .
\end{equation}
We also plot the square of the relative wave functions $| \chi _{\rm
  R} ( r ) |^{2}$ in Fig.~\ref{fig:chi2}.  As shown by the root mean
squared distances in Table~\ref{tab:BS} and the squared wave functions
in Fig.~\ref{fig:chi2}, the spatial extension of the bound states
largely exceeds the typical size of hadrons of $\SI{1}{fm}$.  This
strongly suggests that these dibaryon states are hadronic molecules
rather than compact hexaquark states.  The $\Delta \Delta ( 3 , 0 )$
bound state can be interpreted as the $d^{\ast} (2380)$ recently
confirmed in experiments~\cite{WASA-at-COSY:2011bjg}.  The $\Delta
\Sigma ^{\ast} ( 3, 1/2 )$ bound state was predicted in
Ref.~\cite{Li:2000cb} in the chiral SU(3) quark model.  Furthermore,
the wave functions of the $\Delta \Delta ( 3 , 0 )$ and $\Delta \Sigma
^{\ast} ( 3, 1/2 )$ are very similar to each other.  Indeed, both the
normalization kernel $N ( \bm{r} , \bm{r}^{\prime} )$ and interaction
term $V_{\rm int} ( \bm{r} , \bm{r}^{\prime} )$ are very similar in
the $\Delta \Delta ( 3 , 0 )$ and $\Delta \Sigma ^{\ast} ( 3, 1/2 )$
systems, because they are members of the flavor antidecuplet generated
by two decuplet baryons.  This fact implies that bound states of the
$\Delta \Xi ^{\ast}$-$\Sigma ^{\ast} \Sigma ^{\ast} ( 3 , 1 )$ and
$\Delta \Omega$-$\Sigma ^{\ast} \Xi ^{\ast} ( 3 , 3/2 )$ systems in
coupled channels exist as the other members of the flavor
antidecuplet, although the attraction is not sufficient in these
systems in the single-channel cases.  On the other hand, the
 $\Delta
\Sigma ( 2 , 1/2 )$ and $\Sigma \Sigma ^{\ast} ( 2 , 0 )$ systems
couple to lower baryon-baryon channels $N \Sigma ^{\ast}$ and $N \Xi
^{\ast}$, respectively, in $S$ wave.  Therefore, it is necessary to
take into account the coupled-channels effects to determine the
properties of these bound states.

\subsection{\boldmath Coupled-channels cases}
\label{sec:3B}

In the previous subsection, we only considered the single-channel
cases where transitions between inelastic channels were neglected.
However, in several systems, such transitions may play a significant
role.  Therefore, in this subsection, we examine the coupled-channels
effects.

\subsubsection{Flavor antidecuplet states with $J = 3$}

\begin{figure}[!t]
  \centering
  \Psfig{8.6cm}{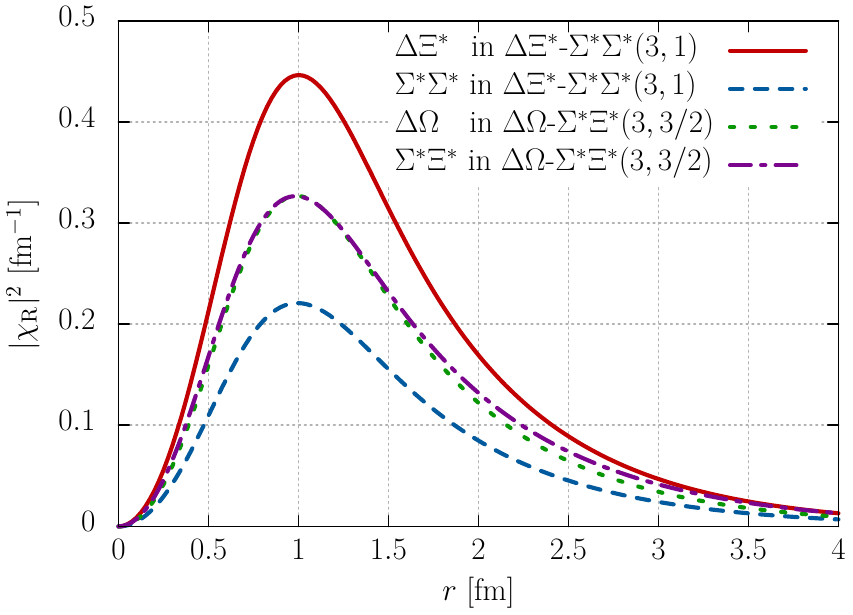}
  \caption{Square of the relative wave functions for the dibaryon
    bound states in coupled channels.}
  \label{fig:chi2_CC}
\end{figure}

Firstly, we consider the coupled channels of $\Delta \Xi
^{\ast}$-$\Sigma ^{\ast} \Sigma ^{\ast} ( 3 , 1 )$ and $\Delta
\Omega$-$\Sigma ^{\ast} \Xi ^{\ast} ( 3 , 3/2 )$.  They are important
because they belong to the flavor antidecuplet generated by two
decuplet baryons, together with the $\Delta \Delta ( 3 , 0 )$ and
$\Delta \Sigma ^{\ast} ( 3 , 1/2 )$ states, which are bound as in the
previous subsection.  We allow transitions between inelastic channels
via $N ( \bm{r} , \bm{r}^{\prime} )$ and $V_{\rm int} ( \bm{r} ,
\bm{r}^{\prime} )$ and solve the RGM equation~\eqref{eq:RGM}.  As a
result, we find bound states below the lower thresholds,
\textit{i.e.}, the $\Sigma ^{\ast} \Sigma ^{\ast}$ and $\Sigma ^{\ast}
\Xi ^{\ast}$, respectively, in our model\footnote{Because we use the
baryon masses in the constituent quark model, a reversal of the
thresholds occurs.  In experiments, the lower thresholds correspond to
$\Delta \Xi ^{\ast}$ and $\Delta \Omega$, respectively.}.  The $\Delta
\Omega$ bound state was predicted in Ref.~\cite{Li:1999bc} in the
chiral SU(3) quark model.  In the strangeness $S = -2$ sector, the
binding energy of the bound state measured from the $\Delta \Xi
^{\ast}$ ($\Sigma ^{\ast} \Sigma ^{\ast}$) threshold is
$\SI{11.7}{MeV}$ ($\SI{11.1}{MeV}$).  In the strangeness $S = -3$
sector, the binding energy is $\SI{11.1}{MeV}$ ($\SI{9.0}{MeV}$) from
the $\Delta \Omega$ ($\Sigma ^{\ast} \Xi ^{\ast}$) threshold.  In
Fig.~\ref{fig:chi2_CC} we plot the squared wave functions of the bound
states with the normalization
\begin{equation}
  \sum _{i} \int _{0}^{\infty} d r | \chi _{\mathrm{R}, i} ( r ) |^{2} = 1 ,
  \label{eq:chi2_CC}
\end{equation}
where $i$ denotes the channels.  As one can see, the squared wave
functions have nonzero values even above the typical size of hadrons
of $\SI{1}{fm}$, indicating that these dibaryon states are hadronic
molecules rather than compact hexaquark states, as the $\Delta \Delta
( 3 , 0 )$ and $\Delta \Sigma ^{\ast} ( 3 , 1/2 )$ bound states.
Furthermore, we can evaluate the fractions of the $\Delta \Xi
^{\ast}$, $\Sigma ^{\ast} \Sigma ^{\ast}$, $\Delta \Omega$, and
$\Sigma ^{\ast} \Xi ^{\ast}$ components in the bound states by
calculating each term of the summation in Eq.~\eqref{eq:chi2_CC}.  The
resulting fractions are $\SI{67}{\percent}$ ($\SI{33}{\percent}$) for
the $\Delta \Xi ^{\ast}$ ($\Sigma ^{\ast} \Sigma ^{\ast}$) component
in the $\Delta \Xi ^{\ast}$-$\Sigma ^{\ast} \Sigma ^{\ast} ( 3 , 1 )$
bound state, and $\SI{49}{\percent}$ ($\SI{51}{\percent}$) for the
$\Delta \Omega$ ($\Sigma ^{\ast} \Xi ^{\ast}$) component in the
$\Delta \Omega$-$\Sigma ^{\ast} \Xi ^{\ast} ( 3 , 3/2 )$ bound state.
We summarize the properties of the bound states in Table~\ref{tab:CC}.

\begin{table}[!t]
  \caption{Properties of the bound and resonance states in coupled
    channels.  In the $\Delta \Xi ^{\ast}$-$\Sigma ^{\ast} \Sigma
    ^{\ast} ( 3 , 1 )$ and $\Delta \Omega$-$\Sigma ^{\ast} Xi ^{\ast}
    ( 3 , 3/2 )$ cases, the binding energy $B$ is measured from the
    $\Delta \Xi ^{\ast}$ and $\Delta \Omega$ thresholds in our model,
    respectively.  The decay width $\Gamma$ is also included in the $N
    \Omega ( 2 , 1/2 )$ and $\Delta \Sigma ( 2 , 1/2 )$ channels.}
  \label{tab:CC}
  \centering
  \begin{ruledtabular}
    \begin{tabular}{lccc}
      System & $B$ [MeV] & $\Gamma$ [MeV] & Note
      \\
      \hline
      $\Delta \Xi ^{\ast}$-$\Sigma ^{\ast} \Sigma ^{\ast} ( 3 , 1 )$
      & 11.7 & --- 
      & $\Delta \Xi ^{\ast}$ $\SI{67}{\percent}$, 
      $\Sigma ^{\ast} \Sigma ^{\ast}$ $\SI{33}{\percent}$
      \\
      $\Delta \Omega$-$\Sigma ^{\ast} \Xi ^{\ast} ( 3 , 3/2 )$
      & 11.1 & ---
      & $\Delta \Omega$ $\SI{49}{\percent}$, 
      $\Sigma ^{\ast} \Xi ^{\ast}$ $\SI{51}{\percent}$
      \\
      $N \Omega ( 2 , 1/2 )$ & 10.3 & 4.6 & With meson exchanges
      \\
      $\Delta \Sigma ( 2 , 1/2 )$ & $-40.6$ & 76.0 & Decay to $N \Sigma ^{\ast}$
    \end{tabular}
  \end{ruledtabular}
\end{table}

\begin{figure*}[!t]
  \centering
  \Psfig{8.6cm}{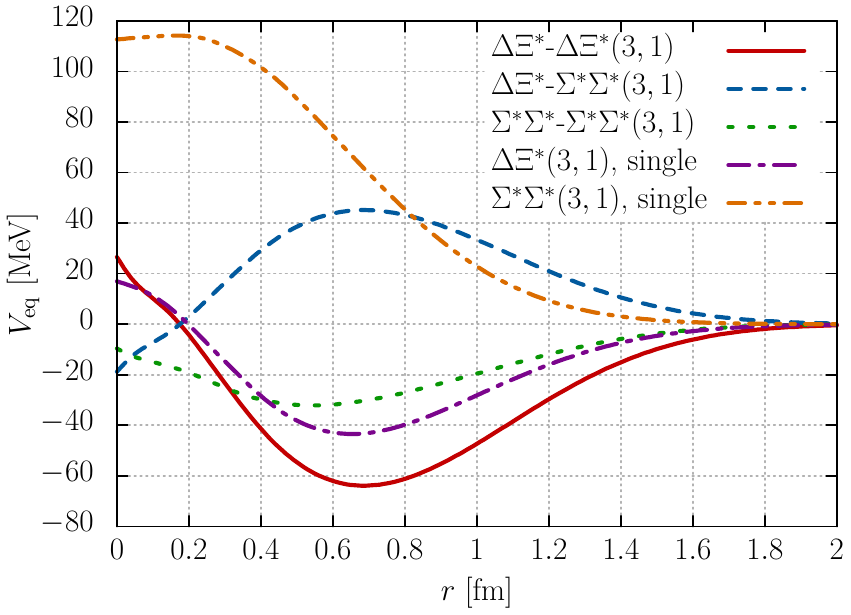}
  ~ ~
  \Psfig{8.6cm}{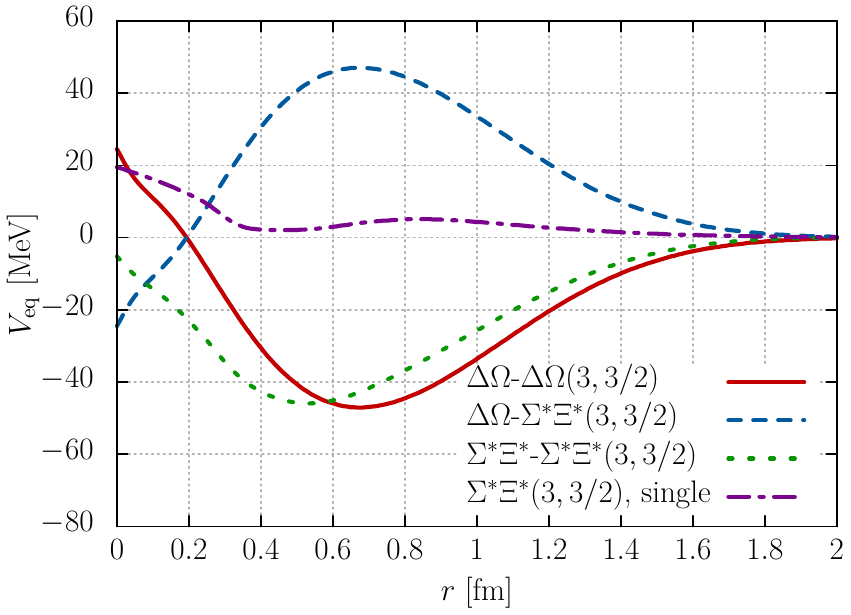}
  \caption{Equivalent local potentials for the baryon-baryon systems:
    flavor antidecuplet states in coupled channels.}
  \label{fig:V_DeltaXiS_CC}
\end{figure*}

From the relative wave functions of the bound state $\chi
_{\mathrm{R}, i}$, we would like to extract the local coupled-channels
potentials.  However, in the coupled-channels cases, this task is not
straightforward in contrast to the single-channel cases, because the
number of coupled-channels potentials is generally larger than the
number of wave equations.  In particular, the wave equations in a
two-channel problem become
\begin{align}
  & \begin{pmatrix}
    \displaystyle
    - \frac{1}{2 \mu _{1}} \frac{d^{2}}{d r^{2}} + V_{11} ( r )
    & V_{12} ( r )
    \\
    V_{21} ( r )
    & \displaystyle
    \Delta - \frac{1}{2 \mu _{2}} \frac{d^{2}}{d r^{2}} + V_{22} ( r )
  \end{pmatrix}
  \begin{pmatrix}
    \displaystyle \vphantom{\frac{d^{2}}{d r^{2}}}
    \chi _{\mathrm{R}, 1} 
    \\
    \displaystyle \vphantom{\frac{d^{2}}{d r^{2}}}
    \chi _{\mathrm{R}, 2} 
  \end{pmatrix}
  \notag \\
  & = - B 
  \begin{pmatrix}
    \chi _{\mathrm{R}, 1}
    \\
    \chi _{\mathrm{R}, 2}
  \end{pmatrix}
\end{align}
where $\Delta$ is the difference of the two threshold values and $B$
is the binding energy measured from the threshold of the first
channel.  We have obtained the relative wave functions $\chi
_{\mathrm{R}, 1}$ and $\chi _{\mathrm{R}, 2}$ by solving the RGM
equation, but they are not sufficient to uniquely determine the
coupled-channels potentials $V_{11}$, $V_{12}$, $V_{21}$, and
$V_{22}$.  To solve this problem, we make three assumptions: 1) the
potential is dominated by the antidecuplet contribution, 2) inelastic
potentials are symmetric, \textit{i.e.}, $V_{12} = V_{21}$, and 3) the
weight of each component in the antidecuplet is fixed purely by the
Clebsch--Gordan coefficient.  For example, in the $\Delta \Xi
^{\ast}$-$\Sigma ^{\ast} \Sigma ^{\ast} ( 3 , 1 )$ case, because the
antidecuplet state $\overline{\bm{10}}$ has the relation
\begin{equation}
  \ket{\overline{\bm{10}} ( 3 , 1 )}
  = \sqrt{\frac{2}{3}} \ket{ \Delta \Xi ^{\ast} ( 3 , 1 ) }
  - \sqrt{\frac{1}{3}} \ket{ \Sigma ^{\ast} \Sigma ^{\ast} ( 3 , 1 ) } ,
  \label{eq:10bar1}
\end{equation}
the equivalent local coupled-channels potentials can be evaluated by
the formulae:
\begin{equation}
  V_{\rm eq}^{\Delta \Xi ^{\ast}\textrm{-}\Sigma ^{\ast} \Sigma ^{\ast}} ( r )
  = V_{\rm eq}^{\Sigma ^{\ast} \Sigma ^{\ast}\textrm{-}\Delta \Xi ^{\ast}} ( r )
  = - \frac{1}{\sqrt{2}}
  V_{\rm eq}^{\Delta \Xi ^{\ast}\textrm{-}\Delta \Xi ^{\ast}} ( r ) ,
\end{equation}
\begin{equation}
  V_{\rm eq}^{\Delta \Xi ^{\ast}\textrm{-}\Delta \Xi ^{\ast}} ( r )
  = \frac{\displaystyle \frac{1}{2 \mu _{\Delta \Xi ^{\ast}}}
    \frac{d^{2} \chi _{\mathrm{R} , \Delta \Xi ^{\ast}}}{d r^{2}}
    - B \chi _{\mathrm{R} , \Delta \Xi ^{\ast}} ( r )}
  {\displaystyle \chi _{\mathrm{R}, \Delta \Xi ^{\ast}} ( r )
    - \frac{1}{\sqrt{2}} \chi _{\mathrm{R}, \Sigma ^{\ast} \Sigma ^{\ast}} ( r )} ,
\end{equation}
\begin{align}
  & V_{\rm eq}^{\Sigma ^{\ast} \Sigma ^{\ast}\textrm{-}\Sigma ^{\ast} \Sigma ^{\ast}} ( r )
  = \frac{1}{\chi _{\mathrm{R}, \Sigma ^{\ast} \Sigma ^{\ast}} ( r )}
  \left [ \frac{1}{2 \mu _{\Sigma ^{\ast} \Sigma ^{\ast}}}
  \frac{d^{2} \chi _{\mathrm{R} , \Sigma ^{\ast} \Sigma ^{\ast}}}{d r^{2}} \right .
  \notag \\
  & \left . \phantom{\frac{d^{2}}{d r^{2}}} 
    - ( B + \Delta ) \chi _{\mathrm{R} , \Sigma ^{\ast} \Sigma ^{\ast}} ( r )
    - V_{\rm eq}^{\Sigma ^{\ast} \Sigma ^{\ast}\textrm{-}\Delta \Xi ^{\ast}} ( r )
    \chi _{\mathrm{R}, \Delta \Xi ^{\ast}} ( r )
    \right ] .
\end{align}
We plot the calculated local coupled-channels potentials for the
$\Delta \Xi ^{\ast}$-$\Sigma ^{\ast} \Sigma ^{\ast} ( 3 , 1 )$ bound
state in the left panel of Fig.~\ref{fig:V_DeltaXiS_CC} together with
the potentials in the single-channel case denoted as ``single''.  The
elastic $\Delta \Xi ^{\ast}$ potential (the solid line in the left
panel of Fig.~\ref{fig:V_DeltaXiS_CC}) becomes more attractive
compared to the single-channel case (the dot-dashed line), and the
elastic $\Sigma ^{\ast} \Sigma ^{\ast}$ potential (the dotted line)
changes to attraction.  Similarly, we can evaluate the equivalent
local potentials for the $\Delta \Omega$-$\Sigma ^{\ast} \Xi ^{\ast}$
bound state via the relation
\begin{equation}
  \ket{\overline{\bm{10}} ( 3 , 3/2 )}
  = \frac{1}{\sqrt{2}} \ket{ \Delta \Omega ( 3 , 3/2 ) }
  - \frac{1}{\sqrt{2}} \ket{ \Sigma ^{\ast} \Xi ^{\ast} ( 3 , 3/2 ) } .
  \label{eq:10bar2}
\end{equation}
The result is plotted in the right panel of
Fig.~\ref{fig:V_DeltaXiS_CC}.  Interestingly, while the $\Delta
\Omega$ interaction does not occur in the single-channel case, as the
quark shuffle associated with the quark-quark interaction inevitably
leads to the transition to inelastic channels, the elastic $\Delta
\Omega$ interaction emerges via the coupling to the $\Sigma ^{\ast}
\Xi ^{\ast}$ and is attractive.  In addition, the attraction of the
elastic $\Sigma ^{\ast} \Xi ^{\ast}$ potential grows when the coupled
channels are taken into account.

We note that, for the flavor antidecuplet $\Delta \Xi ^{\ast}$-$\Sigma
^{\ast} \Sigma ^{\ast} ( 3 , 1 )$ state with the weight in
Eq.~\eqref{eq:10bar1} and $\Delta \Omega$-$\Sigma ^{\ast} \Xi ^{\ast}
( 3 , 3/2 )$ state with the weight in Eq.~\eqref{eq:10bar2}, the
normalization kernel $N ( \bm{r} , \bm{r}^{\prime} )$ and interaction
term $V_{\rm int} ( \bm{r} , \bm{r}^{\prime} )$ are again similar to
those of the $\Delta \Delta ( 3 , 0 )$ state.  Therefore, we extend
the discussion on the $\Delta \Delta ( 3 , 0 )$ bound state and
conclude that both the Pauli exclusion principle for quarks and color
magnetic interactions are essential for the generation of the bound
states in the flavor antidecuplet.

\subsubsection{Flavor octet states with $J = 2$}

Next, we consider the $N \Omega ( 2 , 1/2 )$ system.  In the
single-channel case, the $N \Omega$ interaction is absent because the
shuffling of quarks associated with the quark-quark interaction
inevitably leads to the transition to inelastic channels, which is the
same as the $\Delta \Omega$ interaction.  However, the coupled
channels $\Lambda \Xi ^{\ast}$, $\Sigma ^{\ast} \Xi$, and $\Sigma \Xi
^{\ast}$, whose thresholds are above but close to the $N \Omega$
threshold, may bring attraction to the $N \Omega ( 2 , 1/2 )$ system.
Indeed, the HAL QCD method has predicted a strong attraction in the $N
\Omega ( 2 , 1/2 )$ system~\cite{HALQCD:2018qyu}.  As discussed in
Ref.~\cite{Sekihara:2018tsb}, such attraction cannot be provided by
conventional meson exchanges, so it is natural to examine the $N
\Omega$ interaction in terms of quark degrees of freedom.  Previous
research on the $N \Omega$ interaction in the constituent quark model
can be found in, \textit{e.g.}, Ref.~\cite{Oka:1988yq}, and we revisit
this using more precise wave functions.

\begin{figure}[!t]
  \centering
  \Psfig{8.6cm}{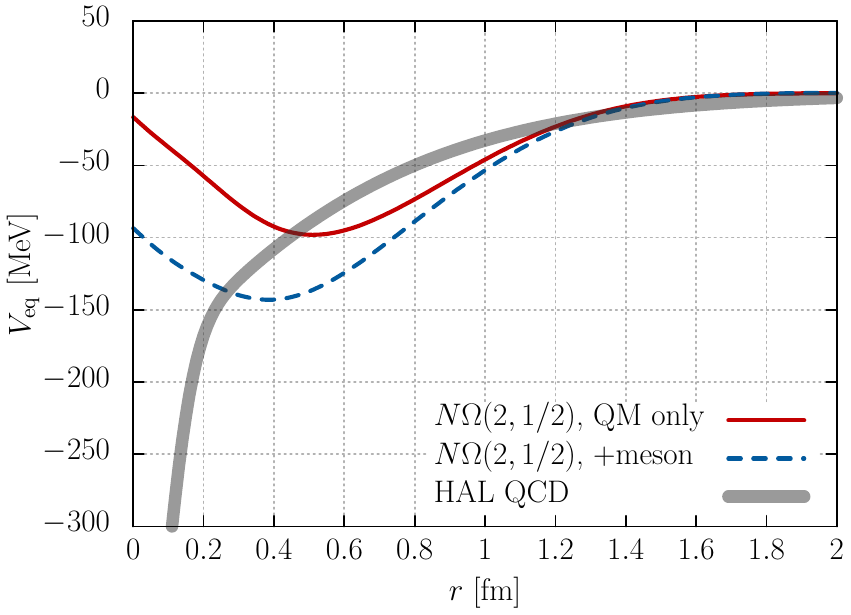}
  \caption{Equivalent local potential for the $N \Omega$ system in our
    model (solid line).  We also plot the real part of the potential
    with the meson exchange contributions added (dashed line) and the
    HAL QCD potential for the $N \Omega ( 2, 1/2 )$
    system~\cite{HALQCD:2018qyu} (thick line).}
  \label{fig:V_NOmega_CC}
\end{figure}

By solving the RGM equation in the coupled channels of $N
\Omega$-$\Lambda \Xi ^{\ast}$-$\Sigma ^{\ast} \Xi$-$\Sigma \Xi ^{\ast}
( 2 , 1/2 )$, we find a bound state with a binding energy $B =
\SI{2.1}{MeV}$ below the $N \Omega$ threshold.  Then, assuming that
the $N \Omega$ contribution is dominant near the $N \Omega$ threshold
in the coupled channels, we evaluate the equivalent local potential
for the elastic $N \Omega ( 2 , 1/2 )$ only from the $N \Omega$ wave
function at the $N \Omega$ bound state energy $\chi _{\mathrm{R}, N
  \Omega} ( r )$:
\begin{equation}
  V_{\rm eq}^{N \Omega \textrm{-} N \Omega} ( r )
  = - B + \frac{1}{2 \mu _{N \Omega} \chi _{\mathrm{R}, N \Omega} ( r )}
  \frac{d^{2} \chi _{\mathrm{R}, N \Omega}}{d r^{2}} .
\end{equation}
The resulting potential is shown as the solid line in
Fig.~\ref{fig:V_NOmega_CC}.  This indicates that the $N \Omega ( 2 ,
1/2 )$ interaction is attractive via the coupled channels.  The
strength of the potential $\beta$~\eqref{eq:beta} is $\beta = 1.29$ in
our model.

Additionally, we can include the meson exchange potential calculated
in Ref.~\cite{Sekihara:2018tsb}, in which the $\eta$ meson, correlated
two mesons in the scalar-isoscalar channel, and $K$ meson in a box
diagram were taken into account.  As a result, we obtain a bound state
with a binding energy $\SI{10.3}{MeV}$ and a decay width
$\SI{4.6}{MeV}$, which arises from the decay to the $\Lambda \Xi$ and
$\Sigma \Xi$ channels in $D$ wave in the box diagram.  We plot the
real part of the $N \Omega ( 2 , 1/2 )$ potential with the meson
exchange contributions added as the dashed line in
Fig.~\ref{fig:V_NOmega_CC}, and also present the simulation results in
the HAL QCD method~\cite{HALQCD:2018qyu} as the thick line.  Comparing
the potentials in the present study and in the HAL QCD method, the
shape is different at the range $r \lesssim \SI{0.4}{fm}$, which was
also observed in the $\Delta \Delta ( 3 , 0 )$ system in
Fig.~\ref{fig:V_DeltaDelta_HAL}.  Although the potential is not
observable, understanding the origin of the discrepancy at the range
$r \lesssim \SI{0.4}{fm}$ may be important.  In contrast, attraction
in the longer range is similar to each other.  The strength of the
potential $\beta$ amounts to $\beta = 1.69 + 0.11 i$ ($\beta = 1.47$)
in our model (HAL QCD method).

\begin{figure}[!t]
  \centering
  \Psfig{8.6cm}{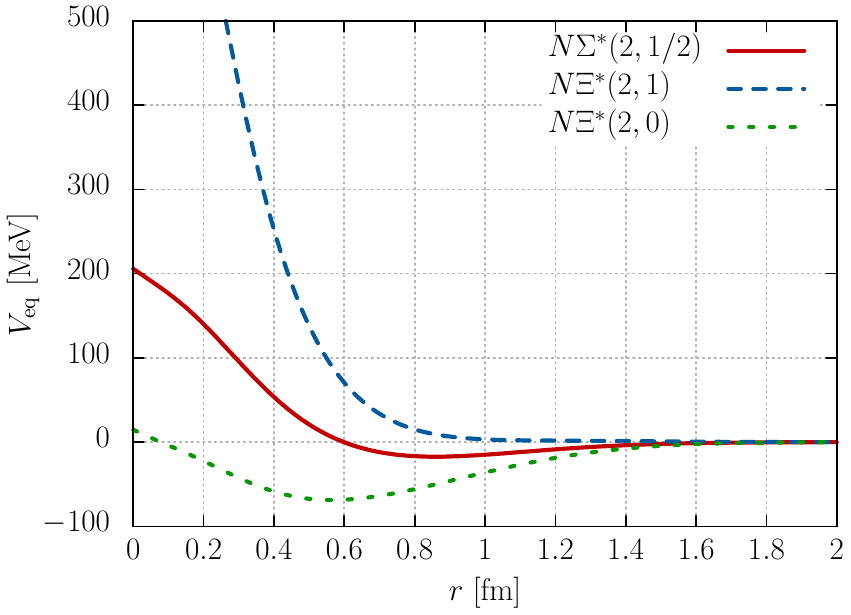}
  \caption{Equivalent local potentials for the baryon-baryon systems:
    flavor octet states in coupled channels.}
  \label{fig:V_NSigmaS_CC}
\end{figure}

Similarly, we calculate the relative wave functions of the $N \Sigma
^{\ast} ( 2 , 1/2 )$, $N \Xi ^{\ast} ( 2 , 1 )$, and $N \Xi ^{\ast} (
2 , 0 )$ states in the coupled-channels problems $N \Sigma
^{\ast}$-$\Delta \Sigma ( 2 , 1/2 )$, $N \Xi ^{\ast}$-$\Lambda \Sigma
^{\ast}$-$\Delta \Xi$-$\Sigma \Sigma ^{\ast} ( 2 , 1 )$, and $N \Xi
^{\ast}$-$\Sigma \Sigma ^{\ast} ( 2 , 0 )$, respectively.  They are of
interest, because they belong to the flavor octet of the two-baryon
states together with the $N \Omega$-$\Lambda \Xi ^{\ast}$-$\Sigma
^{\ast} \Xi$-$\Sigma \Xi ^{\ast} ( 2 , 1/2 )$ coupled
channels~\cite{Oka:1988yq}, and hence we expect attractive interaction
due to the coupled channels.  By solving the RGM equation, we find no
bound states below the $N \Sigma ^{\ast}$ and $N \Xi ^{\ast}$
thresholds.  We calculate the equivalent local potentials for the $N
\Sigma ^{\ast}$ and $N \Xi ^{\ast}$ systems at the $N \Sigma ^{\ast}$
and $N \Xi ^{\ast}$ threshold energies, respectively, and show the
results in Fig.~\ref{fig:V_NSigmaS_CC}.  As we can see, compared to
the single-channel cases, the repulsion becomes moderate in the $N
\Sigma ^{\ast} ( 2 , 1/2 )$ and $N \Xi ^{\ast} ( 2 , 1 )$ systems, and
the attraction grows in the $N \Xi ^{\ast} ( 2 , 0 )$ system.  To
conclude whether these systems are bound or not, we have to evaluate
the contributions from the meson exchanges and add them to the present
potentials.

\subsubsection{Bound states coupling to decay channels}

Among the bound states listed in Table~\ref{tab:BS}, the
 $\Delta
\Sigma ( 2 , 1/2 )$ and $\Sigma \Sigma ^{\ast} ( 2 , 0 )$ bound states
exist above the lowest thresholds with the same quantum numbers,
 $N
\Sigma ^{\ast}$ and $N \Xi ^{\ast}$, respectively.  Therefore, we aim
to evaluate the impact of the decay channels on these bound states by
tracing the bound state poles in the complex energy plane.  However,
solving the fully coupled-channels RGM equation~\eqref{eq:RGM} for the
complex eigenenergy above the lowest threshold is not feasible.  To
circumvent this problem, we incorporate the decay channels
perturbatively.  Specifically, we explicitly consider the bound-state
channels, \textit{i.e.}, $\Delta \Sigma (
2, 1/2 )$ and $\Sigma \Sigma ^{\ast} ( 2 , 0 )$, as in the
single-channel cases, while implicitly accounting for the decay
channels by replacing the interaction term $V_{\rm int} ( \bm{r} ,
\bm{r}^{\prime} )$ with
\begin{align}
  V_{\rm int} ( \bm{r} , \bm{r}^{\prime} )
  \to & V_{\rm int} ( \bm{r} , \bm{r}^{\prime} )
  + \int d^{3} r_{1} d^{3} r_{2} d^{3} r_{3}
  V_{\rm int} ( \bm{r} , \bm{r}_{1} )
  \notag \\
  & \times G ( \mathcal{E} , \bm{r}_{1} , \bm{r}_{2} )
  N^{-1} ( \bm{r}_{2} , \bm{r}_{3} )
  V_{\rm int} ( \bm{r}_{3} , \bm{r}^{\prime} )
  \label{eq:Vint_perturb}
\end{align}
where $G ( \mathcal{E} , \bm{r}_{1} , \bm{r}_{2} )$ is the
loop function of the decay channel
\begin{equation}
  G ( \mathcal{E} , \bm{r}_{1} , \bm{r}_{2} )
  \equiv \int \frac{d^{3} p}{( 2 \pi )^{3}}
  \frac{e^{i \bm{p} \cdot ( \bm{r}_{1} - \bm{r}_{2} )}}
  {\displaystyle \mathcal{E} + \Delta - p^{2} / ( 2 \mu ^{\prime} )}
\end{equation}
with $\Delta$ denoting the difference of the two threshold values and
$\mu ^{\prime}$ the reduced mass of the decay channel.  In
Eq.~\eqref{eq:Vint_perturb}, transitions between the bound-state channel
and the decay channel occur at $V_{\rm int} ( \bm{r} , \bm{r}_{1} )$
and $N^{-1} ( \bm{r}_{2} , \bm{r}_{3} ) V_{\rm int} ( \bm{r}_{3} ,
\bm{r}^{\prime} )$ of the second term.

As a result of including the decay channel, the bound state pole of
the $\Sigma \Sigma ^{\ast} ( 2 , 0 )$ system disappears due to the
repulsion from the inelastic-channel contributions, while the $\Delta
\Sigma ( 2 , 1/2 )$ bound state becomes a resonance with an
eigenenergy of $\mathcal{E} = 40.6 - 38.0 i \, \si{MeV}$, which
corresponds to the binding energy $B = \SI{-40.6}{MeV}$ and decay
width $\Gamma = \SI{76.0}{MeV}$.  Note that the real part of the
resonance pole position is above the $\Delta \Sigma$ threshold, but
the pole exists in the same Riemann sheet as the bound state in the
single-channel case.  Therefore, if the $\Delta \Sigma ( 2 , 1/2 )$
resonance exists as predicted in our calculation, it will be observed
as a cusp structure at the $\Delta \Sigma$ threshold in experiments.

\begin{figure}[!t]
  \centering
  \Psfig{8.6cm}{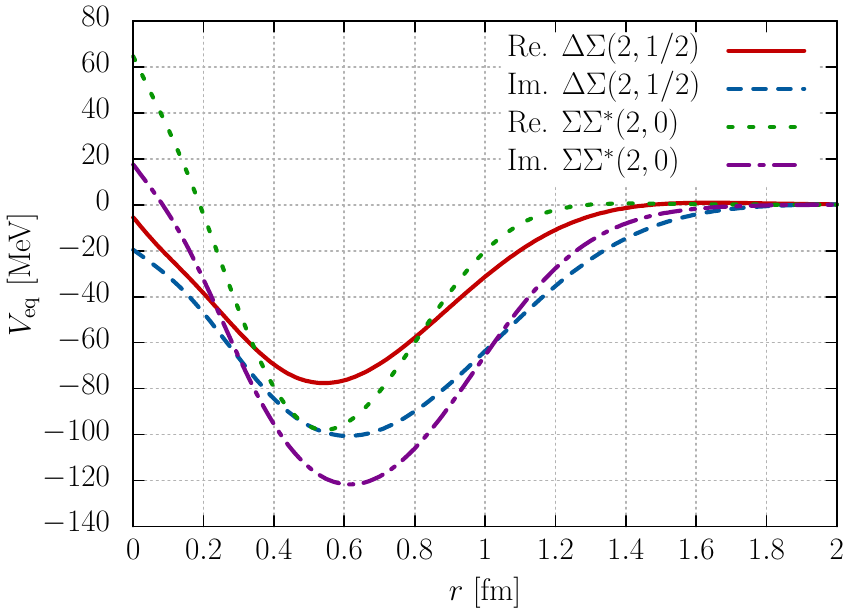}
  \caption{Equivalent local potentials for the baryon-baryon systems:
    inclusion of decay channels.}
  \label{fig:V_DeltaSigma_CC}
\end{figure}

\begin{figure*}[!t]
  \centering
  \Psfig{\textwidth}{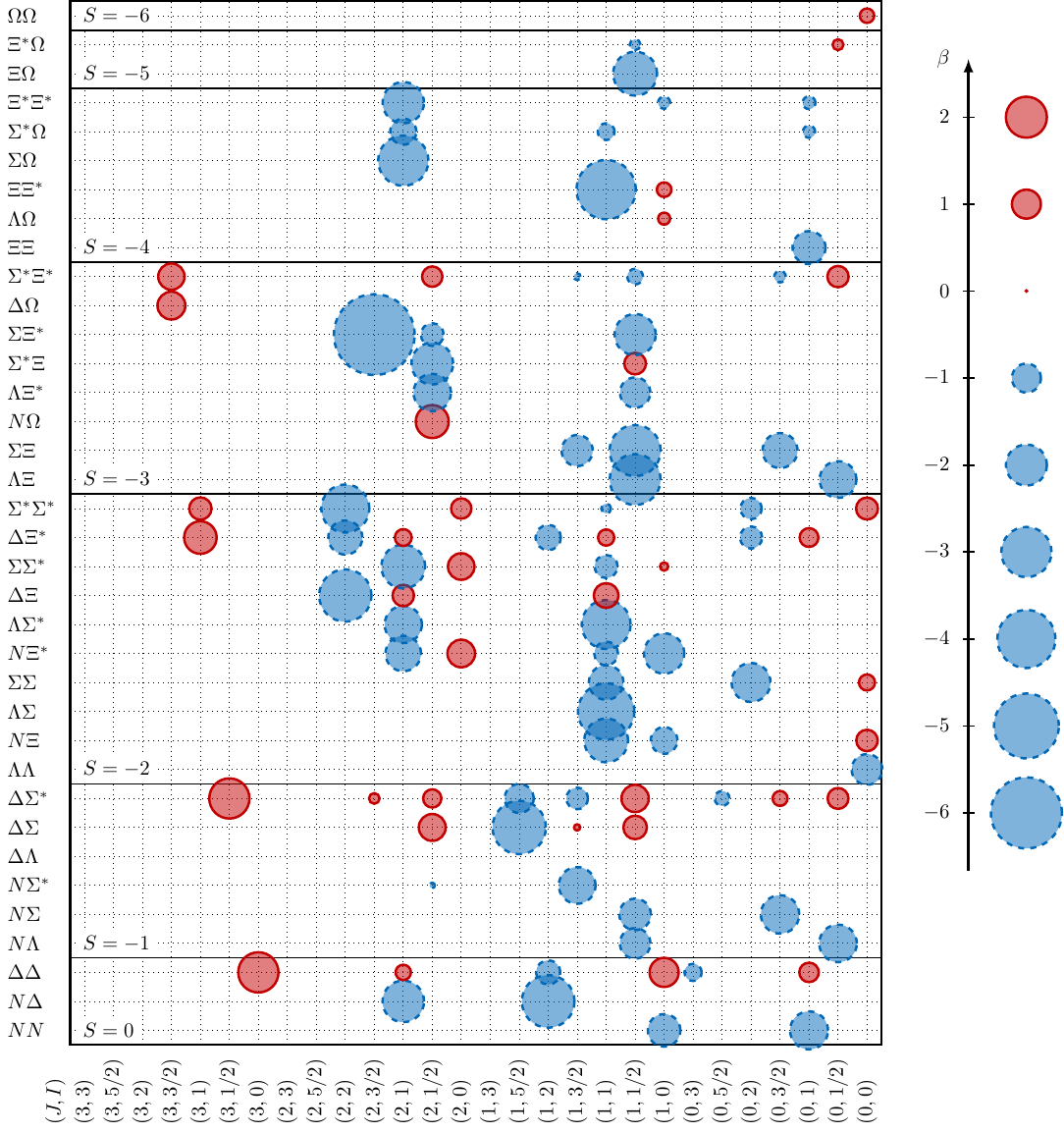}
  \caption{Strength of the potential $\beta$~\eqref{eq:beta} for the
    baryon-baryon systems.  Red circles with solid lines represent
    attractive interactions, while blue circles with dashed lines
    represent repulsive interactions.  The area of the circles
    corresponds to the absolute values of $\beta$.}
  \label{fig:beta}
\end{figure*}

We extract the equivalent local potential for the $\Delta \Sigma ( 2,
1/2 )$ and $\Sigma \Sigma ^{\ast} ( 2 , 0 )$ systems from the relative
wave function $\chi _{\mathrm{R}} ( r )$, where the wave function is
evaluated at the threshold $\mathcal{E} = 0$ for the $\Sigma \Sigma
^{\ast} ( 2 , 0 )$ system and at the resonance eigenenergy
$\mathcal{E} = 40.6 - 38.0 i \, \si{MeV}$ for the $\Delta \Sigma ( 2,
1/2 )$ system.  We plot the equivalent local potentials for the
$\Delta \Sigma ( 2, 1/2 )$ and $\Sigma \Sigma ^{\ast} ( 2 , 0 )$
systems in Fig.~\ref{fig:V_DeltaSigma_CC}.  The equivalent local
potentials contain imaginary parts due to the decay channel, and the
attraction of the real parts becomes moderate in both the $\Delta
\Sigma ( 2, 1/2 )$ and $\Sigma \Sigma ^{\ast} ( 2 , 0 )$ systems.  We
note that, if we assume that the equivalent local potential for the
$\Sigma \Sigma ^{\ast} ( 2 , 0 )$ in Fig.~\ref{fig:V_DeltaSigma_CC} is
valid for any energy, the $\Sigma \Sigma ^{\ast} ( 2 , 0 )$ potential
generates a resonance state at $\mathcal{E} = 46.4 - 51.2 i \,
\si{MeV}$.

\subsubsection{Strength of the potentials}

Finally, we calculate the strength of the potential
$\beta$~\eqref{eq:beta} for all baryon-baryon systems.  The results
are shown in Fig.~\ref{fig:beta} as a bubble chart, where red circles
with solid lines (blue circles with dashed lines) represent attractive
(repulsive) interactions and the area of the circles corresponds to
the absolute values of $\beta$.  We use the elastic parts of the
potentials for the $N \Sigma ^{\ast} ( 2 , 1/2 )$, $\Delta \Sigma ( 2
, 1/2 )$, $N \Xi ^{\ast} ( 2 , 1 )$, $N \Xi ^{\ast} ( 2 , 0 )$,
$\Sigma \Sigma ^{\ast} ( 2 , 0 )$, $\Delta \Xi ^{\ast} ( 3 , 1 )$,
$\Sigma ^{\ast} \Sigma ^{\ast} ( 3 , 1 )$, $N \Omega ( 2, 1/2 )$,
$\Delta \Omega ( 3, 3/2 )$, and $\Sigma ^{\ast} \Xi ^{\ast} ( 3 , 3/2
)$ systems in the coupled-channels cases, while using the
single-channel cases for the others. 
We do not take meson exchange
contributions into account in the $N \Omega ( 2 , 1/2 )$ potential.

As shown in Fig.~\ref{fig:beta}, the strongest attractions can be seen
in the $\Delta \Delta ( 3 , 0 )$, $\Delta \Sigma ^{\ast} ( 3 , 1/2 )$,
$\Delta \Xi ^{\ast}$-$\Sigma ^{\ast} \Sigma ^{\ast} ( 3 , 1 )$, and
$\Delta \Omega$-$\Sigma ^{\ast} \Xi ^{\ast} ( 3 , 3/2 )$ systems,
which are members of the flavor antidecuplet two-baryon states with $J
= 3$.  The constituent quark model suggests that they are attractive
enough to generate bound states.  The explicit values of the strength
$\beta$ are: $\beta = 1.93$ for the $\Delta \Delta ( 3 , 0 )$, $1.92$
for the $\Delta \Sigma ^{\ast} ( 3 , 1/2 )$, $1.26$ for the $\Delta
\Xi ^{\ast} ( 3 , 1 )$, $0.594$ for the $\Sigma ^{\ast} \Sigma ^{\ast}
( 3 , 1 )$, $0.923$ for the $\Delta \Omega ( 3 , 3/2 )$, and $0.825$
for the $\Sigma ^{\ast} \Xi ^{\ast} ( 3 , 3/2 )$ systems.  Therefore,
it would be interesting to perform systematic studies of the
two-baryon interactions in the flavor antidecuplet states with $J = 3$
in lattice QCD simulations and relativistic ion collisions as well as
scattering experiments.  Additionally, strong attraction $\beta > 1$
can be found in the $\Delta \Delta ( 1 , 0 )$ ($\beta = 1.01$) and $N
\Omega ( 2 , 1/2 )$ ($\beta = 1.29$) systems.  While the $N \Omega ( 2
, 1/2 )$ system generates a bound state, the attraction in the $\Delta
\Delta ( 1 , 0 )$ system is slightly insufficient to generate a bound
state.

We would like to point out that, even if the potential is not
attractive in the constituent quark model, meson exchange
contributions, which are not taken into account in the present study
except for the $N \Omega ( 2 , 1/2 )$ system, may help in generating a
bound state, as in the case of the deuteron in the $N N ( 1 , 0 )$
channel.  The exchanges of scalar and pseudoscalar mesons can be
superposed on the quark-model potential without introducing a
double-counting problem~\cite{Yazaki:1989rh}.  In any case, the
results in Fig.~\ref{fig:beta} will serve as a guideline to search for
attractive interactions between two baryons and shed light on the
quark dynamics inside baryons.

\section{Conclusion}
\label{sec:4}

In this study, we investigated the short-range baryon-baryon
interactions in the flavor SU(3) sector within the constituent quark
model.  We employed the color Coulomb, linear confining, and color
magnetic forces between two constituent quarks.  The wave functions of
the ground-state baryons, \textit{i.e.}, the octet $( N , \Lambda ,
\Sigma , \Xi )$ and decuplet $( \Delta , \Sigma ^{\ast} , \Xi ^{\ast}
, \Omega )$ baryons, were described using the Gaussian expansion
method.  The model parameters were determined by fitting the masses of
the ground-state baryons.  We used the forces between constituent
quarks and baryon wave functions to systematically calculate the
relative wave functions of two baryons in the resonating group method.
We then evaluated the equivalent local potentials between two baryons
which reproduce the relative wave functions of two baryons in the
resonating group method.

The most interesting finding was the existence of two-baryon bound
states with a binding energy of approximately $\SI{10}{MeV}$ in the
flavor antidecuplet with total spin $J = 3$, namely, $\Delta \Delta (
J = 3 , I = 0 )$, $\Delta \Sigma ^{\ast} ( 3 , 1/2 )$, $\Delta \Xi
^{\ast}$-$\Sigma ^{\ast} \Sigma ^{\ast} ( 3 , 1 )$, and $\Delta
\Omega$-$\Sigma ^{\ast} \Xi ^{\ast} ( 3 , 3/2 )$.  By decomposing the
potentials for these systems, we confirmed that the contribution from
the color Coulomb plus linear confining force is attractive enough to
produce a bound state, and the color magnetic force brings even more
attractions.  We also checked that a strong attraction associated with
the color Coulomb plus linear force is correlated to the spin-flavor
[33] component $N_{33}$: a larger $N_{33}$ generates a stronger
attraction.  The spin-flavor [33] component measures the contribution
of totally antisymmetric states of six quarks for two ground-state
baryons in $S$ wave.  In this sense, both the Pauli exclusion
principle for quarks and color magnetic interactions are essential for
the generation of the bound states in the flavor antidecuplet.
Because the spatial extension of the bound states in the flavor
antidecuplet largely exceeds the typical size of hadrons $\SI{1}{fm}$,
we concluded that these bound states are hadronic molecules rather
than compact hexaquark states.  In particular, the $\Delta \Delta ( 3
, 0 )$ bound state can be interpreted as the $d^{\ast} (2380)$ state
recently confirmed in experiments.  Therefore, to understand the
mechanism of the quark dynamics on the baryon-baryon interaction, the
experimental search for the other members belonging to the
antidecuplet will be helpful.

Another interesting finding was made in the $N \Omega ( 2 , 1/2 )$
interaction.  When we restrict the model space to the elastic channel,
the $N \Omega$ interaction is absent in our model because the
shuffling of quarks associated with the quark-quark interaction
inevitably leads to a transition to inelastic channels.  On the other
hand, by including the coupling to inelastic channels $\Lambda \Xi
^{\ast}$, $\Sigma ^{\ast} \Xi$, and $\Sigma \Xi ^{\ast}$, attraction
in the $N \Omega ( 2 , 1/2 )$ system emerges, and it is sufficient to
generate a bound state with a binding energy $\SI{2.1}{MeV}$.  Then,
assistance comes from the meson exchange, resulting in a bound state
with a binding energy $\SI{10.3}{MeV}$ and a decay width
$\SI{4.6}{MeV}$.  In the coupled-channels cases, we also found a
resonance state $\Delta \Sigma ( 2 , 1/2 )$ with the eigenenergy
$\mathcal{E} = 40.6 - 38.0 i \, \si{MeV}$.

The calculated equivalent local potentials between two baryons will
not only be useful in further studies on dibaryon states but also
provide clues to understand the mechanism of baryon-baryon
interactions.  When comparing the equivalent local potentials with
those in the HAL QCD method for the $\Delta \Delta ( 3 , 0 )$, $\Omega
\Omega ( 0 , 0 )$, and $N \Omega ( 2 , 1/2 )$ systems, we found that,
while these potentials are attractive in both approaches, their
detailed shapes differ.  In particular, our model provides weak
repulsion at the origin in the $\Delta \Delta ( 3 , 0 )$ and $N \Omega
( 2 , 1/2 )$ systems, in contrast to the HAL QCD potentials.  In the
$\Delta \Delta ( 3 , 0 )$ system, this weak repulsion at the origin
comes from the color Coulomb plus linear confining force.  Although
the potential is not observable, understanding the origin of the
discrepancy at short distances may be important.  Besides, the
$\Omega \Omega ( 0 , 0 )$ potential in our model is weaker and not
sufficiently attractive to generate a bound state, which implies that
meson exchange contributions may assist the attraction.  We also
evaluated the strength of the potentials, which will be a guideline to
search for the attractive interactions between two baryons and shed
light on the quark dynamics inside baryons.

\begin{acknowledgments}

  The authors acknowledge M.~Oka for helpful discussions on the
  baryon-baryon interactions in quark models.

\end{acknowledgments}

\appendix

\section{Weights for baryons}

We summarize the weights $w_{\vec{\mu}}$ for the ground-state baryons
in Table~\ref{tab:weight}.

\begin{table*}[!t]
  \caption{Weights for the octet and decuplet baryons.  The index $s$
    is the third component of the spin.  ``RGB cyclic'' means that the
    terms of color permutation $( \mathrm{R} , \mathrm{G} , \mathrm{B}
    ) \to ( \mathrm{G} , \mathrm{B} , \mathrm{R} )$, $( \mathrm{B} ,
    \mathrm{R} , \mathrm{G} )$ follow with the same weight, flavor,
    and spin.}
  \label{tab:weight}
  \centering
  \begin{tabular}{ccc}
    \hline \hline
    \begin{minipage}[t]{0.33\hsize}
      \begin{tabular}{lccc}
        \multicolumn{4}{c}{Octet baryons}
        \\
        \multicolumn{4}{c}{$p (s = 1/2 )$}
        \\
        $w_{\vec{\mu}}$ & $\mu _{1}$ & $\mu _{2}$ & $\mu _{3}$
        \\
        \hline
        $2/3\sqrt{2}$ & $( u , \uparrow , \mathrm{R} )$
        & $( u , \uparrow , \mathrm{G} )$ & $( d , \downarrow , \mathrm{B} )$
        \\
        $-1/3\sqrt{2}$ & $( u , \uparrow , \mathrm{R} )$
        & $( u , \downarrow , \mathrm{G} )$ & $( d , \uparrow , \mathrm{B} )$
        \\
        $-1/3\sqrt{2}$ & $( u , \downarrow , \mathrm{R} )$
        & $( u , \uparrow , \mathrm{G} )$ & $( d , \uparrow , \mathrm{B} )$
        \\
        \multicolumn{4}{c}{RGB cyclic}
        \\
        \hline      
        \\
        \multicolumn{4}{c}{$p ( s = -1/2 )$}
        \\
        $w_{\vec{\mu}}$ & $\mu _{1}$ & $\mu _{2}$ & $\mu _{3}$
        \\
        \hline
        $-2/3\sqrt{2}$ & $( u , \downarrow , \mathrm{R} )$
        & $( u , \downarrow , \mathrm{G} )$ & $( d , \uparrow , \mathrm{B} )$
        \\
        $1/3\sqrt{2}$ & $( u , \uparrow , \mathrm{R} )$
        & $( u , \downarrow , \mathrm{G} )$ & $( d , \downarrow , \mathrm{B} )$
        \\
        $1/3\sqrt{2}$ & $( u , \downarrow , \mathrm{R} )$
        & $( u , \uparrow , \mathrm{G} )$ & $( d , \downarrow , \mathrm{B} )$
        \\
        \multicolumn{4}{c}{RGB cyclic}
        \\
        \hline      
        \\
        \multicolumn{4}{c}{$n ( s = 1/2 )$}
        \\
        $w_{\vec{\mu}}$ & $\mu _{1}$ & $\mu _{2}$ & $\mu _{3}$
        \\
        \hline
        $- 2/3\sqrt{2}$ & $( d , \uparrow , \mathrm{R} )$
        & $( d , \uparrow , \mathrm{G} )$ & $( u , \downarrow , \mathrm{B} )$
        \\
        $1/3\sqrt{2}$ & $( d , \uparrow , \mathrm{R} )$
        & $( d , \downarrow , \mathrm{G} )$ & $( u , \uparrow , \mathrm{B} )$
        \\
        $1/3\sqrt{2}$ & $( d , \downarrow , \mathrm{R} )$
        & $( d , \uparrow , \mathrm{G} )$ & $( u , \uparrow , \mathrm{B} )$
        \\
        \multicolumn{4}{c}{RGB cyclic}
        \\
        \hline      
        \\
        \multicolumn{4}{c}{$n ( s = -1/2 )$}
        \\
        $w_{\vec{\mu}}$ & $\mu _{1}$ & $\mu _{2}$ & $\mu _{3}$
        \\
        \hline
        $2/3\sqrt{2}$ & $( d , \downarrow , \mathrm{R} )$
        & $( d , \downarrow , \mathrm{G} )$ & $( u , \uparrow , \mathrm{B} )$
        \\
        $- 1/3\sqrt{2}$ & $( d , \uparrow , \mathrm{R} )$
        & $( d, \downarrow , \mathrm{G} )$ & $( u, \downarrow , \mathrm{B} )$
        \\
        $- 1/3\sqrt{2}$ & $( d , \downarrow , \mathrm{R} )$
        & $( d , \uparrow , \mathrm{G} )$ & $( u , \downarrow , \mathrm{B} )$
        \\
        \multicolumn{4}{c}{RGB cyclic}
        \\
        \hline      
        \\
        \multicolumn{4}{c}{$\Lambda ( s = 1/2 )$}
        \\
        $w_{\vec{\mu}}$ & $\mu _{1}$ & $\mu _{2}$ & $\mu _{3}$
        \\
        \hline
        $1/2\sqrt{3}$ & $( u , \uparrow , \mathrm{R} )$
        & $( d , \downarrow , \mathrm{G} )$ & $( s , \uparrow , \mathrm{B} )$
        \\
        $-1/2\sqrt{3}$ & $( u , \downarrow , \mathrm{R} )$
        & $( d , \uparrow , \mathrm{G} )$ & $( s , \uparrow , \mathrm{B} )$
        \\
        $-1/2\sqrt{3}$ & $( u , \uparrow , \mathrm{G} )$
        & $( d , \downarrow , \mathrm{R} )$ & $( s , \uparrow , \mathrm{B} )$
        \\
        $1/2\sqrt{3}$ & $( u , \downarrow , \mathrm{G} )$
        & $( d , \uparrow , \mathrm{R} )$ & $( s , \uparrow , \mathrm{B} )$
        \\
        \multicolumn{4}{c}{RGB cyclic}
        \\
        \hline      
        \\
        \multicolumn{4}{c}{$\Lambda ( s = -1/2 )$}
        \\
        $w_{\vec{\mu}}$ & $\mu _{1}$ & $\mu _{2}$ & $\mu _{3}$
        \\
        \hline
        $1/2\sqrt{3}$ & $( u , \uparrow , \mathrm{R} )$
        & $( d , \downarrow , \mathrm{G} )$ & $( s , \downarrow , \mathrm{B} )$
        \\
        $-1/2\sqrt{3}$ & $( u , \downarrow , \mathrm{R} )$
        & $( d , \uparrow , \mathrm{G} )$ & $( s , \downarrow , \mathrm{B} )$
        \\
        $-1/2\sqrt{3}$ & $( u , \uparrow , \mathrm{G} )$
        & $( d , \downarrow , \mathrm{R} )$ & $( s , \downarrow , \mathrm{B} )$
        \\
        $1/2\sqrt{3}$ & $( u , \downarrow , \mathrm{G} )$
        & $( d , \uparrow , \mathrm{R} )$ & $( s , \downarrow , \mathrm{B} )$
        \\
        \multicolumn{4}{c}{RGB cyclic}
      \\
      \hline
      \end{tabular}
    \end{minipage}
    & 
    \begin{minipage}[t]{0.33\hsize}
      \begin{tabular}{lccc}
        \multicolumn{4}{c}{$\Sigma^{+} ( s = 1/2 )$}
        \\
        $w_{\vec{\mu}}$ & $\mu _{1}$ & $\mu _{2}$ & $\mu _{3}$
        \\
        \hline
        $2/3\sqrt{2}$ & $( u , \uparrow , \mathrm{R} )$
        & $( u , \uparrow , \mathrm{G} )$ & $( s , \downarrow , \mathrm{B} )$
        \\
        $-1/3\sqrt{2}$ & $( u , \uparrow , \mathrm{R} )$
        & $( u , \downarrow , \mathrm{G} )$ & $( s , \uparrow , \mathrm{B} )$
        \\
        $-1/3\sqrt{2}$ & $( u , \downarrow , \mathrm{R} )$
        & $( u , \uparrow , \mathrm{G} )$ & $( s , \uparrow , \mathrm{B} )$
        \\
        \multicolumn{4}{c}{RGB cyclic}
        \\
        \hline
        \\
        \multicolumn{4}{c}{$\Sigma ^{+} ( s = -1/2 )$}
        \\
        $w_{\vec{\mu}}$ & $\mu _{1}$ & $\mu _{2}$ & $\mu _{3}$
        \\
        \hline
        $-2/3\sqrt{2}$ & $( u , \downarrow , \mathrm{R} )$
        & $( u , \downarrow , \mathrm{G} )$ & $( s , \uparrow , \mathrm{B} )$
        \\
        $1/3\sqrt{2}$ & $( u , \uparrow , \mathrm{R} )$
        & $( u , \downarrow , \mathrm{G} )$ & $( s , \downarrow , \mathrm{B} )$
        \\
        $1/3\sqrt{2}$ & $( u , \downarrow , \mathrm{R} )$
        & $( u , \uparrow , \mathrm{G} )$ & $( s , \downarrow , \mathrm{B} )$
        \\
        \multicolumn{4}{c}{RGB cyclic}
        \\
        \hline      
        \\
        \multicolumn{4}{c}{$\Sigma ^{0} ( s = 1/2 )$}
        \\
        $w_{\vec{\mu}}$ & $\mu _{1}$ & $\mu _{2}$ & $\mu _{3}$
        \\
        \hline
        $1/3$ & $( u , \uparrow , \mathrm{R} )$
        & $( d , \uparrow , \mathrm{G} )$ & $( s , \downarrow , \mathrm{B} )$
        \\
        $-1/6$ & $( u , \uparrow , \mathrm{R} )$
        & $( d , \downarrow , \mathrm{G} )$ & $( s , \uparrow , \mathrm{B} )$
        \\
        $-1/6$ & $( u , \downarrow , \mathrm{R} )$
        & $( d , \uparrow , \mathrm{G} )$ & $( s , \uparrow , \mathrm{B} )$
        \\
        $-1/3$ & $( u , \uparrow , \mathrm{G} )$
        & $( d , \uparrow , \mathrm{R} )$ & $( s , \downarrow , \mathrm{B} )$
        \\
        $1/6$ & $( u , \uparrow , \mathrm{G} )$
        & $( d , \downarrow , \mathrm{R} )$ & $( s , \uparrow , \mathrm{B} )$
        \\
        $1/6$ & $( u , \downarrow , \mathrm{G} )$
        & $( d , \uparrow , \mathrm{R} )$ & $( s , \uparrow , \mathrm{B} )$
        \\
        \multicolumn{4}{c}{RGB cyclic}
        \\
        \hline      
        \\
        \multicolumn{4}{c}{$\Sigma ^{0} ( s = -1/2 )$}
        \\
        $w_{\vec{\mu}}$ & $\mu _{1}$ & $\mu _{2}$ & $\mu _{3}$
        \\
        \hline
        $-1/3$ & $( u , \downarrow , \mathrm{R} )$
        & $( d , \downarrow , \mathrm{G} )$ & $( s , \uparrow , \mathrm{B} )$
        \\
        $1/6$ & $( u , \uparrow , \mathrm{R} )$
        & $( d , \downarrow , \mathrm{G} )$ & $( s , \downarrow , \mathrm{B} )$
        \\
        $1/6$ & $( u , \downarrow , \mathrm{R} )$
        & $( d , \uparrow , \mathrm{G} )$ & $( s , \downarrow , \mathrm{B} )$
        \\
        $1/3$ & $( u , \downarrow , \mathrm{G} )$
        & $( d , \downarrow , \mathrm{R} )$ & $( s , \uparrow , \mathrm{B} )$
        \\
        $-1/6$ & $( u , \uparrow , \mathrm{G} )$
        & $( d , \downarrow , \mathrm{R} )$ & $( s , \downarrow , \mathrm{B} )$
        \\
        $-1/6$ & $( u , \downarrow , \mathrm{G} )$
        & $( d , \uparrow , \mathrm{R} )$ & $( s , \downarrow , \mathrm{B} )$
        \\
        \multicolumn{4}{c}{RGB cyclic}
        \\
        \hline      
        \\
        \\
        \\
        \\
        \\
        \\
        \\
        \\
        \\
        \\
      \end{tabular}
    \end{minipage}
    &
    \begin{minipage}[t]{0.33\hsize}
      \begin{tabular}{lccc}
        \multicolumn{4}{c}{$\Sigma ^{-} ( s = 1/2 )$}
        \\
        $w_{\vec{\mu}}$ & $\mu _{1}$ & $\mu _{2}$ & $\mu _{3}$
        \\
        \hline
        $2/3\sqrt{2}$ & $( d , \uparrow , \mathrm{R} )$
        & $( d , \uparrow , \mathrm{G} )$ & $( s , \downarrow , \mathrm{B} )$
        \\
        $-1/3\sqrt{2}$ & $( d , \uparrow , \mathrm{R} )$
        & $( d , \downarrow , \mathrm{G} )$ & $( s , \uparrow , \mathrm{B} )$
        \\
        $-1/3\sqrt{2}$ & $( d , \downarrow , \mathrm{R} )$
        & $( d , \uparrow , \mathrm{G} )$ & $( s , \uparrow , \mathrm{B} )$
        \\ 
        \multicolumn{4}{c}{RGB cyclic}
        \\
        \hline      
        \\
        \multicolumn{4}{c}{$\Sigma ^{-} ( s = -1/2 )$}
        \\
        $w_{\vec{\mu}}$ & $\mu _{1}$ & $\mu _{2}$ & $\mu _{3}$
        \\
        \hline
        $-2/3\sqrt{2}$ & $( d , \downarrow , \mathrm{R} )$
        & $( d , \downarrow , \mathrm{G} )$ & $( s , \uparrow , \mathrm{B} )$
        \\
        $1/3\sqrt{2}$ & $( d , \uparrow , \mathrm{R} )$
        & $( d , \downarrow , \mathrm{G} )$ & $( s , \downarrow , \mathrm{B} )$
        \\
        $1/3\sqrt{2}$ & $( d , \downarrow , \mathrm{R} )$
        & $( d , \uparrow , \mathrm{G} )$ & $( s , \downarrow , \mathrm{B} )$
        \\
        \multicolumn{4}{c}{RGB cyclic}
        \\
        \hline      
        \\
        \multicolumn{4}{c}{$\Xi^{0} ( s = 1/2 )$}
        \\
        $w_{\vec{\mu}}$ & $\mu _{1}$ & $\mu _{2}$ & $\mu _{3}$
        \\
        \hline
        $-2/3\sqrt{2}$ & $( u , \downarrow , \mathrm{R} )$
        & $( s , \uparrow , \mathrm{G} )$ & $( s , \uparrow , \mathrm{B} )$
        \\
        $1/3\sqrt{2}$ & $( u , \uparrow , \mathrm{R} )$
        & $( s , \uparrow , \mathrm{G} )$ & $( s , \downarrow , \mathrm{B} )$
        \\
        $1/3\sqrt{2}$ & $( u , \uparrow , \mathrm{R} )$
        & $( s , \downarrow , \mathrm{G} )$ & $( s , \uparrow , \mathrm{B} )$
        \\
        \multicolumn{4}{c}{RGB cyclic}
        \\
        \hline      
        \\
        \multicolumn{4}{c}{$\Xi^{0} ( s = -1/2 )$}
        \\
        $w_{\vec{\mu}}$ & $\mu _{1}$ & $\mu _{2}$ & $\mu _{3}$
        \\
        \hline
        $2/3\sqrt{2}$ & $( u , \uparrow , \mathrm{R} )$
        & $( s , \downarrow , \mathrm{G} )$ & $( s , \downarrow , \mathrm{B} )$
        \\
        $-1/3\sqrt{2}$ & $( u , \downarrow , \mathrm{R} )$
        & $( s , \uparrow , \mathrm{G} )$ & $( s , \downarrow , \mathrm{B} )$
        \\
        $-1/3\sqrt{2}$ & $( u , \downarrow , \mathrm{R} )$
        & $( s , \downarrow , \mathrm{G} )$ & $( s , \uparrow , \mathrm{B} )$
        \\
        \multicolumn{4}{c}{RGB cyclic}
        \\
        \hline      
        \\
        \multicolumn{4}{c}{$\Xi ^{-} ( s = 1/2 )$}
        \\
        $w_{\vec{\mu}}$ & $\mu _{1}$ & $\mu _{2}$ & $\mu _{3}$
        \\
        \hline
        $-2/3\sqrt{2}$ & $( d , \downarrow , \mathrm{R} )$
        & $( s , \uparrow , \mathrm{G} )$ & $( s , \uparrow , \mathrm{B} )$
        \\
        $1/3\sqrt{2}$ & $( d , \uparrow , \mathrm{R} )$
        & $( s , \uparrow , \mathrm{G} )$ & $( s , \downarrow , \mathrm{B} )$
        \\
        $1/3\sqrt{2}$ & $( d , \uparrow , \mathrm{R} )$
        & $( s , \downarrow , \mathrm{G} )$ & $( s , \uparrow , \mathrm{B} )$
        \\
        \multicolumn{4}{c}{RGB cyclic}
        \\
        \hline      
        \\
        \multicolumn{4}{c}{$\Xi ^{-} ( s = -1/2 )$}
        \\
        $w_{\vec{\mu}}$ & $\mu _{1}$ & $\mu _{2}$ & $\mu _{3}$
        \\
        \hline
        $2/3\sqrt{2}$ & $( d , \uparrow , \mathrm{R} )$
        & $( s , \downarrow , \mathrm{G} )$ & $( s , \downarrow , \mathrm{B} )$
        \\
        $-1/3\sqrt{2}$ & $( d , \downarrow , \mathrm{R} )$
        & $( s , \uparrow , \mathrm{G} )$ & $( s , \downarrow , \mathrm{B} )$
        \\
        $-1/3\sqrt{2}$ & $( d , \downarrow , \mathrm{R} )$
        & $( s , \downarrow , \mathrm{G} )$ & $( s , \uparrow , \mathrm{B} )$
        \\
        \multicolumn{4}{c}{RGB cyclic}
        \\
        \hline      
        \\
      \end{tabular}
    \end{minipage}
    \\
    \\
    \hline
  \end{tabular}
\end{table*}
\addtocounter{table}{-1}

\begin{table*}[!t]
  \caption{(continued)}
  \centering
  \begin{tabular}{ccc}
    \hline
    \begin{minipage}[t]{0.33\hsize} 
      \begin{tabular}{lccc}
        \multicolumn{4}{c}{Decuplet baryons}
        \\
        \multicolumn{4}{c}{$\Delta ^{++} (s = 3/2 )$}
        \\
        $w_{\vec{\mu}}$ & $\mu _{1}$ & $\mu _{2}$ & $\mu _{3}$
        \\
        \hline
        $1$ & $( u , \uparrow , \mathrm{R} )$
        & $( u , \uparrow , \mathrm{G} )$ & $( u , \uparrow , \mathrm{B} )$
        \\
        \hline
        \\
        \multicolumn{4}{c}{$\Delta ^{++} (s = 1/2 )$}
        \\
        $w_{\vec{\mu}}$ & $\mu _{1}$ & $\mu _{2}$ & $\mu _{3}$
        \\
        \hline
        $1/\sqrt{3}$ & $( u , \uparrow , \mathrm{R} )$
        & $( u , \uparrow , \mathrm{G} )$ & $( u , \downarrow , \mathrm{B} )$
        \\
        \multicolumn{4}{c}{RGB cyclic}
        \\
        \hline
        \\
        \multicolumn{4}{c}{$\Delta ^{++} (s = -1/2 )$}
        \\
        $w_{\vec{\mu}}$ & $\mu _{1}$ & $\mu _{2}$ & $\mu _{3}$
        \\
        \hline
        $1/\sqrt{3}$ & $( u , \uparrow , \mathrm{R} )$
        & $( u, \downarrow , \mathrm{G} )$ & $( u, \downarrow , \mathrm{B} )$
        \\
        \multicolumn{4}{c}{RGB cyclic}
        \\
        \hline
        \\
        \multicolumn{4}{c}{$\Delta ^{++} (s = - 3/2 )$}
        \\
        $w_{\vec{\mu}}$ & $\mu _{1}$ & $\mu _{2}$ & $\mu _{3}$
        \\
        \hline
        $1$ & $( u , \downarrow , \mathrm{R} )$
        & $( u, \downarrow , \mathrm{G} )$ & $( u, \downarrow , \mathrm{B} )$
        \\
        \hline
        \\
        \multicolumn{4}{c}{$\Delta ^{+} (s = 3/2 )$}
        \\
        $w_{\vec{\mu}}$ & $\mu _{1}$ & $\mu _{2}$ & $\mu _{3}$
        \\
        \hline
        $1/\sqrt{3}$ & $( u , \uparrow , \mathrm{R} )$
        & $( u , \uparrow , \mathrm{G} )$ & $( d , \uparrow , \mathrm{B} )$
        \\
        \multicolumn{4}{c}{RGB cyclic}
        \\
        \hline
        \\
        \multicolumn{4}{c}{$\Delta ^{+} (s = 1/2 )$}
        \\
        $w_{\vec{\mu}}$ & $\mu _{1}$ & $\mu _{2}$ & $\mu _{3}$
        \\
        \hline
        $1/3$ & $( u , \uparrow , \mathrm{R} )$
        & $( u , \uparrow , \mathrm{G} )$ & $( d , \downarrow , \mathrm{B} )$
        \\
        $1/3$ & $( u , \uparrow , \mathrm{R} )$
        & $( u , \downarrow , \mathrm{G} )$ & $( d , \uparrow , \mathrm{B} )$
        \\
        $1/3$ & $( u , \downarrow , \mathrm{R} )$
        & $( u , \uparrow , \mathrm{G} )$ & $( d , \uparrow , \mathrm{B} )$
        \\
        \multicolumn{4}{c}{RGB cyclic}
        \\
        \hline
        \\
        \multicolumn{4}{c}{$\Delta ^{+} (s = -1/2 )$}
        \\
        $w_{\vec{\mu}}$ & $\mu _{1}$ & $\mu _{2}$ & $\mu _{3}$
        \\
        \hline
        $1/3$ & $( u , \uparrow , \mathrm{R} )$
        & $( u, \downarrow , \mathrm{G} )$ & $( d, \downarrow , \mathrm{B} )$
        \\
        $1/3$ & $( u , \downarrow , \mathrm{R} )$
        & $( u , \uparrow , \mathrm{G} )$ & $( d , \downarrow , \mathrm{B} )$
        \\
        $1/3$ & $( u , \downarrow , \mathrm{R} )$
        & $( u , \downarrow , \mathrm{G} )$ & $( d, \uparrow , \mathrm{B} )$
        \\
        \multicolumn{4}{c}{RGB cyclic}
        \\
        \hline
        \\
        \multicolumn{4}{c}{$\Delta ^{+} (s = - 3/2 )$}
        \\
        $w_{\vec{\mu}}$ & $\mu _{1}$ & $\mu _{2}$ & $\mu _{3}$
        \\
        \hline
        $1/\sqrt{3}$ & $( u , \downarrow , \mathrm{R} )$
        & $( u, \downarrow , \mathrm{G} )$ & $( d, \downarrow , \mathrm{B} )$
        \\
        \multicolumn{4}{c}{RGB cyclic}
        \\
        \hline
        \\
        \multicolumn{4}{c}{$\Delta ^{0} (s = 3/2 )$}
        \\
        $w_{\vec{\mu}}$ & $\mu _{1}$ & $\mu _{2}$ & $\mu _{3}$
        \\
        \hline
        $1/\sqrt{3}$ & $( u , \uparrow , \mathrm{R} )$
        & $( d , \uparrow , \mathrm{G} )$ & $( d , \uparrow , \mathrm{B} )$
        \\
        \multicolumn{4}{c}{RGB cyclic}
        \\
        \hline
        \\
        \multicolumn{4}{c}{$\Delta ^{0} (s = 1/2 )$}
        \\
        $w_{\vec{\mu}}$ & $\mu _{1}$ & $\mu _{2}$ & $\mu _{3}$
        \\
        \hline
        $1/3$ & $( u , \uparrow , \mathrm{R} )$
        & $( d , \uparrow , \mathrm{G} )$ & $( d , \downarrow , \mathrm{B} )$
        \\
        $1/3$ & $( u , \uparrow , \mathrm{R} )$
        & $( d , \downarrow , \mathrm{G} )$ & $( d , \uparrow , \mathrm{B} )$
        \\
        $1/3$ & $( u , \downarrow , \mathrm{R} )$
        & $( d , \uparrow , \mathrm{G} )$ & $( d , \uparrow , \mathrm{B} )$
        \\
        \multicolumn{4}{c}{RGB cyclic}
        \\
        \hline
        \\
        \\
        \\
        \\
      \end{tabular}
    \end{minipage}
    &
    \begin{minipage}[t]{0.33\hsize} 
      \begin{tabular}{lccc}
        \multicolumn{4}{c}{$\Delta ^{0} (s = -1/2 )$}
        \\
        $w_{\vec{\mu}}$ & $\mu _{1}$ & $\mu _{2}$ & $\mu _{3}$
        \\
        \hline
        $1/3$ & $( u , \uparrow , \mathrm{R} )$
        & $( d, \downarrow , \mathrm{G} )$ & $( d, \downarrow , \mathrm{B} )$
        \\
        $1/3$ & $( u , \downarrow , \mathrm{R} )$
        & $( d , \uparrow , \mathrm{G} )$ & $( d , \downarrow , \mathrm{B} )$
        \\
        $1/3$ & $( u , \downarrow , \mathrm{R} )$
        & $( d , \downarrow , \mathrm{G} )$ & $( d, \uparrow , \mathrm{B} )$
        \\
        \multicolumn{4}{c}{RGB cyclic}
        \\
        \hline
        \\
        \multicolumn{4}{c}{$\Delta ^{0} (s = - 3/2 )$}
        \\
        $w_{\vec{\mu}}$ & $\mu _{1}$ & $\mu _{2}$ & $\mu _{3}$
        \\
        \hline
        $1/\sqrt{3}$ & $( u , \downarrow , \mathrm{R} )$
        & $( d, \downarrow , \mathrm{G} )$ & $( d, \downarrow , \mathrm{B} )$
        \\
        \multicolumn{4}{c}{RGB cyclic}
        \\
        \hline
        \\
        \multicolumn{4}{c}{$\Delta ^{-} (s = 3/2 )$}
        \\
        $w_{\vec{\mu}}$ & $\mu _{1}$ & $\mu _{2}$ & $\mu _{3}$
        \\
        \hline
        $1$ & $( d , \uparrow , \mathrm{R} )$
        & $( d , \uparrow , \mathrm{G} )$ & $( d , \uparrow , \mathrm{B} )$
        \\
        \hline
        \\
        \multicolumn{4}{c}{$\Delta ^{-} (s = 1/2 )$}
        \\
        $w_{\vec{\mu}}$ & $\mu _{1}$ & $\mu _{2}$ & $\mu _{3}$
        \\
        \hline
        $1/\sqrt{3}$ & $( d , \uparrow , \mathrm{R} )$
        & $( d , \uparrow , \mathrm{G} )$ & $( d , \downarrow , \mathrm{B} )$
        \\
        \multicolumn{4}{c}{RGB cyclic}
        \\
        \hline
        \\
        \multicolumn{4}{c}{$\Delta ^{-} (s = -1/2 )$}
        \\
        $w_{\vec{\mu}}$ & $\mu _{1}$ & $\mu _{2}$ & $\mu _{3}$
        \\
        \hline
        $1/\sqrt{3}$ & $( d , \uparrow , \mathrm{R} )$
        & $( d, \downarrow , \mathrm{G} )$ & $( d, \downarrow , \mathrm{B} )$
        \\
        \multicolumn{4}{c}{RGB cyclic}
        \\
        \hline
        \\
        \multicolumn{4}{c}{$\Delta ^{-} (s = - 3/2 )$}
        \\
        $w_{\vec{\mu}}$ & $\mu _{1}$ & $\mu _{2}$ & $\mu _{3}$
        \\
        \hline
        $1$ & $( d , \downarrow , \mathrm{R} )$
        & $( d, \downarrow , \mathrm{G} )$ & $( d, \downarrow , \mathrm{B} )$
        \\
        \hline
        \\
        \multicolumn{4}{c}{$\Sigma ^{\ast +} (s = 3/2 )$}
        \\
        $w_{\vec{\mu}}$ & $\mu _{1}$ & $\mu _{2}$ & $\mu _{3}$
        \\
        \hline
        $1/\sqrt{3}$ & $( u , \uparrow , \mathrm{R} )$
        & $( u , \uparrow , \mathrm{G} )$ & $( s , \uparrow , \mathrm{B} )$
        \\
        \multicolumn{4}{c}{RGB cyclic}
        \\
        \hline
        \\
        \multicolumn{4}{c}{$\Sigma ^{\ast +} (s = 1/2 )$}
        \\
        $w_{\vec{\mu}}$ & $\mu _{1}$ & $\mu _{2}$ & $\mu _{3}$
        \\
        \hline
        $1/3$ & $( u , \uparrow , \mathrm{R} )$
        & $( u , \uparrow , \mathrm{G} )$ & $( s , \downarrow , \mathrm{B} )$
        \\
        $1/3$ & $( u , \uparrow , \mathrm{R} )$
        & $( u , \downarrow , \mathrm{G} )$ & $( s , \uparrow , \mathrm{B} )$
        \\
        $1/3$ & $( u , \downarrow , \mathrm{R} )$
        & $( u , \uparrow , \mathrm{G} )$ & $( s , \uparrow , \mathrm{B} )$
        \\
        \multicolumn{4}{c}{RGB cyclic}
        \\
        \hline
        \\
        \multicolumn{4}{c}{$\Sigma ^{\ast +} (s = -1/2 )$}
        \\
        $w_{\vec{\mu}}$ & $\mu _{1}$ & $\mu _{2}$ & $\mu _{3}$
        \\
        \hline
        $1/3$ & $( u , \uparrow , \mathrm{R} )$
        & $( u, \downarrow , \mathrm{G} )$ & $( s, \downarrow , \mathrm{B} )$
        \\
        $1/3$ & $( u , \downarrow , \mathrm{R} )$
        & $( u , \uparrow , \mathrm{G} )$ & $( s , \downarrow , \mathrm{B} )$
        \\
        $1/3$ & $( u , \downarrow , \mathrm{R} )$
        & $( u , \downarrow , \mathrm{G} )$ & $( s, \uparrow , \mathrm{B} )$
        \\
        \multicolumn{4}{c}{RGB cyclic}
        \\
        \hline
        \\
        \multicolumn{4}{c}{$\Sigma ^{\ast +} (s = - 3/2 )$}
        \\
        $w_{\vec{\mu}}$ & $\mu _{1}$ & $\mu _{2}$ & $\mu _{3}$
        \\
        \hline
        $1/\sqrt{3}$ & $( u , \downarrow , \mathrm{R} )$
        & $( u, \downarrow , \mathrm{G} )$ & $( s, \downarrow , \mathrm{B} )$
        \\
        \multicolumn{4}{c}{RGB cyclic}
        \\
        \hline
        %
        \\
        \\
        \\
      \end{tabular}
    \end{minipage}
    &
    \begin{minipage}[t]{0.33\hsize}
      \begin{tabular}{lccc}
        \multicolumn{4}{c}{$\Sigma ^{\ast 0} (s = 3/2 )$}
        \\
        $w_{\vec{\mu}}$ & $\mu _{1}$ & $\mu _{2}$ & $\mu _{3}$
        \\
        \hline
        $1/\sqrt{6}$ & $( u , \uparrow , \mathrm{R} )$
        & $( d, \uparrow , \mathrm{G} )$ & $( s, \uparrow , \mathrm{B} )$
        \\
        $- 1/\sqrt{6}$ & $( u , \uparrow , \mathrm{G} )$
        & $( d, \uparrow , \mathrm{R} )$ & $( s, \uparrow , \mathrm{B} )$
        \\
        \multicolumn{4}{c}{RGB cyclic}
        \\
        \hline
        \\
        \multicolumn{4}{c}{$\Sigma ^{\ast 0} (s = 1/2 )$}
        \\
        $w_{\vec{\mu}}$ & $\mu _{1}$ & $\mu _{2}$ & $\mu _{3}$
        \\
        \hline
        $1/3 \sqrt{2}$ & $( u , \uparrow , \mathrm{R} )$
        & $( d, \uparrow , \mathrm{G} )$ & $( s, \downarrow , \mathrm{B} )$
        \\
        $1/3 \sqrt{2}$ & $( u , \uparrow , \mathrm{R} )$
        & $( d, \downarrow , \mathrm{G} )$ & $( s, \uparrow , \mathrm{B} )$
        \\
        $1/3 \sqrt{2}$ & $( u , \downarrow , \mathrm{R} )$
        & $( d, \uparrow , \mathrm{G} )$ & $( s, \uparrow , \mathrm{B} )$
        \\
        $- 1/3 \sqrt{2}$ & $( u , \uparrow , \mathrm{G} )$
        & $( d, \uparrow , \mathrm{R} )$ & $( s, \downarrow , \mathrm{B} )$
        \\
        $- 1/3 \sqrt{2}$ & $( u , \uparrow , \mathrm{G} )$
        & $( d, \downarrow , \mathrm{R} )$ & $( s, \uparrow , \mathrm{B} )$
        \\
        $- 1/3 \sqrt{2}$ & $( u , \downarrow , \mathrm{G} )$
        & $( d, \uparrow , \mathrm{R} )$ & $( s, \uparrow , \mathrm{B} )$
        \\
        \multicolumn{4}{c}{RGB cyclic}
        \\
        \hline
        \\
        \multicolumn{4}{c}{$\Sigma ^{\ast 0} (s = - 1/2 )$}
        \\
        $w_{\vec{\mu}}$ & $\mu _{1}$ & $\mu _{2}$ & $\mu _{3}$
        \\
        \hline
        $1/3 \sqrt{2}$ & $( u , \uparrow , \mathrm{R} )$
        & $( d, \downarrow , \mathrm{G} )$ & $( s, \downarrow , \mathrm{B} )$
        \\
        $1/3 \sqrt{2}$ & $( u , \downarrow , \mathrm{R} )$
        & $( d, \uparrow , \mathrm{G} )$ & $( s, \downarrow , \mathrm{B} )$
        \\
        $1/3 \sqrt{2}$ & $( u , \downarrow , \mathrm{R} )$
        & $( d, \downarrow , \mathrm{G} )$ & $( s, \uparrow , \mathrm{B} )$
        \\
        $- 1/3 \sqrt{2}$ & $( u , \uparrow , \mathrm{G} )$
        & $( d, \downarrow , \mathrm{R} )$ & $( s, \downarrow , \mathrm{B} )$
        \\
        $- 1/3 \sqrt{2}$ & $( u , \downarrow , \mathrm{G} )$
        & $( d, \uparrow , \mathrm{R} )$ & $( s, \downarrow , \mathrm{B} )$
        \\
        $- 1/3 \sqrt{2}$ & $( u , \downarrow , \mathrm{G} )$
        & $( d, \downarrow , \mathrm{R} )$ & $( s, \uparrow , \mathrm{B} )$
        \\
        \multicolumn{4}{c}{RGB cyclic}
        \\
        \hline
        \\
        \multicolumn{4}{c}{$\Sigma ^{\ast 0} (s = - 3/2 )$}
        \\
        $w_{\vec{\mu}}$ & $\mu _{1}$ & $\mu _{2}$ & $\mu _{3}$
        \\
        \hline
        $1/\sqrt{6}$ & $( u , \downarrow , \mathrm{R} )$
        & $( d, \downarrow , \mathrm{G} )$ & $( s, \downarrow , \mathrm{B} )$
        \\
        $- 1/\sqrt{6}$ & $( u , \downarrow , \mathrm{G} )$
        & $( d, \downarrow , \mathrm{R} )$ & $( s, \downarrow , \mathrm{B} )$
        \\
        \multicolumn{4}{c}{RGB cyclic}
        \\
        \hline
        \\
        \multicolumn{4}{c}{$\Sigma ^{\ast -} (s = 3/2 )$}
        \\
        $w_{\vec{\mu}}$ & $\mu _{1}$ & $\mu _{2}$ & $\mu _{3}$
        \\
        \hline
        $1/\sqrt{3}$ & $( d , \uparrow , \mathrm{R} )$
        & $( d , \uparrow , \mathrm{G} )$ & $( s , \uparrow , \mathrm{B} )$
        \\
        \multicolumn{4}{c}{RGB cyclic}
        \\
        \hline
        \\
        \multicolumn{4}{c}{$\Sigma ^{\ast -} (s = 1/2 )$}
        \\
        $w_{\vec{\mu}}$ & $\mu _{1}$ & $\mu _{2}$ & $\mu _{3}$
        \\
        \hline
        $1/3$ & $( d , \uparrow , \mathrm{R} )$
        & $( d , \uparrow , \mathrm{G} )$ & $( s , \downarrow , \mathrm{B} )$
        \\
        $1/3$ & $( d , \uparrow , \mathrm{R} )$
        & $( d , \downarrow , \mathrm{G} )$ & $( s , \uparrow , \mathrm{B} )$
        \\
        $1/3$ & $( d , \downarrow , \mathrm{R} )$
        & $( d , \uparrow , \mathrm{G} )$ & $( s , \uparrow , \mathrm{B} )$
        \\
        \multicolumn{4}{c}{RGB cyclic}
        \\
        \hline
        \\
        \multicolumn{4}{c}{$\Sigma ^{\ast -} (s = -1/2 )$}
        \\
        $w_{\vec{\mu}}$ & $\mu _{1}$ & $\mu _{2}$ & $\mu _{3}$
        \\
        \hline
        $1/3$ & $( d , \uparrow , \mathrm{R} )$
        & $( d, \downarrow , \mathrm{G} )$ & $( s, \downarrow , \mathrm{B} )$
        \\
        $1/3$ & $( d , \downarrow , \mathrm{R} )$
        & $( d , \uparrow , \mathrm{G} )$ & $( s , \downarrow , \mathrm{B} )$
        \\
        $1/3$ & $( d , \downarrow , \mathrm{R} )$
        & $( d , \downarrow , \mathrm{G} )$ & $( s, \uparrow , \mathrm{B} )$
        \\
        \multicolumn{4}{c}{RGB cyclic}
        \\
        \hline
        \\
        \multicolumn{4}{c}{$\Sigma ^{\ast -} (s = - 3/2 )$}
        \\
        $w_{\vec{\mu}}$ & $\mu _{1}$ & $\mu _{2}$ & $\mu _{3}$
        \\
        \hline
        $1/\sqrt{3}$ & $( d , \downarrow , \mathrm{R} )$
        & $( d, \downarrow , \mathrm{G} )$ & $( s, \downarrow , \mathrm{B} )$
        \\
        \multicolumn{4}{c}{RGB cyclic}
        \\
        \hline
      \end{tabular}
    \end{minipage}
    \\
    \hline
  \end{tabular}
\end{table*}
\addtocounter{table}{-1}

\begin{table*}[!t]
  \caption{(continued)}
  \centering
  \begin{tabular}{ccc}
    \hline
    \begin{minipage}[t]{0.33\hsize} 
      \begin{tabular}{lccc}
        \\
        \multicolumn{4}{c}{$\Xi ^{\ast 0} (s = 3/2 )$}
        \\
        $w_{\vec{\mu}}$ & $\mu _{1}$ & $\mu _{2}$ & $\mu _{3}$
        \\
        \hline
        $1/\sqrt{3}$ & $( u , \uparrow , \mathrm{R} )$
        & $( s , \uparrow , \mathrm{G} )$ & $( s , \uparrow , \mathrm{B} )$
        \\
        \multicolumn{4}{c}{RGB cyclic}
        \\
        \hline
        \\
        \multicolumn{4}{c}{$\Xi ^{\ast 0} (s = 1/2 )$}
        \\
        $w_{\vec{\mu}}$ & $\mu _{1}$ & $\mu _{2}$ & $\mu _{3}$
        \\
        \hline
        $1/3$ & $( u , \uparrow , \mathrm{R} )$
        & $( s , \uparrow , \mathrm{G} )$ & $( s , \downarrow , \mathrm{B} )$
        \\
        $1/3$ & $( u , \uparrow , \mathrm{R} )$
        & $( s , \downarrow , \mathrm{G} )$ & $( s , \uparrow , \mathrm{B} )$
        \\
        $1/3$ & $( u , \downarrow , \mathrm{R} )$
        & $( s , \uparrow , \mathrm{G} )$ & $( s , \uparrow , \mathrm{B} )$
        \\
        \multicolumn{4}{c}{RGB cyclic}
        \\
        \hline
        \\
        \multicolumn{4}{c}{$\Xi ^{\ast 0} (s = -1/2 )$}
        \\
        $w_{\vec{\mu}}$ & $\mu _{1}$ & $\mu _{2}$ & $\mu _{3}$
        \\
        \hline
        $1/3$ & $( u , \uparrow , \mathrm{R} )$
        & $( s, \downarrow , \mathrm{G} )$ & $( s, \downarrow , \mathrm{B} )$
        \\
        $1/3$ & $( u , \downarrow , \mathrm{R} )$
        & $( s , \uparrow , \mathrm{G} )$ & $( s , \downarrow , \mathrm{B} )$
        \\
        $1/3$ & $( u , \downarrow , \mathrm{R} )$
        & $( s , \downarrow , \mathrm{G} )$ & $( s, \uparrow , \mathrm{B} )$
        \\
        \multicolumn{4}{c}{RGB cyclic}
        \\
        \hline
        \\
        \multicolumn{4}{c}{$\Xi ^{\ast 0} (s = - 3/2 )$}
        \\
        $w_{\vec{\mu}}$ & $\mu _{1}$ & $\mu _{2}$ & $\mu _{3}$
        \\
        \hline
        $1/\sqrt{3}$ & $( u , \downarrow , \mathrm{R} )$
        & $( s, \downarrow , \mathrm{G} )$ & $( s, \downarrow , \mathrm{B} )$
        \\
        \multicolumn{4}{c}{RGB cyclic}
        \\
        \hline
      \end{tabular}
    \end{minipage}
    &
    \begin{minipage}[t]{0.33\hsize} 
      \begin{tabular}{lccc}
        \\
        \multicolumn{4}{c}{$\Xi ^{\ast -} (s = 3/2 )$}
        \\
        $w_{\vec{\mu}}$ & $\mu _{1}$ & $\mu _{2}$ & $\mu _{3}$
        \\
        \hline
        $1/\sqrt{3}$ & $( d , \uparrow , \mathrm{R} )$
        & $( s , \uparrow , \mathrm{G} )$ & $( s , \uparrow , \mathrm{B} )$
        \\
        \multicolumn{4}{c}{RGB cyclic}
        \\
        \hline
        \\
        \multicolumn{4}{c}{$\Xi ^{\ast -} (s = 1/2 )$}
        \\
        $w_{\vec{\mu}}$ & $\mu _{1}$ & $\mu _{2}$ & $\mu _{3}$
        \\
        \hline
        $1/3$ & $( d , \uparrow , \mathrm{R} )$
        & $( s , \uparrow , \mathrm{G} )$ & $( s , \downarrow , \mathrm{B} )$
        \\
        $1/3$ & $( d , \uparrow , \mathrm{R} )$
        & $( s , \downarrow , \mathrm{G} )$ & $( s , \uparrow , \mathrm{B} )$
        \\
        $1/3$ & $( d , \downarrow , \mathrm{R} )$
        & $( s , \uparrow , \mathrm{G} )$ & $( s , \uparrow , \mathrm{B} )$
        \\
        \multicolumn{4}{c}{RGB cyclic}
        \\
        \hline
        \\
        \multicolumn{4}{c}{$\Xi ^{\ast -} (s = -1/2 )$}
        \\
        $w_{\vec{\mu}}$ & $\mu _{1}$ & $\mu _{2}$ & $\mu _{3}$
        \\
        \hline
        $1/3$ & $( d , \uparrow , \mathrm{R} )$
        & $( s, \downarrow , \mathrm{G} )$ & $( s, \downarrow , \mathrm{B} )$
        \\
        $1/3$ & $( d , \downarrow , \mathrm{R} )$
        & $( s , \uparrow , \mathrm{G} )$ & $( s , \downarrow , \mathrm{B} )$
        \\
        $1/3$ & $( d , \downarrow , \mathrm{R} )$
        & $( s , \downarrow , \mathrm{G} )$ & $( s, \uparrow , \mathrm{B} )$
        \\
        \multicolumn{4}{c}{RGB cyclic}
        \\
        \hline
        \\
        \multicolumn{4}{c}{$\Xi ^{\ast -} (s = - 3/2 )$}
        \\
        $w_{\vec{\mu}}$ & $\mu _{1}$ & $\mu _{2}$ & $\mu _{3}$
        \\
        \hline
        $1/\sqrt{3}$ & $( d , \downarrow , \mathrm{R} )$
        & $( s, \downarrow , \mathrm{G} )$ & $( s, \downarrow , \mathrm{B} )$
        \\
        \multicolumn{4}{c}{RGB cyclic}
        \\
        \hline
      \end{tabular}
    \end{minipage}
    &
    \begin{minipage}[t]{0.33\hsize} 
      \begin{tabular}{lccc}
        \\
        \multicolumn{4}{c}{$\Omega (s = 3/2 )$}
        \\
        $w_{\vec{\mu}}$ & $\mu _{1}$ & $\mu _{2}$ & $\mu _{3}$
        \\
        \hline
        $1$ & $( s , \uparrow , \mathrm{R} )$
        & $( s , \uparrow , \mathrm{G} )$ & $( s , \uparrow , \mathrm{B} )$
        \\
        \hline
        \\
        \multicolumn{4}{c}{$\Omega (s = 1/2 )$}
        \\
        $w_{\vec{\mu}}$ & $\mu _{1}$ & $\mu _{2}$ & $\mu _{3}$
        \\
        \hline
        $1/\sqrt{3}$ & $( s , \uparrow , \mathrm{R} )$
        & $( s , \uparrow , \mathrm{G} )$ & $( s , \downarrow , \mathrm{B} )$
        \\
        \multicolumn{4}{c}{RGB cyclic}
        \\
        \hline
        \\
        \multicolumn{4}{c}{$\Omega (s = -1/2 )$}
        \\
        $w_{\vec{\mu}}$ & $\mu _{1}$ & $\mu _{2}$ & $\mu _{3}$
        \\
        \hline
        $1/\sqrt{3}$ & $( s , \uparrow , \mathrm{R} )$
        & $( s, \downarrow , \mathrm{G} )$ & $( s, \downarrow , \mathrm{B} )$
        \\
        \multicolumn{4}{c}{RGB cyclic}
        \\
        \hline
        \\
        \multicolumn{4}{c}{$\Omega (s = - 3/2 )$}
        \\
        $w_{\vec{\mu}}$ & $\mu _{1}$ & $\mu _{2}$ & $\mu _{3}$
        \\
        \hline
        $1$ & $( s , \downarrow , \mathrm{R} )$
        & $( s, \downarrow , \mathrm{G} )$ & $( s, \downarrow , \mathrm{B} )$
        \\
        \hline
        \\
        \\
        \\
        \\
        \\
        \\
      \end{tabular}
    \end{minipage}
    \\
    \\
    \hline \hline
  \end{tabular}
\end{table*}

\end{document}